\pdfoutput=1
\documentclass[cernpreprint,atlasdraft=true,coverpage=false,texlive=2016,UKenglish,texmf]{atlasdoc}

\usepackage{atlaspackage}
\usepackage{siunitx}
 
\usepackage{atlasbiblatex}
 
\usepackage{atlasphysics}
 
\graphicspath{{logos/}{figures/}}
 
\usepackage{ANA-HIGG-2019-02-PAPER-defs}
\usepackage{PMGRefs-defs}

\AtlasTitle{Search for dark matter in events with missing transverse momentum and a Higgs
boson decaying into two photons in $pp$ collisions at $\sqrt{s} = 13$ TeV with the ATLAS detector}
 
\AtlasAbstract{
A search for dark-matter particles in events with large missing transverse momentum and a Higgs boson candidate decaying into two photons is reported.
The search uses $139$ fb$^{-1}$ of proton--proton collision data collected at $\sqrt{s}=13$ TeV with the ATLAS detector at the CERN LHC between
2015 and 2018. No significant excess of events over the Standard Model predictions is observed. The results are interpreted by extracting limits on three
simplified models that include  either vector or pseudoscalar mediators and predict a final state
with a pair of dark-matter candidates and a Higgs boson decaying into two photons.
}
 
\author{The ATLAS Collaboration}
 
\AtlasRefCode{HIGG-2019-02}
 
\PreprintIdNumber{CERN-EP-2021-012}
 
\AtlasJournalRef{JHEP 10 (2021) 13}
 
\AtlasDOI{10.1007/JHEP10(2021)013}

\AtlasJournal{JHEP}

\hypersetup{pdftitle={ATLAS document},pdfauthor={The ATLAS Collaboration}}
 
\begin{document}
 
\maketitle
 
\tableofcontents

\section{Introduction}
\label{sec:intro}
 
The discovery of a particle exhibiting the expected properties of a Standard
Model (SM) Higgs boson in 2012 by the ATLAS~\cite{Aad:2012tfa} and
CMS~\cite{Chatrchyan:2012xdj} Collaborations has opened up new possibilities in
searches for physics beyond the SM (BSM). The precision reached since then in the production
cross-section and mass measurements of the observed Higgs boson with the additional data collected by the LHC
experiments~\cite{PhysRevD.101.012002, cmscollaboration2021measurements} provides sharp tools to probe the possible existence of
new physics.
 
Astrophysical data~\cite{Zyla:2020zbs,Bertone:2004pz} support
the existence of dark matter (DM) in our universe, while there is yet no evidence of
a non-gravitational interaction between DM and SM particles, nor any indication of the
microscopic nature of any possible DM particles.
 
Assuming the weakly interacting nature of DM, DM
particles ($\chi$) are expected to escape detection at the LHC. For this reason,
searches concentrate on final states with missing transverse momentum ($\etmiss$)
produced in association with detectable particles ($X$) complementing the undetectable particles' signatures, giving rise to $X$+\met final states.
 
The detectable particle $X$ is usually chosen to be a photon~\cite{Aaboud:2017dor},
a $W$ or $Z$ boson~\cite{Aaboud:2018xdl}, a jet~\cite{Aaboud:2017phn}, a single top quark~\cite{atlas_tMET}, or a pair of top quarks~\cite{atlas_ttbarMET}, all emitted from a light quark or gluon as
initial-state radiation through the usual SM gauge interactions. However, $X$ can also be the neutral Higgs boson ($h$) observed at the LHC with a mass of about 125~\GeV\, whose radiative production is highly suppressed. In that case, $pp$ collision
events with $h$ produced in association with some \etmiss can be very sensitive probes
of the structure of the BSM physics responsible for producing DM~\cite{monoh}.
 
Both the ATLAS and CMS Collaborations have previously searched for such $h$+\met final states using $20.3~\mathrm{fb}^{-1}$
of $pp$ collision data at $\sqrt{s} =8$~\TeV~\cite{Aad:2015yga,Aad:2015dva}, and up to $36.1~\mathrm{fb}^{-1}$
of $pp$ collision data at $\sqrt{s}=13$~\TeV~\cite{Aaboud:2016obm,Sirunyan:2017hnk,Aaboud:2017uak,Aaboud:2017yyg,Sirunyan_2019,Sirunyan:2018fpy,Sirunyan_2020},
considering the decay of $h$ into a pair of photons, $b$-quarks, $\tau$ leptons, $W$ bosons, or $Z$ bosons.
In principle, the diphoton channel has lower sensitivity than the $b\bar{b}$ channel because the $h \rightarrow \gamma \gamma$ branching fraction is smaller than that of $h \rightarrow b\bar{b}$ by two orders of magnitude.
However, if a possible excess were to be seen in the $b$-quarks final state, the diphoton channel could offer a way to cross-check this evidence for a sizeable fraction of the new-physics
parameter space covered by the $b\bar{b}$ channel~\cite{Aaboud:2019yqu}. It also provides complementary sensitivity at lower \met where the $b\bar{b}$ final state is limited by the experimental
\met trigger threshold, since the diphoton channel is triggered using the photon pair, allowing for much lower and better resolved \met in the event.
This paper presents an updated search for DM particles produced in association with the decay of $h$ into a pair of photons using the full LHC Run~2 $pp$ collision dataset collected at $\sqrt{s}=13~\TeV$ by ATLAS from 2015 to 2018. This
corresponds to an integrated luminosity of $139~\mathrm{fb}^{-1}$, four times higher than that used in the previous analysis published by the ATLAS Collaboration~\cite{Aaboud:2017yyg}.
 
\begin{figure}[!htp]
\centering
\subfloat[]{\includegraphics[width=0.3\textwidth]{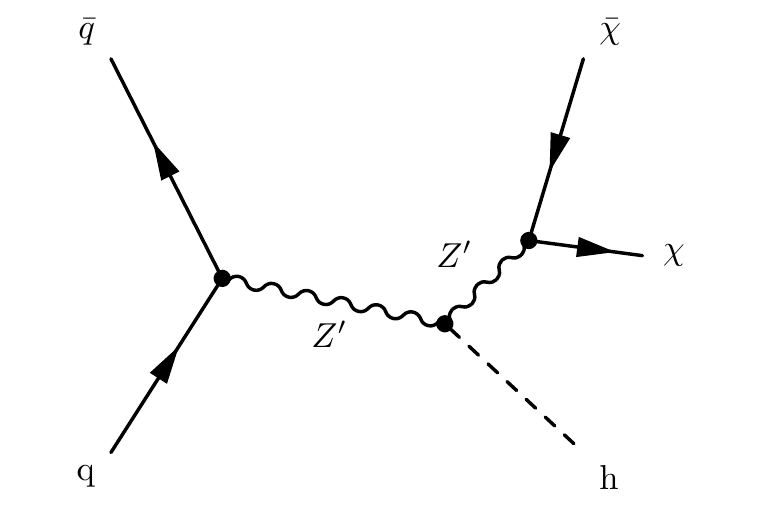}}
\subfloat[]{\includegraphics[width=0.3\textwidth]{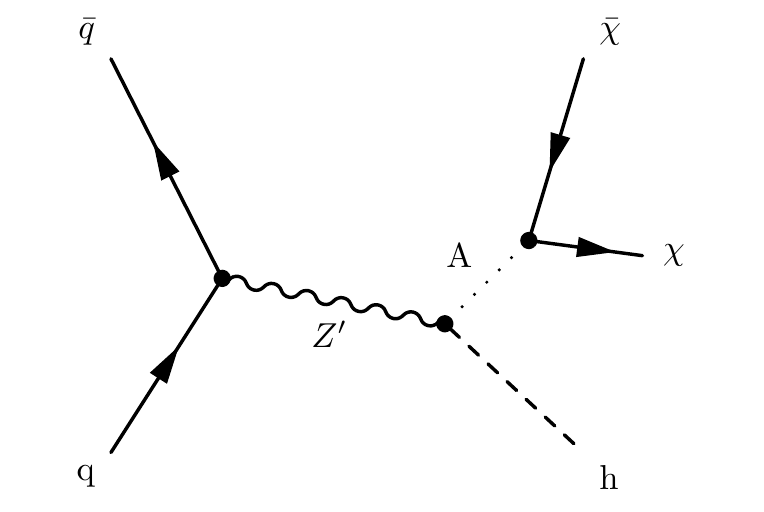}} \\
\subfloat[]{\includegraphics[width=0.3\textwidth]{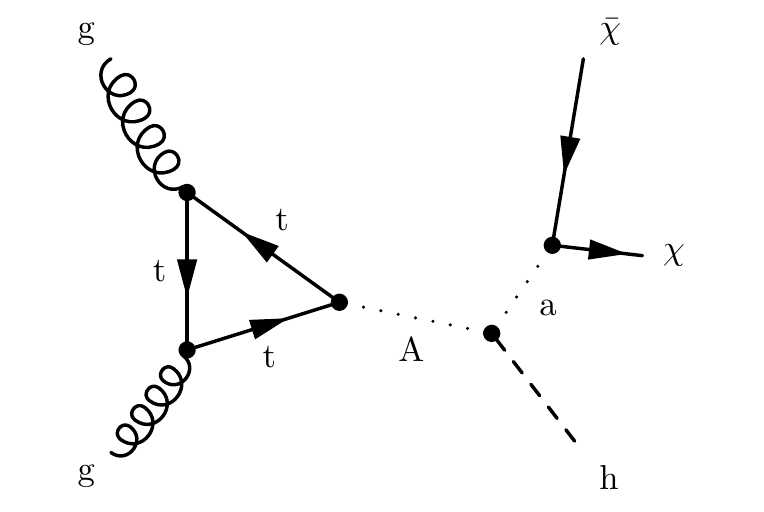} \label{feyn:2HDM-a-1}}
\subfloat[]{\includegraphics[width=0.3\textwidth]{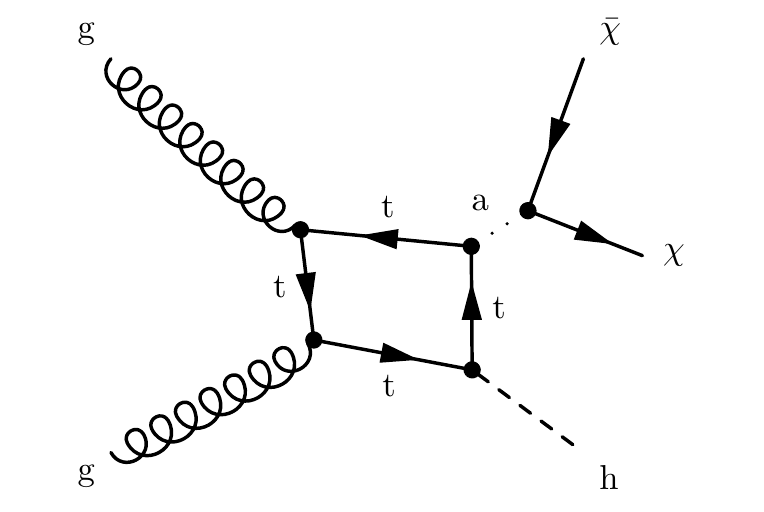} \label{feyn:2HDM-a-2}}
\subfloat[]{\includegraphics[width=0.3\textwidth]{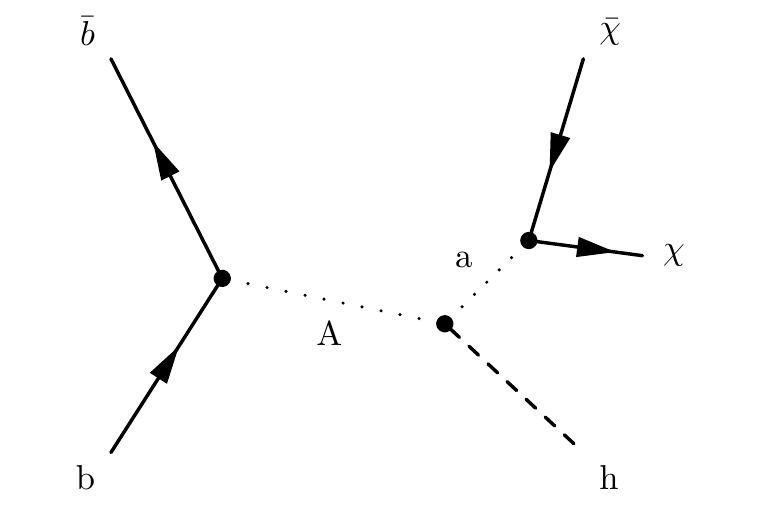}}
\caption{The Feynman diagrams of DM ($\chi$) production in association with the observed Higgs boson ($h$) arising from three theoretical models
considered in this paper: (a) the $Z'_{B}$ model, (b) the $Z'$-2HDM
model, and (c), (d), and (e) the 2HDM+a model.}
\label{fig:feyn}
\end{figure}
 
Three theoretical benchmark models are considered in this analysis. The leading-order (LO) Feynman diagrams
representing the production of $h$+\met in these three simplified models~\cite{lhcdmf} are shown in
Figure~\ref{fig:feyn}.
 
In the first model, called $Z'_{B}$~\cite{monoh}, a massive vector mediator $Z'$ emits a Higgs boson and subsequently decays into a pair of Dirac
fermionic DM candidates. DM couples to SM particles only via the new $Z'$ boson, and the new associated $U(1)_B$ baryon number symmetry ensures the stability of the DM particle in a natural way. An additional scalar particle (referred to as a baryonic Higgs boson) is introduced to break this symmetry spontaneously and generate the $Z'$ boson mass.
The parameters of the $Z'_B$ model are:
\begin{itemize}
\item $m_{Z'}$, the $Z'$ boson mass;
\item $g_\chi$, the coupling of the $Z'$ boson to the DM particle $\chi$;
\item $g_q$, the coupling of the $Z'$ boson to quarks;
\item $g_{hZ'Z'}$, the coupling between the $Z'$ boson and the observed Higgs boson $h$;
\item $\sin \theta$, the mixing angle between the baryonic Higgs boson and the observed Higgs boson; and
\item $m_{\chi}$, the mass of the fermionic dark-matter candidate $\chi$.
\end{itemize}
 
The second and third simplified models implement different possible mediators connecting the SM spectrum to DM particles but are both
derived from a general extension of the SM implementing two Higgs doublets, called 2HDM~\cite{z2hdm}. These models predict the existence
of five Higgs bosons: two scalars, one being the already observed Higgs boson; one heavy pseudoscalar $A$; and two charged Higgs bosons $H^{\pm}$.
The first of these two models is called the $Z'$-2HDM model~\cite{z2hdm}, as it introduces a new vector mediator $Z'$ whose mass is generated
by the extended Higgs sector. In this model, DM particles are produced through the decay of the pseudoscalar $A$, giving rise to an \met distribution
that becomes harder as the mass difference between the $Z'$ and $A$ bosons increases. The parameters of the $Z'$-2HDM model are:
\begin{itemize}
\item $m_{A}$, the pseudoscalar boson mass;
\item $m_{Z'}$, the $Z'$ boson mass;
\item $m_{\chi}$, the mass of the  fermionic dark-matter candidate $\chi$;
\item $\tan \beta$, the ratio of the vacuum expectation values of the two Higgs doublets;
\item $g_{Z'}$ the coupling strength of the $Z'$ boson to quarks; and
\item $\alpha$, the mixing angle between the two neutral scalars in the 2HDM model.
\end{itemize}
 
The third model considered in this analysis is called 2HDM+a \cite{Bauer2017} and offers another phenomenological option where a new pseudoscalar mediator $a$ couples directly to both the SM fermions and dark-matter particles~\cite{2018arXiv181009420A}. This model is of particular interest since it allows for gluon--gluon fusion production (Figures~\ref{feyn:2HDM-a-1},~\ref{feyn:2HDM-a-2}), which is forbidden for the $Z'$ mediator. This model has an extended set of 13 parameters, namely:
\begin{itemize}
\item $m_A$ and $m_a$, the pseudoscalar particle masses;
\item $m_H$ and $m_h$, the scalar particle masses;
\item $m_{H^\pm}$, the mass of the charged Higgs bosons;
\item $m_{\chi}$, the fermionic DM particle mass;
\item $y_{\chi}$, the DM Yukawa coupling;
\item $\tan \beta$, the ratio of the vacuum expectation values of the two Higgs doublets;
\item $\alpha$, the mixing angle between the two neutral scalars in the 2HDM model;
\item $\theta$, the mixing angle between the two pseudoscalars; and
\item $\lambda_3$, the quartic coupling of the Higgs potential, and $\lambda_{1P}$ and $\lambda_{2P}$, the quartic couplings of the pseudoscalar potentials.
\end{itemize}
 
By choosing $\cos(\alpha -\beta) = 0$ and a null coupling to quarks and leptons of the $a$ field, the 2HDM+a model can escape experimental constraints from both Higgs precision measurements and DM direct search limits.
 
For the particles mediating the interaction between SM and DM in these three models, it is assumed that only decays which are kinematically accessible and strictly necessary for the self-consistency of the model are considered in the decay widths~\cite{lhcdmf}. In the case of the 2HDM+a model, a SM-like Higgs width is assumed for $h$, which restricts the possible parameter space.
 
The analysis reported in the present paper selects events with two photons and large \met. Searches are performed in different regions of observed diphoton transverse momentum  (\pTgg) and event \met significance, defined in Section~\ref{sec:evtselection}.
 
The main backgrounds in the analysis correspond to either SM Higgs boson production contributions, QCD-induced non-resonant diphoton events ($\gamma\gamma$, $V\gamma\gamma$, where $V$ is a $W$ or $Z$ boson), or to reducible
contributions where an electron or a jet is misidentified as a photon (`fake photons') and \met is generated either by particles escaping the detector acceptance or by neutrinos ($V\gamma$, $\gamma$+jet). An additional background contribution dominating the low \met region is associated with resolution effects when computing the transverse energy from high-energy
objects and softer contributions measured in the ATLAS calorimeters. The \met can therefore be either `fake', i.e.\ spurious values of \met reconstructed in the detector for events with no invisible particles, or `true', i.e.\ genuine \met associated with the presence of particles escaping detection in the event.
 
This paper is organized as follows.
Section~\ref{sec:detector} gives a brief description of the ATLAS detector.
Section~\ref{sec:simulation} describes the dataset and the signal and background Monte Carlo (MC) simulation samples used.
Section~\ref{sec:reconstruction} explains the reconstruction and identification of objects, while
Section~\ref{sec:evtselection} outlines the optimization of the event selection and categorization.
Section~\ref{sec:backgroundest} summarizes the signal and background modelling.
Section~\ref{sec:effsys} discusses the experimental and theoretical systematic uncertainties.
Section~\ref{sec:result} presents the results and their interpretations, and finally
a summary is given in
Section~\ref{sec:summary}.

\section{ATLAS detector}
\label{sec:detector}
 
The ATLAS detector~\cite{Aad:2008zzm,ATLAS-TDR-2010-19,PIX-2018-001} is a multipurpose particle physics detector with approximately forward--backward symmetric cylindrical geometry.\footnote{ATLAS uses a right-handed coordinate system with its origin at the nominal interaction point (IP) in the centre of the detector and the $z$-axis along the beam pipe. The $x$-axis points from the IP to the centre of the LHC ring, and the $y$-axis points upward. Cylindrical coordinates $(r,\phi)$ are used in the transverse plane, $\phi$ being the azimuthal angle around the beam pipe. The pseudorapidity is defined in terms of the polar angle $\theta$ as $\eta=-\ln[\tan(\theta/2)]$. Angular distance is measured in units of \(\Delta R \equiv \sqrt{(\Delta\eta)^{2} + (\Delta\phi)^{2}}\). The photon transverse energy is $\et = E/ \cosh(\eta)$, where $E$ is its energy.} The inner detector (ID) tracking system covers $\abseta$ < 2.5 and consists of a silicon pixel detector, a silicon microstrip detector and a transition radiation tracker (TRT). The ID allows precise
reconstruction of charged-particle trajectories and of decay vertices of long-lived particles. The ID is surrounded by a thin superconducting solenoid providing a $2\:\mathrm{T}$ axial magnetic field. A high-granularity lead/liquid-argon (LAr) sampling calorimeter measures the energy and the position of electromagnetic showers in the central ($\left| \eta \right|<1.475$) and endcap ($1.375 < \left| \eta \right| < 3.2$) regions. It includes a presampler (for $|\eta|<1.8$) and three sampling layers for $|\eta|<2.5$. The longitudinal and lateral segmentation of the calorimeter allows a measurement of the shower direction without assuming that the photon originates from a specific point along the beamline. LAr sampling calorimeters with copper and tungsten absorbers are also used to measure hadronic showers in the endcap ($1.5<\left| \eta \right|<3.2$) and forward ($3.1<\left| \eta \right|<4.9$) regions, while a steel/scintillator-tile
calorimeter measures hadronic showers in the central region ($\left| \eta \right|<1.7$). The muon spectrometer surrounds the calorimeters and consists of three large superconducting air-core toroid magnets, each with eight coils, a system of precision tracking chambers ($\left| \eta \right|<2.7$), and fast tracking chambers for triggering ($|\eta|<2.4$). Reconstructed events are selected by a two-level trigger system. The first-level trigger is hardware-based, while the second-level trigger is implemented in software~\cite{Aaboud:2016leb}.


\section{Data and simulation samples}
\label{sec:simulation}
 
The $pp$ collision data used in the analysis correspond to the full LHC Run 2 dataset
taken by the ATLAS experiment during the period 2015 to 2018 with proton beams colliding at $\sqrt s =13$~\TeV.
The full dataset represents an integrated luminosity of $139.0 \pm 2.4$~fb$^{-1}$~\cite{LUCID2}
after the application of data quality requirements checking that the ATLAS detector was fully
functional and that the LHC was running in stable conditions.
This dataset was recorded with a mean number of about 34 $pp$ interactions per bunch crossing, with a peak value of 60.
Events used in this analysis were selected using a diphoton trigger requiring two reconstructed
photon candidates with minimum transverse energies of 35 and 25~\GeV\ for the leading
and subleading photons, respectively, where leading (subleading) refers to the photon with the highest (second-highest) transverse energy~\cite{Aad:2019wsl}.
Both photons were required by the trigger to fulfil the `Loose' photon identification criteria
in 2015 and 2016, whereas the `Medium' criteria were used in 2017--2018 to handle the higher number of
$pp$ collisions per bunch crossing resulting from the higher instantaneous luminosity in those years.
The trigger selections used during 2015--2018 are estimated to be fully efficient for events satisfying the
offline event selection criteria presented in Section~\ref{sec:evtselection}.
 
The analysis of the data sample requires the use of different types of MC generated
samples to design the event selection, categorize events, and estimate systematic uncertainties
applied to the statistical inference procedure used to estimate the presence of a possible DM signal
in the dataset.
These generated samples include the signal samples for the three models described in Section~\ref{sec:intro} and the background samples for both the non-resonant
contributions ($\gamma\gamma$, $V\gamma$, $V\gamma\gamma$) and the SM Higgs
contributions where the Higgs boson decays into a pair of photons. Due to difficulties in correctly simulating the effect of fake photons and fake \met, the non-resonant $\gamma$+jet background is instead estimated using a data-driven method detailed in Section~\ref{sec:backgroundest}. The simulated non-resonant samples and data-derived $\gamma$+jet sample are only used to calculate uncertainties affecting the background model used in the final statistical fit.
 
Events from gluon--gluon fusion (ggF), vector-boson fusion (VBF), $Wh$, and $Zh$ processes were generated using \POWHEGBOX{v2}~\cite{Hamilton:2013fea,Nason:2004rx,Frixione:2007vw,Alioli:2010xd,Nason:2009ai,Mimasu:2015nqa,Campbell:2012am,Luisoni:2013kna,Hartanto:2015uka} with the \pdflhc\ parton distribution function (PDF) set~\cite{Butterworth:2015oua}, and interfaced to \PYTHIA{8.2}~\cite{Sjostrand:2014zea} for parton showering, hadronization and the underlying event, using a set of parameters tuned to data and called the AZNLO tune~\cite{Aad:2014xaa}.
The \ggF\ simulation achieves next-to-next-to-leading-order (NNLO) accuracy for arbitrary inclusive $gg\to h$ observables by reweighting the Higgs boson rapidity spectrum in {\textsc Hj-MiNLO}~\cite{Hamilton:2012np,Campbell:2012am,Hamilton:2012rf} to that of HNNLO~\cite{Catani:2007vq}. The transverse momentum spectrum of the Higgs boson obtained with this sample is found to be compatible with the fixed-order HNNLO calculation and the \hres calculation~\cite{Bozzi:2005wk,deFlorian:2011xf} performing resummation at next-to-next-to-leading-logarithm accuracy matched to a NNLO fixed-order calculation (NNLL+NNLO).

Events from $t\overline{t}h$ processes were modelled using the
\POWHEGBOX{v2}~\cite{Frixione:2007nw,Nason:2004rx,Frixione:2007vw,Alioli:2010xd,Hartanto:2015uka}
generator, which provides matrix elements at next-to-leading order (NLO) in the strong coupling
constant \alphas in the five-flavour scheme with the \nnpdfnlo~\cite{Ball:2014uwa} PDF set.
The functional form of the renormalization and factorization scales is
set to $\sqrt[3]{m_\text{T}(t)\cdot m_\text{T}(\bar{t}) \cdot m_\text{T}(h)}$.\footnote{$m_\text{T}$ denotes the transverse mass of a particle, defined as $m_\text{T}=\sqrt{m^2 + p_\text{T}^2}$ where $m$ and $p_\text{T}$ are respectively its mass and transverse momentum.}
The generator was interfaced to \PYTHIA{8.2}
using the A14 tune~\cite{ATL-PHYS-PUB-2014-021} and the
\nnpdftwo~\cite{Ball:2014uwa} PDF set. The decays of bottom and charm hadrons were simulated using the \evtgen~v1.6.0 program~\cite{EvtGen}.

Events from $thqb$ ($tWh$) processes were produced with \MGNLO\ in the four-flavour (five-flavour) scheme with the \nnpdfnnlo  PDF~\cite{Ball:2014uwa}. The same flavour scheme is used in the matrix element calculation and the PDF. The top quark and $W$ boson decays were handled by MadSpin~\cite{Artoisenet:2012st} for the correct treatment of the spin correlations of the decay products. In the case of $tWh$ the overlap of this process with $t\overline{t}h$ at NLO was removed following a diagram removal technique~\cite{Frixione:2008yi,Demartin:2016axk}. The simulation of the parton shower, hadronization and underlying event was then performed by \PYTHIA{8.2} with the A14 tune for both the $thqb$ and $tWh$ samples.
 
The cross sections for the SM Higgs boson processes were calculated at NLO in electroweak theory and
NNLO in QCD for the VBF, $Zh$ and $Wh$ samples~\cite{deFlorian:2016spz}, and next-to-next-to-next-to-leading order
plus NNLL (N$^{3}$LO+NNLL) in QCD for the ggF sample~\cite{deFlorian:2016spz}.
The $t\overline{t}h$ cross section was calculated with NLO accuracy in QCD with NLO electroweak corrections~\cite{Beenakker:2002nc}.
The $b\overline{b}h$ cross section was found to be negligible. All samples were normalized to the most precise available theoretical cross sections corresponding to a Higgs boson mass of 125.09~\GeV~\cite{HIGG-2014-14}. The analysis assumes a branching ratio for the Higgs boson decay into two photons
of 0.227\%~\cite{deFlorian:2016spz}.
 
The $\gamma\gamma$, $V\gamma\gamma$, and $V\gamma$ processes were simulated with the \sherpa~v2.2.4~\cite{Bothmann:2019yzt} generator. For the $\gamma\gamma$ and $V\gamma$ processes, QCD NLO-accurate matrix elements for up to one parton, and LO-accurate matrix elements for up to three partons, were calculated with the Comix~\cite{Gleisberg:2008fv} and \openloops~\cite{Buccioni:2019sur,Cascioli:2011va,Denner:2016kdg} libraries, while for the $V\gamma\gamma$ process, QCD LO-accurate matrix elements for up to one additional parton emission were calculated. They were matched and merged with the \sherpa parton shower based on Catani--Seymour dipole factorization~\cite{Gleisberg:2008fv,Schumann:2007mg} using the MEPS@NLO prescription~\cite{Hoeche:2011fd,Hoeche:2012yf,Catani:2001cc,Hoeche:2009rj}. Samples were generated using the \nnpdfnnlo PDF set, along with the dedicated set of tuned parton-shower parameters developed by the \sherpa authors.
 
MC simulated samples for the different signal models were generated using the $\MADGRAPH5$~\cite{Alwall:2014hca} generator
at LO accuracy, using the \nnpdflo PDF set~\cite{Ball:2012cx}
for the $Z'_{B}$ and $Z'$-2HDM signal samples or the \nnpdfnlo PDF set for 2HDM+a signal samples.
Parton showering and hadronization were simulated using the \PYTHIA{8.1}~\cite{pythia8} generator for $Z'_{B}$ or the \PYTHIA{8.2} generator for $Z'$-2HDM and 2HDM+a samples, with
the A14 tune and the \nnpdftwo PDF set~\cite{Ball:2012cx}. Multiple samples were generated in order to scan the mediator masses and the key parameters of each model, while the default values of other parameters were set to fixed values. In general the choice of parameter values follows the recommendations of the LHC DM Forum report~\cite{lhcdmf} and are based on the sensitivities expected at the LHC.
 
For the $Z'_B$ model, the generation scans a wide range of $Z'$ and DM particle masses. The mass $m_{Z'}$ ranges from $10$~\GeV\ to $2000$~\GeV\ while $m_{\chi}$ ranges from $1$~\GeV\ to $1000$~\GeV. The different couplings of this model are required
to fulfil perturbativity bounds that allow the choice of $g_{\chi} = 1.0$, $g_{q} = 1/3$, $g_{hZ'Z'} = m_{Z'}$, and $\sin \theta = 0.3$ to maximize the expected cross section. The couplings choice affects only the magnitude of the cross section and not the shape of the \etmiss and $\pTgg$ distributions, and in particular, a lower $\sin \theta$ value decreases the expected yield. In this model, $\pTgg$  and \etmiss grow with $m_{Z'}$ while the diphoton pair becomes more back-to-back with the
\etmiss vector. Additionally, the photon pair is more collimated for higher $m_{Z'}$ and $m_\chi$, since the Higgs boson is more boosted in that region of the parameter space.
 
For the $Z'$-2HDM model, the cross section and kinematics depend on the $Z'$ and $A$ masses but much less on the DM particle mass while $m_{\chi} < m_{A}/2$ is satisfied.
Consequently, signal samples were generated for different values of $m_{Z'} \in [400, 1600]$~\GeV\ and $m_{A} \in [200, 600]$~\GeV\ and for $m_{\chi} = 100$~\GeV.
The values of other parameters which only affect the cross section were fixed to $\alpha = \beta - \pi/2$, $g_{Z'} = 0.8$, and $\tan \beta = 1$. In this model, \etmiss exhibits a peaked distribution
arising from the fact that DM particles are produced from the pseudoscalar $A$. The peak position depends strongly on the mass difference $m_{Z'}-m_{A}$.
 
For the 2HDM+a model, the parameter space is more complex than for the two first models and different scans were generated to test its richer
phenomenology. This includes two $\sin \theta$ scans corresponding to two sets of ($A$, $a$) masses, a two-dimensional scan of the $m_A$--$m_a$ mass plane, and a two-dimensional scan of the $\tan \beta$--$m_a$ plane. The default parameter values were set to $\sin(\alpha -\beta) = 1$, $\lambda_3 = \lambda_{1P} = \lambda_{2P} = 3$, $y_{\chi} = 1$, $m_{\chi} = 10$~\GeV, and a decoupled Higgs boson spectrum with $m_{A} = m_H = m_{H^\pm}$ sizeably larger than that of the observed scalar Higgs boson with $m_h$ near $125$~\GeV.
 
The effects of additional $pp$ collisions from the same and neighbouring bunch crossings (`pile-up') were simulated by overlaying the hard-scattering event with inelastic $pp$ events generated by \PYTHIA{8.1} using the \nnpdftwo PDF set and the
A3 tune~\cite{ATL-PHYS-PUB-2016-017}. Differences between the simulated and observed distributions of the number of interactions per bunch crossing were removed by applying pile-up weights to simulated events. A full simulation of the ATLAS detector~\cite{SOFT-2010-01} based on \textsc{Geant4}~\cite{Agostinelli:2002hh} was used to reproduce the detector response to SM Higgs boson processes and $V\gamma$ and $V\gamma\gamma$ backgrounds. The $\gamma\gamma$ background and the signal samples were simulated using \atlfasttwo~\cite{ATLAS:1300517}, a fast simulation of the ATLAS detector response which was shown to be able to accurately simulate diphoton events.

As it is not computationally feasible to generate fully simulated samples for all interesting parameter points, an interpolation method was used to efficiently create new samples from existing fully simulated ones. First, a set of base samples covering a wide range of generator-level `truth' values for key variables was chosen from the existing fully simulated samples. For the $Z'_B$ and $Z'$-2HDM models, these variables are $\pTgg$ and \met, while for the 2HDM+a model, these are $\pTgg$, $p^{\chi\chi}_\mathrm{T}$ (the transverse momentum of the $\chi\chi$ system), and $|\Delta\phi(\pTgg,p^{\chi\chi}_\mathrm{T})|$ (the difference in azimuthal angle between the $\gamma\gamma$ and $\chi\chi$ systems). For any desired additional parameter point, a new sample was generated containing only these generator-level key variables. The base samples were then reweighted on an event-by-event basis to the desired new one, using the bin-by-bin ratio of the generator-level distributions from the new sample to the corresponding distribution from the base samples. Finally, for a few validation samples, the differences between the generated and reweighted samples were used to estimate a conservative uncertainty from this procedure.

\section{Event reconstruction}
\label{sec:reconstruction}
 
Within the ATLAS detector, photons are reconstructed from topologically connected clusters~\cite{ATL-PHYS-PUB-2017-022} of energy deposits in the EM calorimeters in the region $|\eta| < 2.37$.
The transition region between the barrel and endcap EM calorimeters, $1.37 < |\eta| < 1.52$, is excluded.
Photon candidates matched to a conversion vertex or a track, which are consistent with originating from a photon conversion, are classified as converted photons.
Those without a matched conversion vertex or track are classified as unconverted photons.
The efficiency of the diphoton trigger used to select the events used in the present analysis is estimated to be greater than 99.2\% on average for events passing the final event selection~\cite{Aad:2019wsl}.
 
The calibration of the photon energy is based on a multivariate regression algorithm trained with MC samples, where the input variables are corrected with data-driven techniques. The calibrated energy is then adjusted by applying energy scale factors derived from
$Z \rightarrow e^+ e^-$ events~\cite{EGAM-2018-01}.
The photon direction is reconstructed using the longitudinal segmentation of the calorimeters and a constraint from the average collision point of the proton beams.
Additionally, the conversion vertex position is included in the case of converted photons.
 
Events are required to have at least one reconstructed primary vertex (PV), defined as a vertex associated with at least two tracks with $\pt > 0.5$~\GeV.
To select the correct PV of a given event, a neural network~\cite{HIGG-2013-08} that uses the pointing information from the selected photons is deployed.
Although this neural-network-selected vertex is taken to be the nominally correct choice, an alternative `hardest vertex', defined as the vertex with the highest sum of squares of the transverse momenta of associated tracks, is also used for certain \etmiss-related calculations described in Section~\ref{sec:evtselection}.
 
Photon identification is based on the lateral shower profile of the energy deposits in the first and second EM calorimeter layers and on the
energy leakage fraction in the hadronic calorimeter. `Tight' identification criteria~\cite{EGAM-2018-01} are applied, after tuning them for converted and unconverted photons separately.
These criteria reduce the misidentification of hadronic jets containing large neutral components, primarily $\pi^0$ particles, which decay into highly collimated photons.
For transverse momenta between $30~\GeV$ and $250~\GeV$, the identification efficiency for unconverted and converted photons ranges from 85\% to 99\%, while fake photons originating from jets have an identification efficiency of 25\% to 40\%~\cite{EGAM-2018-01}.

To further improve fake-photon rejection, two isolation variables are defined to quantify the activity around a photon.
The calorimeter-based isolation $E^{\textrm{iso}}_{\textrm{T}}$ is defined as the sum of the transverse energy in topological clusters of calorimeter cells  within a cone of size $\Delta R = 0.2$ around the photon, correcting for the energy of the photon candidate itself as well as an average expected pile-up contribution.
The track-based isolation $p_{\textrm{T}}^{\textrm{iso}}$ is defined as the scalar sum of the transverse momenta of all tracks with $\pT > 1$~\GeV\ that originate from the neural-network-selected vertex and are within a cone of $\Delta R = 0.2$. Isolated photons must have $E^{\textrm{iso}}_{\textrm{T}} < 0.065\;E_\mathrm{T}$ and $p_{\textrm{T}}^{\textrm{iso}} < 0.05\;E_\mathrm{T}$.
 
Electrons are reconstructed from energy deposits measured in the EM calorimeter which are matched to ID tracks~\cite{EGAM-2018-01}.
They are required to satisfy $|\eta| < 2.47$, excluding the EM calorimeter transition region $1.37 < |\eta| < 1.52$, and have a transverse momentum $\pT > 10$~\GeV.
Electrons are required to satisfy the `Medium' identification criteria based on the use of shower shape, track--cluster matching and TRT parameters~\cite{PERF-2017-01} in a likelihood-based algorithm.
Muons are reconstructed from high-quality track segments found in the muon spectrometer~\cite{Aad:2016jkr}.
A matching of these segments to ID tracks is required in the region $|\eta| < 2.5$. Muons are required to have $|\eta| < 2.7$ and $\pT > 10$~\GeV, and to satisfy the `Medium' identification criteria~\cite{Aad:2016jkr}.
Both the electrons and muons are matched to the PV via requirements on the tracks' longitudinal and transverse impact parameters, $|z_{0}|$ and $|d_{0}|$.
The applied requirements are $|z_{0}| \sin \theta < 0.5$~mm (where $\theta$ is the polar angle of the track) for electrons and muons and  $|d_{0}|/\sigma_{d_{0}} < 5 (3)$ for electrons (muons).
 
Jets are reconstructed by a particle-flow algorithm~\cite{PERF-2015-09} using noise-suppressed positive-energy topological clusters in the calorimeter~\cite{PERF-2014-07} which are formed by the anti-$\kt$ algorithm~\cite{Cacciari:2008gp} with a radius parameter of $R=0.4$.
They are required to have $|\eta|<4.4$ and $\pt>25~\gev$.
To further suppress jets produced in pile-up interactions, each jet within the tracking acceptance, i.e.\ $|\eta|<2.4$, and with $\pT < 60~\GeV$, is required to satisfy jet vertex tagger (JVT)~\cite{JVT} criteria used to identify the jet as originating from the PV.
 
To resolve ambiguities between photon, electron, muon and jet reconstruction, an overlap removal procedure is applied to avoid multiple usage of the same detector signals in the same event.
The prioritization of photons in this analysis requires the removal of electrons, muons and jets within $\Delta R = 0.4$ of a selected photon.
Next, jets within $\Delta R = 0.2$ of electrons are removed.
In the last step, electrons and muons within $\Delta R = 0.4$ of any jet are removed.
 
The missing transverse momentum \etmiss is calculated as the magnitude of the negative vectorial sum of the transverse momenta of all selected and calibrated physics objects of an event that can be matched to the PV. A so-called `soft term' is calculated from the residual tracks that originate from the PV but are not associated with any other object and is added to the \etmiss~\cite{Aaboud:2018tkc}.

\section{Event selection}
\label{sec:evtselection}
 
Events are required to have at least two photon candidates within a fiducial region of the EM calorimeter defined by $|\eta| < 2.37$, excluding the region of $1.37<|\eta|<1.52$. Photon candidates in this fiducial region are ordered
according to their $E_\mathrm{T}$ and only the highest two are considered. These leading and subleading photon candidates must have $E_\mathrm{T}^{\gamma}/m_{\gamma\gamma} > 0.35$ and 0.25, respectively, where $m_{\gamma\gamma}$ is the invariant mass of the two selected photons.
Furthermore, events are required to have $105 < m_{\gamma\gamma} < 160$~\GeV. The data sideband is defined to consist of events in this region but excluding the region $120 < m_{\gamma\gamma} < 130$~\GeV.
 
\begin{figure}[ht]
\centering
\includegraphics[width=0.48\textwidth]{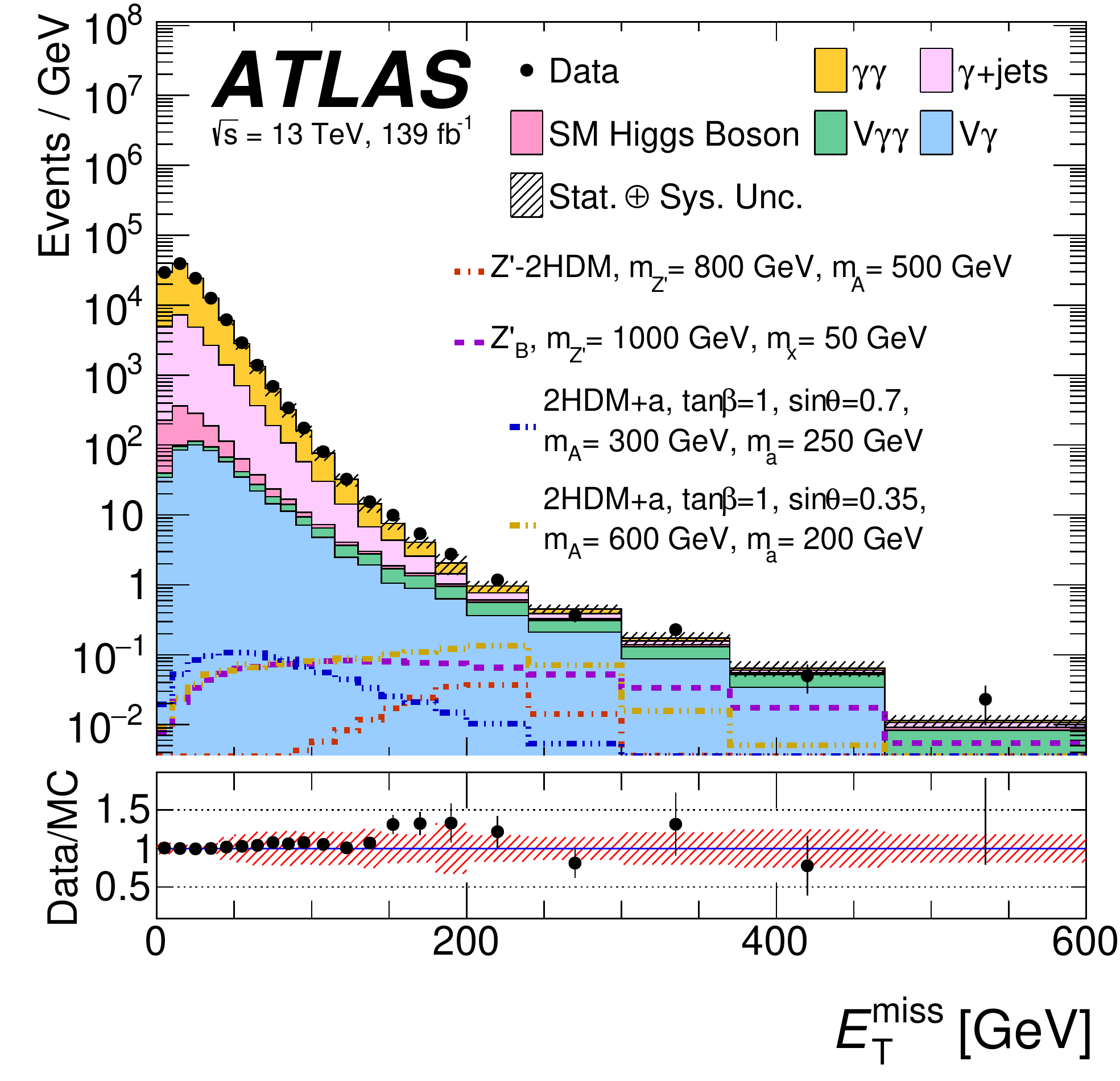}
\caption{Comparison of the \etmiss distribution in MC simulation and data, shown after the photon requirements but before the preselection requirements prior to BDT training. In particular, events are required to have $105 < m_{\gamma\gamma} < 160$~\GeV. In addition to the signal used for training (blue), one representative signal is also shown for each of the three signal models. The $Z'$-2HDM, $Z'_B$, 2HDM+a ($\tan\beta=1, \sin\theta = 0.7, m_A = 300\ \GeV, m_a = 250\ \GeV$), and 2HDM+a ($\tan\beta=1, \sin\theta = 0.35, m_A = 600\ \GeV, m_a = 200\ \GeV$) signal models are normalized to their respective theoretical cross sections times branching ratios of 0.0815 fb, 0.411 fb, 0.269 fb, and 0.533 fb. The lower panel shows the ratio of data to MC. The uncertainty bands indicate the sum of statistical and experimental systematic uncertainties.}
\label{fig:met}
\end{figure}

In the DM production models, the Higgs boson recoils against the DM pair, resulting in large \etmiss\ in the event and large \pT of the diphoton system, denoted by \pTgg.
The \etmiss distribution after the aforementioned photon requirements is shown in Figure~\ref{fig:met}.
 
After the photon requirements, events are required to pass a preselection of $\etmiss > 90$~\GeV\ and $\detmiss < 30$~\GeV, where \detmiss is the \etmiss calculated from the vertex selected by the neural network minus the \etmiss calculated from the hardest vertex. High values of \detmiss indicate that the \etmiss can simply be attributed to the misidentification of the primary vertex. In addition, events with electrons or muons are vetoed to suppress \vy and \vyy backgrounds.
 
Following this preselection, a boosted decision tree (BDT) is trained using XGBoost 0.82~\cite{Chen:2016:XST:2939672.2939785} to discriminate between DM signals and the non-resonant diphoton background, using \pTgg and the \met significance \setmiss as input variables. This latter variable is defined as the ratio of the \etmiss to its expected resolution: $\setmiss = \etmiss / \sqrt{\sum E_\mathrm{T}}$, where the total transverse energy $\sum E_\mathrm{T}$ is calculated from the scalar sum of the transverse momenta of the calibrated photons, electrons, muons, jets, and the soft term used in the \etmiss calculation described in Section~\ref{sec:reconstruction}. \setmiss is used to better control the experimental uncertainties that affect the reconstruction of the different physical objects that enter the calculation of \etmiss and improves the impact of \met on the search sensitivity at low \etmiss. A study showed that very effective discriminating power between signal and backgrounds can be achieved by using only these two variables, considering both increases in sensitivity and in the size of systematic uncertainties associated with the use of additional variables.
 
\begin{figure}[ht]
\centering
\includegraphics[width=0.48\textwidth]{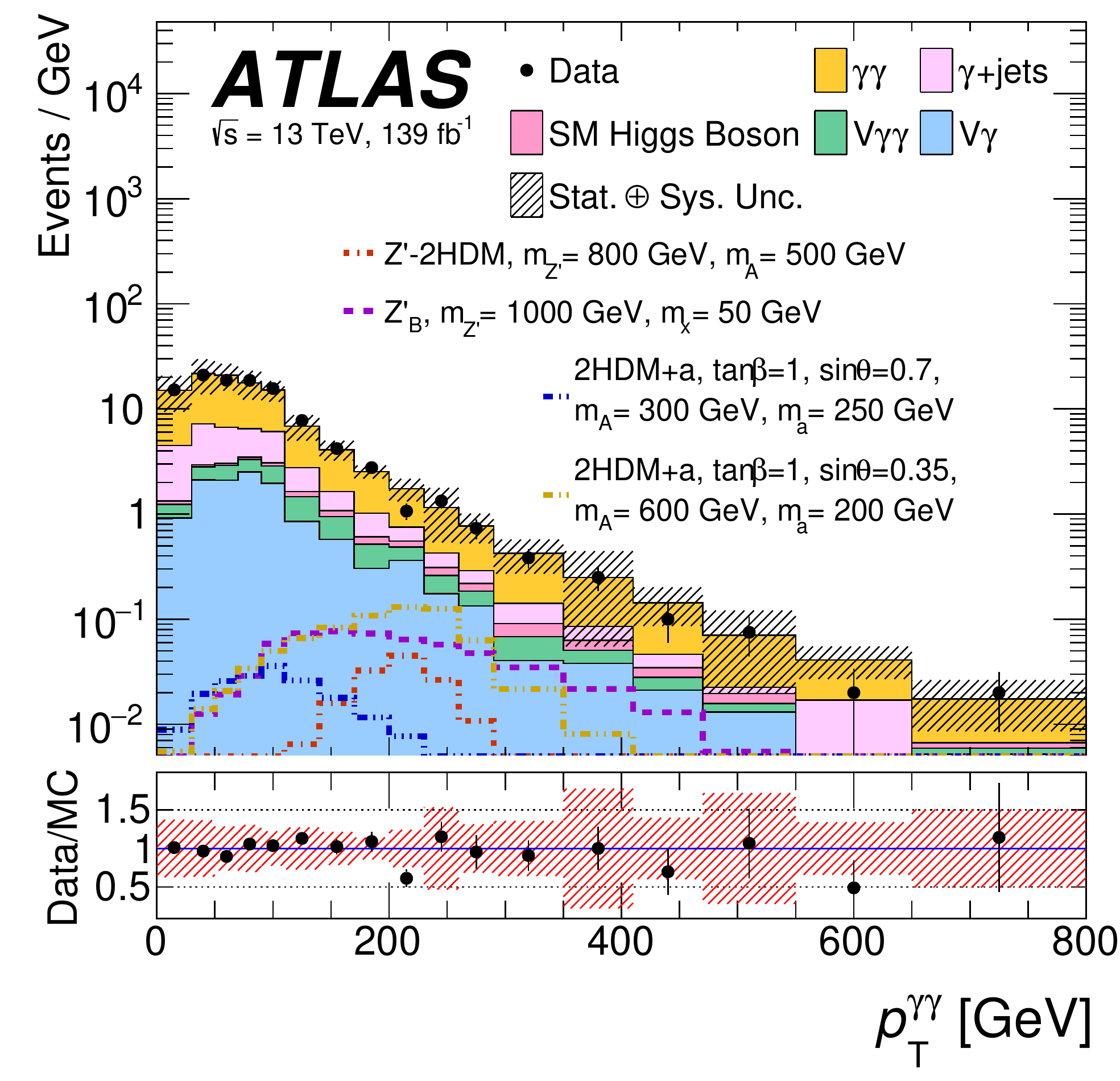}
\includegraphics[width=0.48\textwidth]{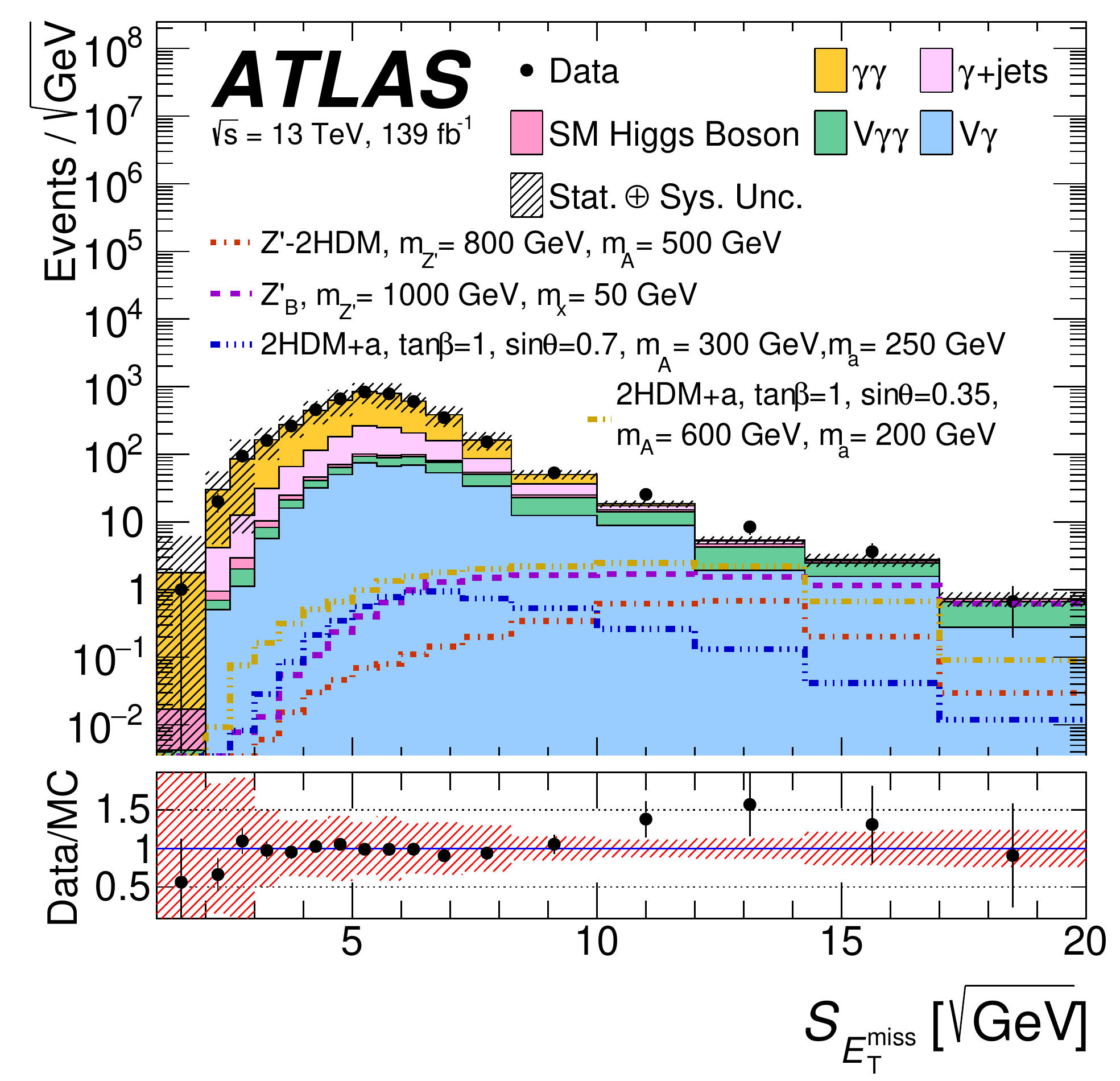}
\caption{Data and MC distributions of the BDT input variables, \pTgg and \setmiss, after the photon requirements and the preselection requirements prior to BDT training. In particular, events are required to have $105 < m_{\gamma\gamma} < 160$~\GeV. In addition to the signal used for training (blue), one representative signal is also shown for each of the three signal models. The $Z'$-2HDM, $Z'_B$, 2HDM+a ($\tan\beta=1, \sin\theta = 0.7, m_A = 300\ \GeV, m_a = 250\ \GeV$), and 2HDM+a ($\tan\beta=1, \sin\theta = 0.35, m_A = 600\ \GeV, m_a = 200\ \GeV$) signal models are normalized to their respective theoretical cross sections times branching ratios of 0.0815 fb, 0.411 fb, 0.269 fb, and 0.533 fb. The lower panel shows the ratio of data to MC. The narrow uncertainty bands (indicating the sum of statistical and experimental systematic uncertainties) at high \setmiss values are due to the absence of $\gamma\gamma$ and $\gamma$+jet contributions, which have large jet and \etmiss-related systematic uncertainties.}
\label{fig:bdt_input_vars}
\end{figure}

Figure \ref{fig:bdt_input_vars} shows how different signal models considered in the present search populate the phase space of these two crucial BDT training variables, after applying both the photon requirements and the preselection. Combined with the \etmiss distribution in Figure~\ref{fig:met}, these figures illustrate that high-\etmiss backgrounds are dominated by events with true \etmiss from $V\gamma$ and $V\gamma\gamma$ processes, while $\gamma\gamma$ and $\gamma$+jet backgrounds contribute significantly across the full diphoton candidate \pT range.  The sum of statistical and experimental systematic uncertainties (indicated by the shaded bands) is relatively flat as a function of \pTgg but decreases in the high \setmiss tail, due to the large jet and \etmiss-related systematic uncertainties associated with the fake \etmiss in the $\gamma\gamma$ and $\gamma$+jet components.
 
\begin{figure}[ht]
\centering
\includegraphics[width=0.68\textwidth]{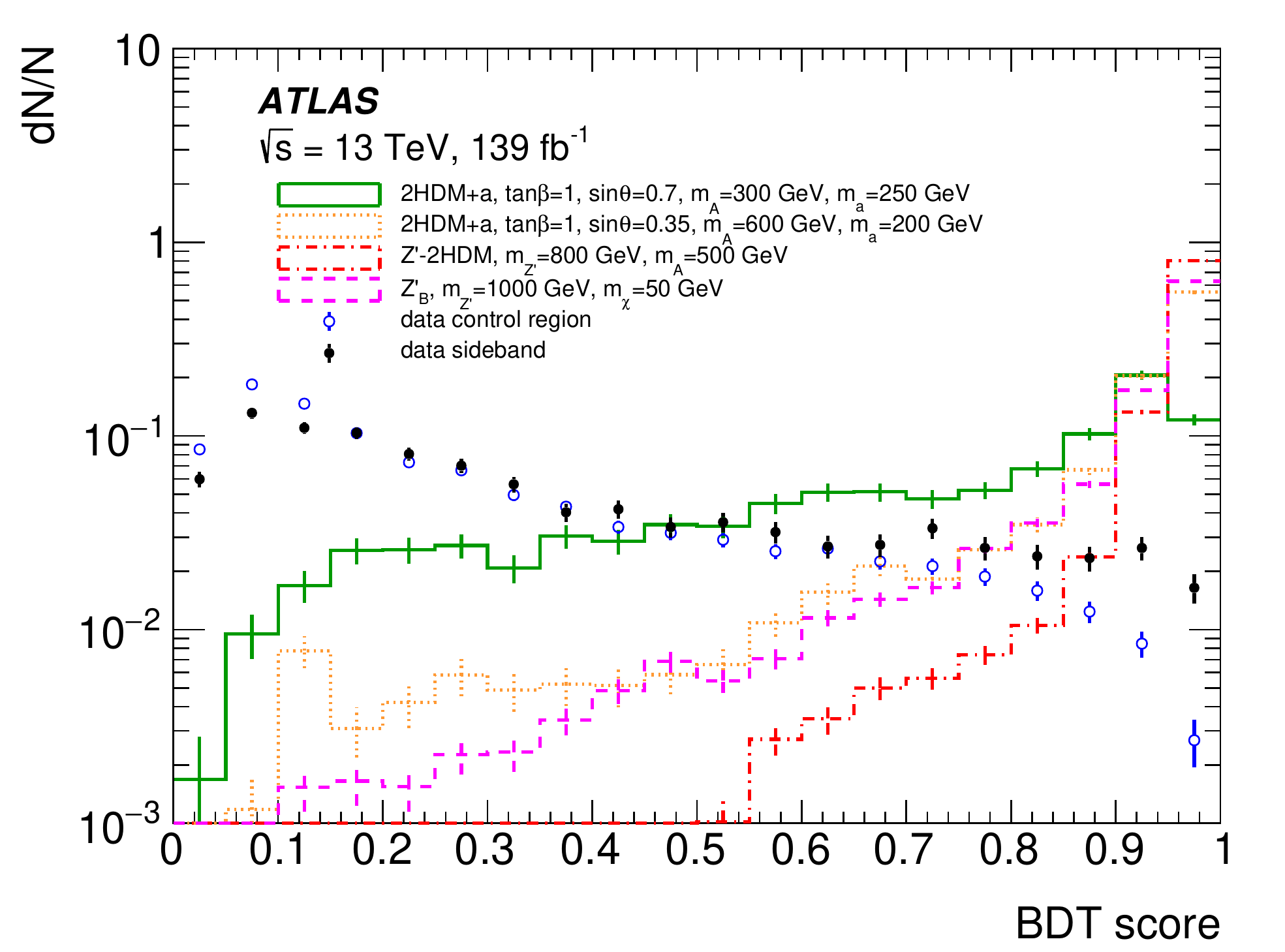}
\caption{The BDT score for selected signals, the data control region, and the data sideband, with 1 being more signal-like and 0 being more background-like. The data sideband consists of events with two `tight' identified and isolated photon candidates, while the data control region consists of diphoton events where at least one photon candidate fails at least one of the identification or isolation requirements. All events are required to have $105 < m_{\gamma\gamma} < 160$~\GeV, and events in the $120 < m_{\gamma\gamma} < 130$~\GeV\ region of the data sideband are vetoed. The error bars represent statistical uncertainties.
}
\label{fig:bdtscore}
\end{figure}

The BDT is trained with signal taken from simulated 2HDM+a ($m_A = 300$~\GeV, $m_a = 250$~\GeV, $\tan{\beta} = 1$, $\sin{\theta} = 0.7$) events, while the background used in training is taken from a data control region that differs from the nominal data selection by requiring that at least one photon fails either the identification or isolation requirements, with the expectation that this will not drastically change the event kinematics. The choice of training signal is motivated by its soft \etmiss distribution, which is relatively close to that of the background. The trained classifier performs well for all signal models, and using a different signal model with soft \etmiss in the training does not appreciably affect the sensitivity. For instance, for each of the four signals shown in Figure~\ref{fig:bdt_input_vars}, the variance in sensitivity from retraining the existing classifier on a different signal is generally on the order of 10\%.  Figure \ref{fig:bdtscore} shows the distribution of the BDT score for various signals and backgrounds after training. The model trained on the data control region (dominated by $\gamma$+jet contributions) performs well when applied to the data sideband (dominated by true diphoton events) as the BDT is able to separate both these backgrounds from expected signals. The disagreement between the data control region and the data sideband at high BDT values is expected due to their different compositions and does not impact the final statistical results as the data control region is only used to train the BDT classifier.
 
\begin{figure}[ht]
\centering
\includegraphics[width=0.68\textwidth]{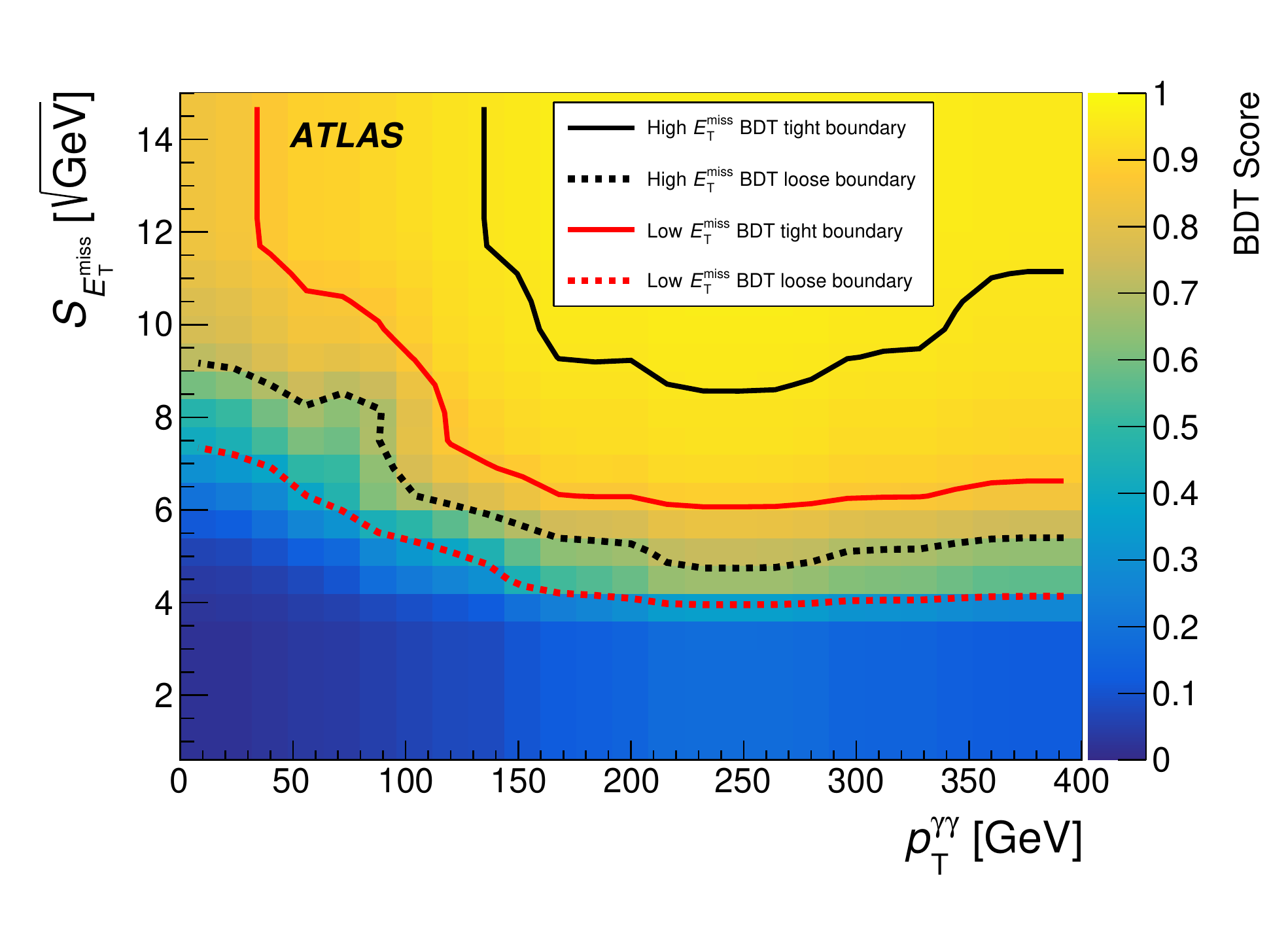}
\caption{The BDT score as a function of the input variables, \pTgg and \setmiss, with 1 being more signal-like and 0 being more background-like. The plot is produced by splitting the plane into equal cells of $\pTgg \times \setmiss$, and evaluating the corresponding BDT score. The black (red) lines indicate the minimum score boundaries for the BDT categories in the high \etmiss (low \etmiss) region, as defined in Table~\ref{tab:category_definition}.
}
\label{fig:bdtscore_2D}
\end{figure}

Finally, events are separated into low \etmiss ($\etmiss < 150$~\GeV) and high \etmiss ($\etmiss > 150$~\GeV) regions. Such a split not only improves overall sensitivity but also complements $h$+\met searches where $h$ decays into a pair of $b$-quarks~\cite{Aaboud:2016obm}, which have low sensitivity in the $\etmiss  < 150$~\GeV\ region due to trigger thresholds. In each region, two categories are defined from two sequential ranges of the BDT score, with the ranges optimized to maximize the combined signal sensitivity in the two chosen categories while discarding the remaining events. The category naming scheme and corresponding ranges are summarized in Table~\ref{tab:category_definition}, with the `tight' categories having a higher training signal purity than the `loose' categories. Figure~\ref{fig:bdtscore_2D} demonstrates that in each of the two regions, the BDT score is a non-trivial function of the input variables. The black and red contours in the figure indicate the separation between categories based on the BDT score for the high \etmiss and low \etmiss regions, respectively.

\begin{table*}[!htbp]
\caption{The definitions of the four signal region categories.
}
\centering
\begin{tabular}{lcc}
\toprule
\toprule
\multicolumn{1}{l}{ Category }  &  \multicolumn{1}{c}{\etmiss requirement} &
\multicolumn{1}{c}{BDT score range} \\
\midrule
High \etmiss BDT tight   & \etmiss $>$ 150 \GeV & 0.950 $<$ BDT score $<$ 1~~~~  \\
High \etmiss BDT loose  & \etmiss $>$ 150 \GeV & ~~~0.694 $<$ BDT score $<$ 0.950 \\
Low \etmiss BDT tight  & \etmiss $<$ 150 \GeV & 0.864 $<$ BDT score $<$ 1~~~~  \\
Low \etmiss BDT loose  & \etmiss $<$ 150 \GeV & ~~~0.386 $<$ BDT score $<$ 0.864   \\
\bottomrule
\bottomrule
\end{tabular}
\label{tab:category_definition}
\end{table*}
 
\section{Signal and background parameterization}
\label{sec:backgroundest}
 
The signal and backgrounds are extracted by fitting analytic functions to the diphoton invariant mass distribution in the range $105 < m_{\gamma\gamma} < 160$~\GeV\ in each category. For the signal and the SM Higgs boson background, the expected normalizations are obtained from their theoretical cross sections multiplied by the product of the acceptance times efficiency from the simulation. The diphoton invariant mass distribution shapes are modelled with a double-sided Crystal Ball function (as defined in Ref.~\cite{HIGG-2013-08}). The shape parameters are determined by fitting the diphoton mass distribution in simulation for each category. The width of the fitted function is largely insensitive to the specific signal model, with maximum variations of approximately 10\%.
 
Both the normalization and the shape of the non-resonant background are obtained by fitting the diphoton invariant mass distribution in data for each category. A variety of analytic functions are considered for the non-resonant background parameterization: exponential functions of different-order polynomials, Bernstein polynomials of different order~\cite{Bernstein}, and an adapted dijet function~\cite{Aaltonen:2008dn}. The potential bias associated with the choice of a specific analytic function to model the continuum background is referred to as the non-resonant background modelling uncertainty, $\Delta N^\textrm{bkg model}_\mathrm{sig}$. It is estimated for each category as the signal event yield extracted from a signal-plus-background maximum-likelihood fit to a background-only diphoton invariant mass distribution, following the procedure in Ref.~\cite{HIGG-2013-08}. The background-only distribution is obtained by combining $\gamma\gamma$, $\gamma$+jet, $V\gamma$, and $V\gamma\gamma$ processes so that the total template is normalized to the data sideband.
 
Around 80\% of sideband events consist of true diphoton events, with the remaining 20\% consisting of $\gamma$+jet events and a negligible number of dijet events. These fractions are determined by a two-dimensional sideband technique which counts the number of events in 16 regions defined by whether each photon passes or fails the identification or isolation requirements~\cite{Aad:2012tba}. The final background modelling shows a negligible dependence on these fractions when varied by their respective uncertainties.
 
The true diphoton component is modelled by simulated $\gamma\gamma$, $V\gamma$, and $V\gamma\gamma$ samples. The $V\gamma$ sample consists mainly of $V$ decays into electrons subsequently misidentified as photons, resulting in an $m_{\gamma\gamma}$ spectrum similar to that of the $V\gamma\gamma$ sample. Both the $V\gamma$ and $V\gamma\gamma$ templates are normalized according to their theoretical cross sections, which accounts for approximately 10\% of the sideband events. The $\gamma\gamma$ template is then normalized to approximately 70\% of the sideband events, so that the sum of the $\gamma\gamma$, $V\gamma$, and $V\gamma\gamma$ templates accounts for the entire true diphoton component. To represent the remaining $\gamma$+jet and dijet components, a $\gamma$+jet template derived from a data control region in which exactly one photon fails the identification requirements is normalized to 20\% of sideband events and combined with the true diphoton template.
 
For a given functional form, several fits are tested by varying the position of the signal peak between $121$ and $129$~\GeV. The largest number of signal events obtained in these fits to the background-only templates is taken as $\Delta N^\textrm{bkg model}_\mathrm{sig}$. Among the different analytic functions that are tested, the one with $\Delta N^\textrm{bkg model}_\mathrm{sig}$ smaller than 50\% of the statistical uncertainty of the fitted signal yield and the least number of free parameters is chosen
as the nominal background parameterization to describe the non-resonant background shape. In each category, an exponential function $\exp(a\cdot \mgg)$ is found to be the best choice.
 
The non-resonant background modelling uncertainty $\Delta N^\textrm{bkg model}_\mathrm{sig}$, which is taken as an estimate of the systematic uncertainty due to the choice of parameterization, is shown in Table~\ref{tab:spurious_signal} for each category. As a comparison of scale, its ratio to the expected non-resonant background $N^{\textrm{non-res.}}_{\mathrm{bkg}}$ in the signal window $120 < m_{\gamma\gamma} < 130$~\GeV\ is also provided. This quantity is less than 7\% in the most sensitive category, showing that the systematic uncertainty is under control in a category already dominated by statistical uncertainties. This uncertainty is implemented in the statistical model described in Section~\ref{sec:effsys} as an additional signal normalized to $\Delta N^\textrm{bkg model}_\mathrm{sig}$, times a Gaussian-constrained nuisance parameter.
 
\begin{table*}[!htbp]
\caption{The uncertainty in the signal due to the non-resonant background
modelling ($\Delta N^\textrm{bkg model}_\mathrm{sig}$) per category. As a comparison of scale, its ratio to the expected non-resonant background in the signal window $120 < m_{\gamma\gamma} < 130$~\GeV\ is also shown. In each category, the non-resonant \mgg distribution is modelled by an exponential function ($\exp(a\cdot \mgg)$), where $a$ is a free parameter.
}
\centering
\begin{tabular}{lcc}
\toprule
\toprule
\multicolumn{1}{l}{ Category }  &  \multicolumn{1}{c}{ $\Delta N^\textrm{bkg model}_\mathrm{sig}$ } &
\multicolumn{1}{c}{ $\Delta N^\textrm{bkg model}_\mathrm{sig}$/$N^{\textrm{non-res.}}_{\mathrm{bkg}}$ [\%]} \\
\midrule
High \etmiss BDT tight   & 0.54 & 6.8  \\
High \etmiss BDT loose  & 1.07 & 4.2 \\
Low \etmiss BDT tight  & 0.62 & 6.3  \\
Low \etmiss BDT loose  & 2.64 & 2.0  \\
\bottomrule
\bottomrule
\end{tabular}
\label{tab:spurious_signal}
\end{table*}

\section{Systematic uncertainties}
\label{sec:effsys}
 
Uncertainties from experimental and theoretical sources affect both the shapes and yields of the signal and the SM Higgs boson background, estimated from the simulated MC samples. Of these, the largest uncertainties are due to the \met reconstruction and jets, pile-up, and signal efficiency interpolation. The theoretical systematic uncertainties include uncertainties on the factorization and renormalization scale and the parton distribution function and \alphas (PDF+\alphas). The non-resonant background is obtained directly from the fit to the data and therefore its only systematic uncertainty is the potential bias $\Delta N^\textrm{bkg model}_\mathrm{sig}$, as described in Section~\ref{sec:backgroundest}. A summary of the experimental and theoretical uncertainties affecting the yields from SM Higgs boson processes, non-resonant background, and
signal production is shown in Table~\ref{tab:unc}.
 
\begin{table*}[ht]
\small
\caption{Breakdown of the dominant systematic uncertainties. The impact of uncertainties on the yield of the SM Higgs boson
processes and signal samples is shown. All production modes of the SM Higgs boson are considered together. Representative values
for the impact on the most sensitive category are shown, unless one of the systematic uncertainties is not applicable to the sample, in which case the value is substituted by a ``-''. The ``<'' on the signal efficiency interpolation uncertainty indicates that the value is estimated from the maximum relative difference between the fully simulated and reweighted samples over all validation points. The impact of theoretical uncertainties is split into their effects on the migration of events between different categories and on the total cross section. Here the signal is a 2HDM+a model with $m_{A}=200$~\GeV, $m_{a}=100$~\GeV, $\tan\beta=1.0$, $\sin\theta = 0.35$, which provides a conservative estimate of the size of the uncertainties for all signal points.
}
\begin{center}
\begin{tabular}{l...}
\toprule
\toprule
\multirow{3}{*}{Source}           & \multicolumn{1}{c}{\multirow{3}{*}{Signals [\%]}}  & \multicolumn{2}{c}{Backgrounds [\%]} \\
\cline{3-4}
&  & \multicolumn{1}{c}{\multirow{2}{*}{SM Higgs boson}} & \multicolumn{1}{c}{Non-resonant} \\
&  &  & \multicolumn{1}{c}{background} \\
\midrule
\multicolumn{3}{l}{Experimental} \\
\midrule
Luminosity & 1.7 & 1.7  & - \\
Trigger efficiency & 1.0 & 1.0 & - \\
Vertex selection (inclusive cat.) & 0.01 & 0.01 & - \\
Photon energy scale & 1.0 & 1.2  & - \\
Photon energy resolution & 0.3 & 0.4  & -\\
Photon identification efficiency & 1.3 & 1.3   & - \\
Photon isolation efficiency & 1.3 & 1.4  & - \\
\atlfasttwo simulation & 2.0 & - & - \\
\etmiss\ reconstruction and jet uncertainty & 2.8  &  1.7 & - \\
Pile-up reweighting & 2.3  & 2.0 & -\\
Signal efficiency interpolation & < 13 & - & - \\
Non-resonant background modelling & - & - &   6.8  \\
\midrule
\multicolumn{3}{l}{Theoretical} \\
\midrule
Factorization and renormalization scale in migration& 1.3 & 3.5  & - \\
PDF+\alphas in migration & 1.2 & 1.0   & - \\
Factorization and renormalization scale in cross section& - & 2.8  & - \\
PDF+\alphas in cross section & - & 2.8   & - \\
Multi-parton interactions, ISR/FSR, hadronization & 3.0  & 3.0 & - \\
$B$($H\rightarrow \gamma\gamma$) & 1.7 & 1.7 & - \\
\bottomrule
\bottomrule
\end{tabular}
\label{tab:unc}
\end{center}
\end{table*}
 
The uncertainty in the integrated luminosity of the full Run~2 dataset is 1.7\%~\cite{ATLAS-CONF-2019-021}, obtained using the LUCID-2 detector~\cite{LUCID2} for the primary luminosity measurements.
 
The efficiency of the diphoton trigger used to select events is evaluated in MC simulation using a trigger matching technique and in data using a bootstrap method~\cite{Aad:2019wsl}. In the diphoton invariant mass window of $105 < m_{\gamma\gamma} < 160$~\GeV, the trigger efficiency uncertainty affects the acceptance by 1\% in each category.
 
The uncertainty in the vertex selection efficiency is assessed by comparing the efficiency of finding photon-pointing vertices in $Z \rightarrow e^+e^-$ events in data with that in MC simulation~\cite{HIGG-2016-08}. The resulting uncertainty is found to be negligible in the inclusive photon selection.
 
The systematic uncertainties due to the photon identification and isolation efficiencies are estimated following the prescriptions in Ref.~\cite{EGAM-2018-01}. They are evaluated by varying the correction factors of photon selection efficiencies in MC simulation by the corresponding uncertainties. In the most sensitive category, the photon identification efficiency uncertainty is $1.3\%$ for the SM Higgs boson and for the signal samples, while the photon isolation efficiency uncertainty is $1.4\%$ for the SM Higgs boson and $1.3\%$ for the signal samples. For the signal samples, an additional 2\% uncertainty is added to account for photon mismodelling by the \atlfasttwo simulations.
 
The experimental uncertainties in photon scale and resolution are obtained from Ref.~\cite{EGAM-2018-01}. In the most sensitive category, the uncertainty in the energy scale has an effect of $1.0\%$ on the normalization of the signals and $1.2\%$ on the normalization of the SM Higgs boson background. The uncertainty in the energy resolution has an effect below $0.3\%$ on the normalization of the signals and $0.4 \%$ on the normalization of the SM Higgs boson background. The effects of photon energy scale and resolution uncertainties on the signal and SM Higgs boson background mass distributions are also evaluated and parameterized in the fit to MC simulation. In particular, the impact of the scale uncertainties on the mean of these mass distributions is $0.3\%$, while the impact of the energy resolution on the width is $6\%$. An additional uncertainty in the Higgs boson mass position of 0.24~\GeV\ from the measurement of the Higgs boson mass is considered in the fit, although this does not significantly impact the final result~\cite{HIGG-2014-14}.
 
The migration of events among categories results from changes in object energies and momenta, due mostly to misreconstruction of jets and \etmiss. The experimental uncertainties in jet energy scale and resolution are propagated to the \etmiss calculation. In addition, the uncertainties in the scale and resolution of the \etmiss\ soft term are evaluated by using the method described in Ref.~\cite{Aaboud:2018tkc}. The uncertainties for the \etmiss reconstructed from the hardest vertex and the neural-network-selected vertex are evaluated independently by using jets reconstructed relative to the respective vertices. In the most sensitive category, the uncertainties due to the \etmiss soft term are 0.6\% for both the signal and background. The migration uncertainties due to the jet energy scale and resolution have an effect of 2.5\% and 1.3\% on the event yield of the signal and SM Higgs boson samples, respectively. The pile-up reweighting uncertainty is taken into account by propagating it through the event selection, and results in a 2.3\% and 2.0\% uncertainty in the event yield of the signal and SM Higgs boson samples, respectively, in the most sensitive category.
 
The uncertainty in the signal efficiency due to the interpolation method detailed in Section~\ref{sec:simulation} is estimated by comparing the yields from fully simulated samples and reweighted samples for certain validation parameter points. The maximum relative difference between the fully simulated and reweighted samples over all these points ranges from 9\% to 13\%, depending on the model, and these numbers are taken as the uncertainty.
 
The effects of theoretical scale uncertainties on the SM Higgs boson and 2HDM+a signal samples are estimated by varying the factorization and renormalization scales up and down from their nominal values by a factor of two, recalculating the cross section in each case, and taking the largest deviation from the nominal cross section as the uncertainty. The scale uncertainties affect the event migration between the categories by 3.5\% for the SM Higgs boson processes and 1.3\% for the signal processes. The uncertainties in the cross section of the SM Higgs boson processes are taken from Ref.~\cite{deFlorian:2016spz}. The effect of the scale uncertainty when selecting ggF Higgs boson events with high \pTgg is estimated to be 20\% for each category, but the effect is negligible in the final fit because the ggF Higgs boson contribution is small.  In general, the effects of theory uncertainties on the SM Higgs boson yield are small compared to the uncertainties in the non-resonant background, and do not significantly impact the sensitivity. For signals, only the uncertainties on the 2HDM+a models are taken into account, because for the $Z^{\prime}_{B}$ and $Z^{\prime}$-2HDM models, no QCD vertex is calculated in the matrix element. The uncertainties in the signal cross section are not used in the fit, but are instead shown as bands on the observed limits.
 
For signals, PDF+\alphas uncertainties are estimated using the SysCalc~\cite{kalogeropoulos2018syscalc} package associated with $\MADGRAPH5$.
The uncertainties in the migration of each category are estimated by varying the parameters of the \nnpdflo PDF set and taking the maximum change as the uncertainty.
For the SM Higgs boson, the effects of PDF+\alphas uncertainties on the cross sections are taken from Ref.~\cite{deFlorian:2016spz}. The uncertainties in the migration of each category are estimated using the recommendations of PDF4LHC~\cite{Butterworth:2015oua}. These uncertainties are 1.0\% for the SM Higgs boson and 1.2\% for signals. The effects of PDF+\alphas uncertainties on the cross section are estimated to be 2.8\% for the SM Higgs boson and up to 32\% for the signals. The effects of these SM Higgs boson uncertainties are once again small compared to the impact of the non-resonant background uncertainty, while the uncertainties in the signal cross sections are not used in the fit, but are instead shown as bands on the observed limits.
 
The uncertainty in the branching ratio ($B$) of $h\rightarrow \gamma\gamma$ is 1.73\%~\cite{deFlorian:2016spz}. The same $h\rightarrow \gamma\gamma$ branching ratio is used for the SM Higgs boson and the signal models when setting limits.  For the 2HDM+a models, the Higgs boson is only allowed to decay into $a\chi\chi$ or through a SM channel, where the $h \to a\chi\chi$ decay branching ratio is smaller than 1\% when $m_{a}$ is greater than 110~\GeV\ for $m_{\chi}=10~\GeV$. This 1\% effect is neglected when setting limits as it is small with respect to the uncertainty in the SM $h\rightarrow \gamma\gamma$ branching ratio. Similarly, the $h\rightarrow Z'Z'$ decay branching ratio is neglected in the $Z'_B$ model for $m_{Z'} < m_h/2$.
 
The effects of multi-parton interactions, parton showering and hadronization are evaluated by varying the eigen-variables in \textsc{Pythia}\xspace for signal, by comparing \herwig~7.1.3~\cite{Bellm:2017jjp} with \textsc{Pythia}\xspace for ggF, VBF, and $Vh$, and by comparing \herwig~7.0.4~\cite{Bellm:2015jjp} with \textsc{Pythia}\xspace for $t\overline{t}h$. These uncertainties are of the order of 3\% for both the signal and SM Higgs boson processes.
 

\section{Results}
\label{sec:result}
\subsection{Statistical framework}
 
The results of the analysis are derived from a likelihood fit of the \mgg distribution in the range $105 < \mgg < 160~\GeV$, performed simultaneously over all four categories. The likelihood function is defined as
 
\begin{equation*}
\mathcal{L} = {\displaystyle\prod_{c=1}^{4}\left(\mathrm{Pois}(n_c|N_c(\theta))\cdot  {\displaystyle\prod_{i=1}^{n_c} f_c(m_{\gamma\gamma}^i,\theta)} \cdot G(\theta)\right)}
\end{equation*}
 
where for each event $i$ in a category $c$, $n_c$ is the observed number of events, $N_c$ is the expected number of events, $f_c$ is the value of the probability density function, $\theta$ are nuisance parameters, and $G(\theta)$ are Gaussian constraints on the nuisance parameters.
 
The expected number of events $N_c$ is the sum of the expected yields from BSM signals, SM Higgs boson processes, the non-resonant background modelling uncertainty, and the non-resonant background:
 
\begin{equation*}
N_c(\theta) = \mu \cdot N_{\mathrm{BSM}, c}(\theta_\mathrm{yield}) + N_{\mathrm{Higgs}, c}(\theta_\mathrm{yield}) + \Delta N^\textrm{bkg model}_{\mathrm{sig},c} \cdot \theta^\textrm{bkg model}_{\mathrm{sig}, c} + N^\textrm{non-res.}_{\mathrm{bkg}, c}
\end{equation*}
 
where $\mu$ is the signal strength, and $\theta_\mathrm{yield}$ and $\theta^\textrm{bkg model}_{\mathrm{sig}, c}$ represent systematic uncertainties in the resonant and non-resonant yields, as detailed in Section~\ref{sec:effsys}.
 
The nominal Higgs boson mass is set to $125.09$~\GeV~\cite{HIGG-2014-14} and the nominal resonant yields are fixed to values from simulation. The signal strength, non-resonant background shape parameters, and the nuisance parameters representing the systematic uncertainties are free parameters in the fit.
 
The test statistic is based on the likelihood ratio approach, as presented in Ref.~\cite{Cowan:2010js}.
 
The event yields and uncertainties in the observed data, signal, and backgrounds in the four categories within a window of \mbox{$120 < \mgg < 130~\GeV$} after the fit are shown in
Table~\ref{tab:cut}, for representative $Z'_B$, $Z'$-2HDM, and 2HDM+a signals. While the signal yields in each category depend on the specific model, the SM Higgs boson contribution comprises approximately 30\% of the total background in the High \etmiss BDT tight and Low \etmiss BDT tight categories, due primarily to the presence of $Vh$ events with high genuine \etmiss. The uncertainties in the SM Higgs backgrounds reflect the impact of systematic uncertainties, while the uncertainties in the non-resonant background are constrained by the full diphoton mass range and the given analytic form, giving a lower uncertainty relative to the background contribution. On the other hand, the uncertainties in the signal yields reflect the fact that the fitted signal is driven mainly by the High \etmiss BDT tight category, where there is a large statistical uncertainty in the data.
 
\begin{table*}[!htbp]
\footnotesize
\sisetup{separate-uncertainty}
\caption{Event yields and uncertainties after a fit to data in the range of 120 $<\mgg < 130$~\GeV\  for
data, signal models, the SM Higgs boson background, and non-resonant background
in each analysis category, for an integrated luminosity of $139$~fb$^{-1}$.
The signal samples shown correspond to a $Z'_B$ signal with $m_{Z'}=1000~\GeV$ and $m_{\chi}=50~\GeV$, a $Z'$-2HDM signal with $m_{A} = 800~\GeV$ and $m_{\chi}=500~\GeV$, and a 2HDM+a signal with $m_{A}=600~\GeV$, $m_{a} = 200~\GeV$, $\tan{\beta}=1.0$, and $\sin{\theta}=0.35$. The uncertainties
correspond to the quadrature sum of the statistical and systematic uncertainties.}
\begin{center}
\begin{tabular}{lS[table-format = 3.2(2)]S[table-format = 3.2(2)]S[table-format = 3.2(2)]S[table-format = 3.2(2)]}
\toprule \toprule
\multicolumn{1}{l}{Category} & \multicolumn{1}{c}{High \etmiss BDT tight} & \multicolumn{1}{c}{High \etmiss BDT loose} & \multicolumn{1}{c}{Low \etmiss BDT tight} & \multicolumn{1}{c}{Low \etmiss BDT loose} \\
\midrule
Data  & 12 & 29 & 11 & 143 \\
\midrule
\multicolumn{5}{c}{Backgrounds} \\
\midrule
SM Higgs boson &  3.74 \pm 0.25 &  3.40 \pm 0.28 &  3.12 \pm 0.23 &   9.9 \pm 1.5 \\
Non-resonant   &  7.8  \pm 1.3  & 25.3  \pm 2.3  &  9.8  \pm 1.5  & 130 \pm 5 \\
Total          & 11.6  \pm 1.3  & 28.7  \pm 2.3  & 12.9  \pm 1.5  & 140 \pm 5 \\
\midrule
\multicolumn{5}{c}{$Z'_B$ model, $m_{Z'}=1000$~\GeV, $m_{\chi}=50$~\GeV} \\
\midrule
Signal yields & 0.7 \pm 3.1 & 0.1 \pm 0.6 & 0.1 \pm 0.6 & 0.1 \pm 0.6 \\
\midrule
\multicolumn{5}{c}{$Z'$-2HDM model, $m_{A} = 800$~\GeV\ and $m_{\chi}=500$~\GeV}\\
\midrule
Signal yields & 0.6 \pm 3.1 & 0.1 \pm 0.4 & 0.05 \pm 0.26 & 0.03 \pm 0.17 \\
\midrule
\multicolumn{5}{c}{2HDM+a model, $m_{A}=600$~\GeV, $m_{a} = 200$~\GeV, $\tan{\beta}=1.0$, $\sin{\theta}=0.35$}\\
\midrule
Signal yields & 0.6 \pm 3.1 & 0.2 \pm 1.2 & 0.1 \pm 0.5 & 0.1 \pm 0.7 \\
\bottomrule \bottomrule
\end{tabular}
\label{tab:cut}
\end{center}
\end{table*}
 
\begin{figure}[!htbp]
\centering
\includegraphics[width=0.48\textwidth]{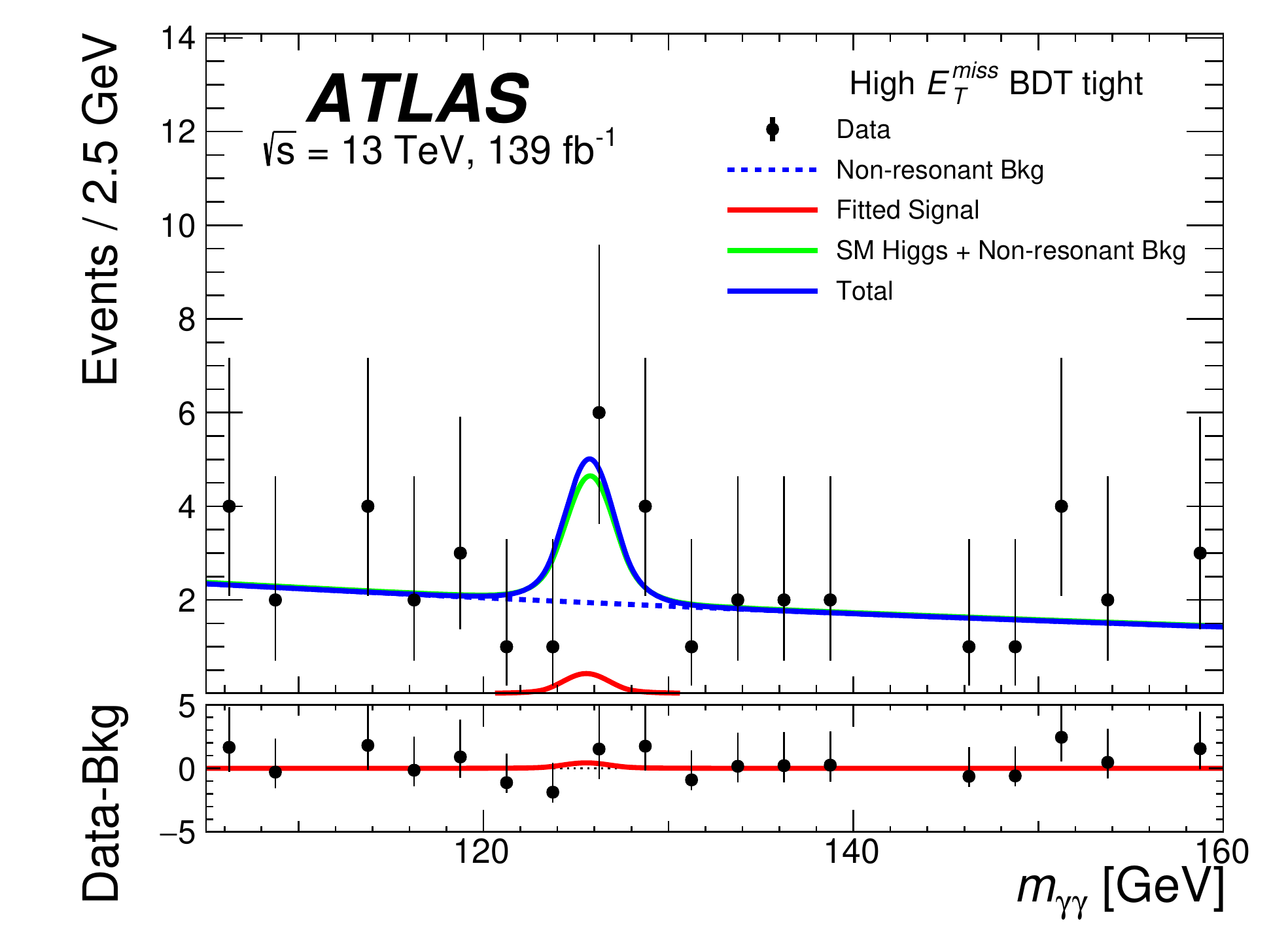}
\includegraphics[width=0.48\textwidth]{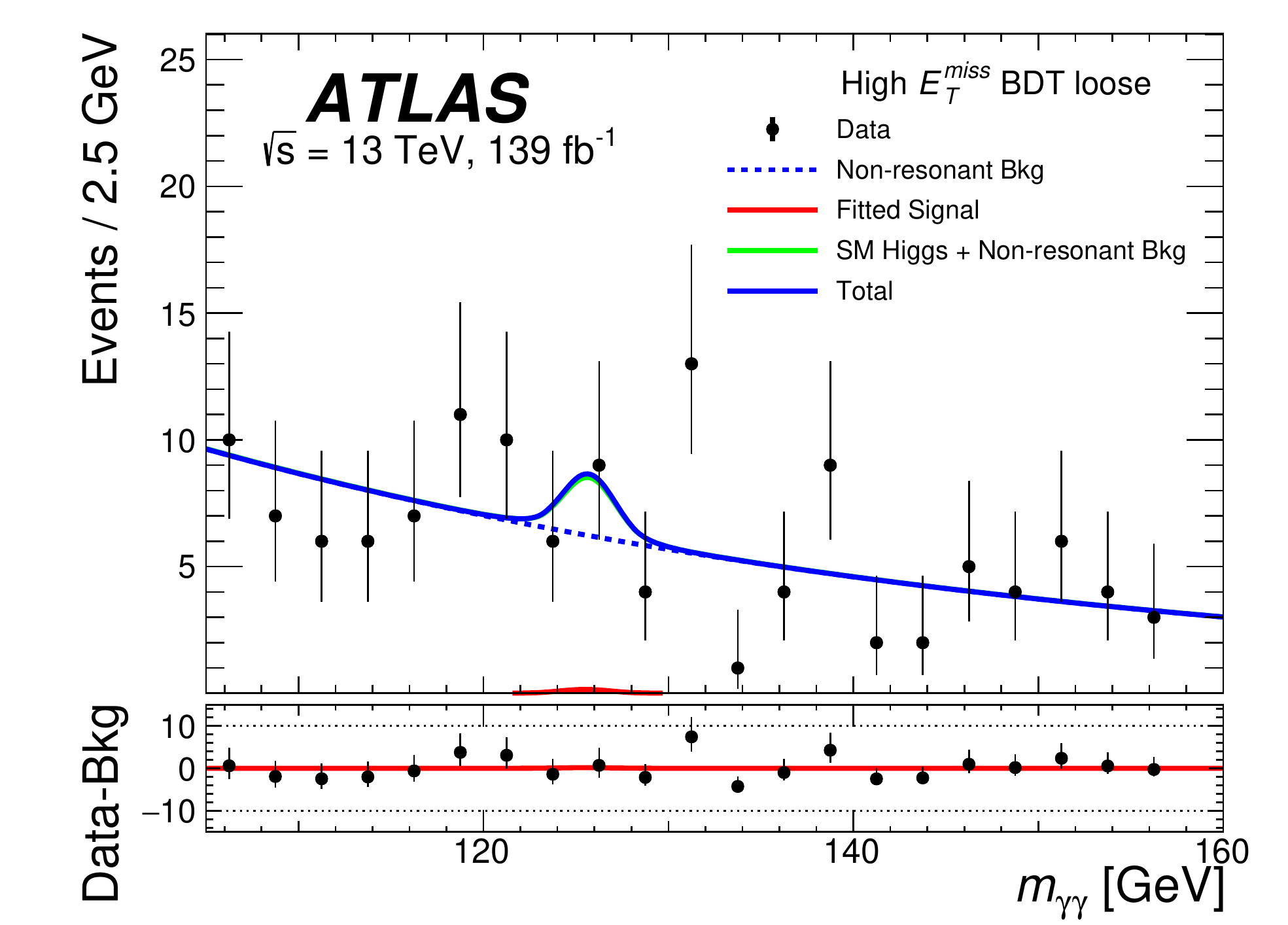}
\includegraphics[width=0.48\textwidth]{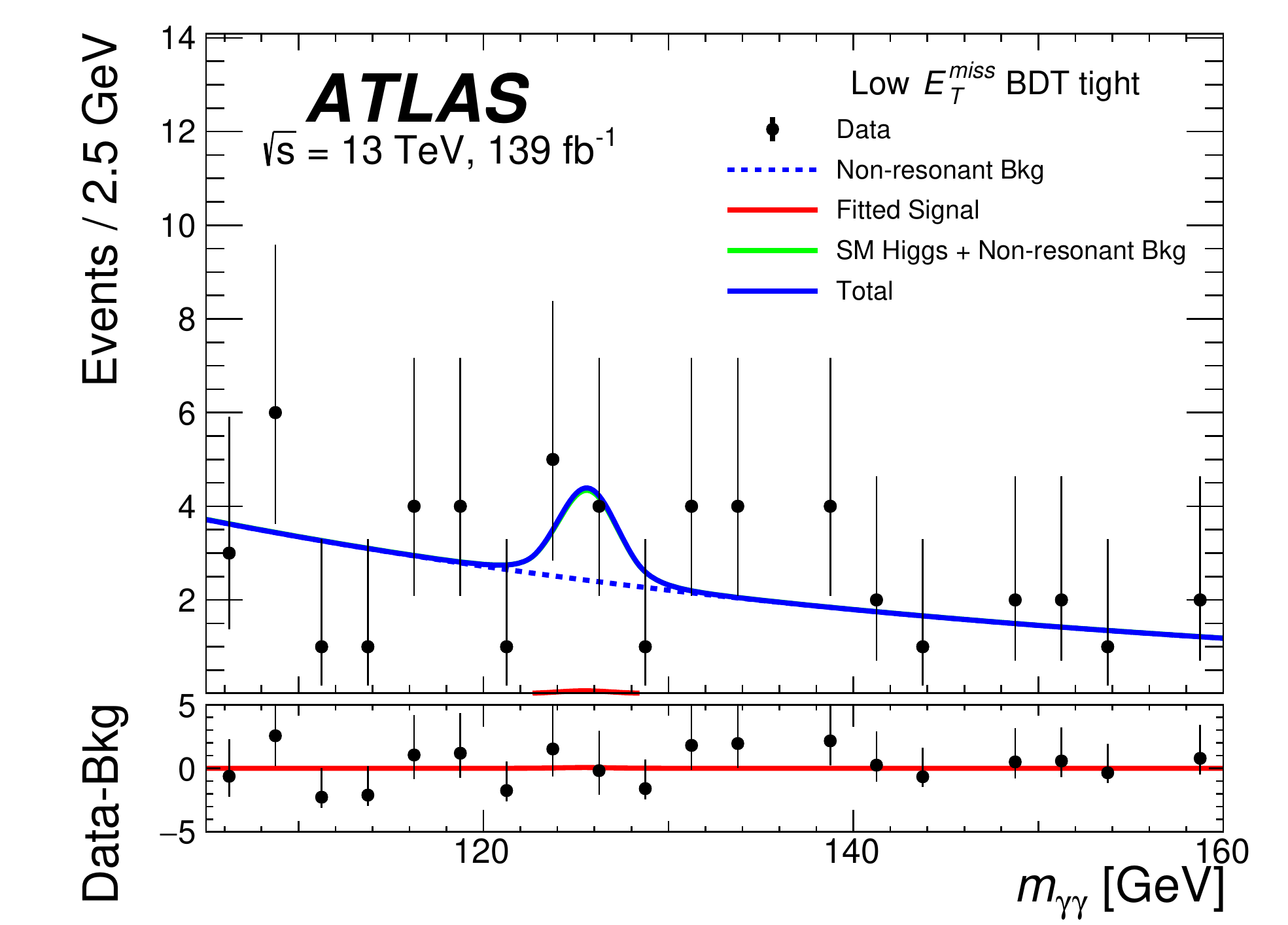}
\includegraphics[width=0.48\textwidth]{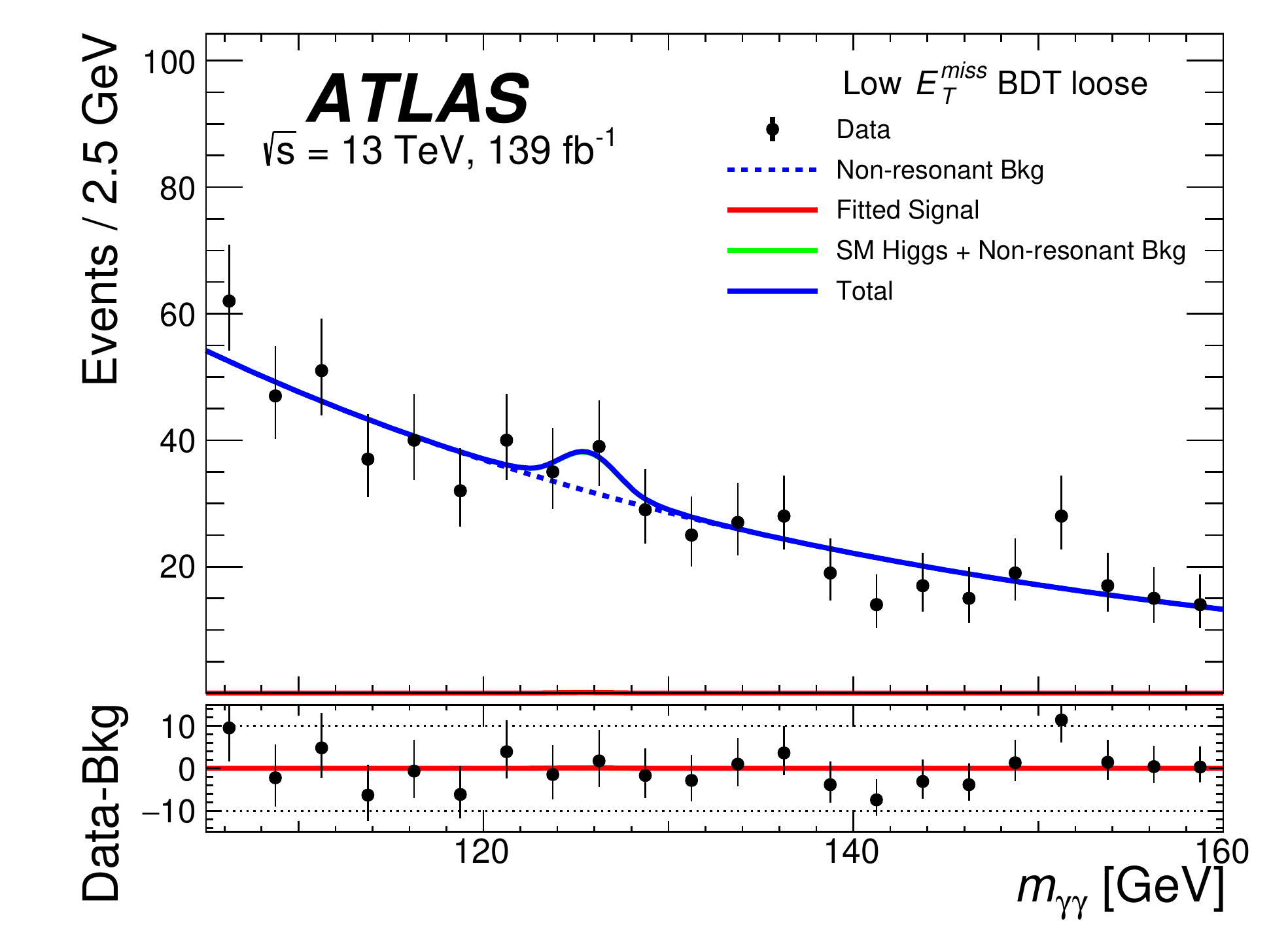}
\caption{The diphoton invariant mass spectra from the data and the corresponding fitted signal and background in each BDT category. The signal is the 2HDM+a model with $\tan\beta=1.0$, $\sin\theta=0.35$, $m_{A}=600$~\GeV, and $m_{a}=200$~\GeV. Different signal models have similar \mgg shapes and therefore will mainly differ in the relative signal contributions in the various categories.
The non-resonant background and the predicted SM Higgs boson contribution are shown. The blue curve shows the sum of the signal,
SM Higgs boson, and non-resonant background after the fit. The subplot shows the residual between the observed number of events and the fitted non-resonant and SM Higgs boson backgrounds. The error bars represent statistical uncertainties.
}
\label{fig:fit_data}
\end{figure}

Figure~\ref{fig:fit_data} shows the \mgg distribution in each BDT category as well as the analytic fits to the data for the 2HDM+a model with $\tan\beta=1.0$, $\sin\theta=0.35$, $m_{A}=600$~\GeV, and $m_{a}=200$~\GeV. Different signal models have similar \mgg shapes and therefore will mainly differ in the relative signal contributions in the various categories. The fit demonstrates sensitivity to the SM Higgs boson expected production yield, with the $Vh$ channel providing sensitivity in the high \etmiss region. For each \mgg distribution, the residual between data observed in each bin and the fitted non-resonant and SM Higgs boson backgrounds is shown in the lower panel. The experimental data distribution clearly exhibits no excess with respect to the total background and the fit result is then interpreted to set 95\% confidence level (CL) limits on the different models discussed in Section~\ref{sec:intro}.
 
\subsection{Interpretation}
 
\begin{figure}[!htbp]
\centering
\includegraphics[width=0.68\textwidth]{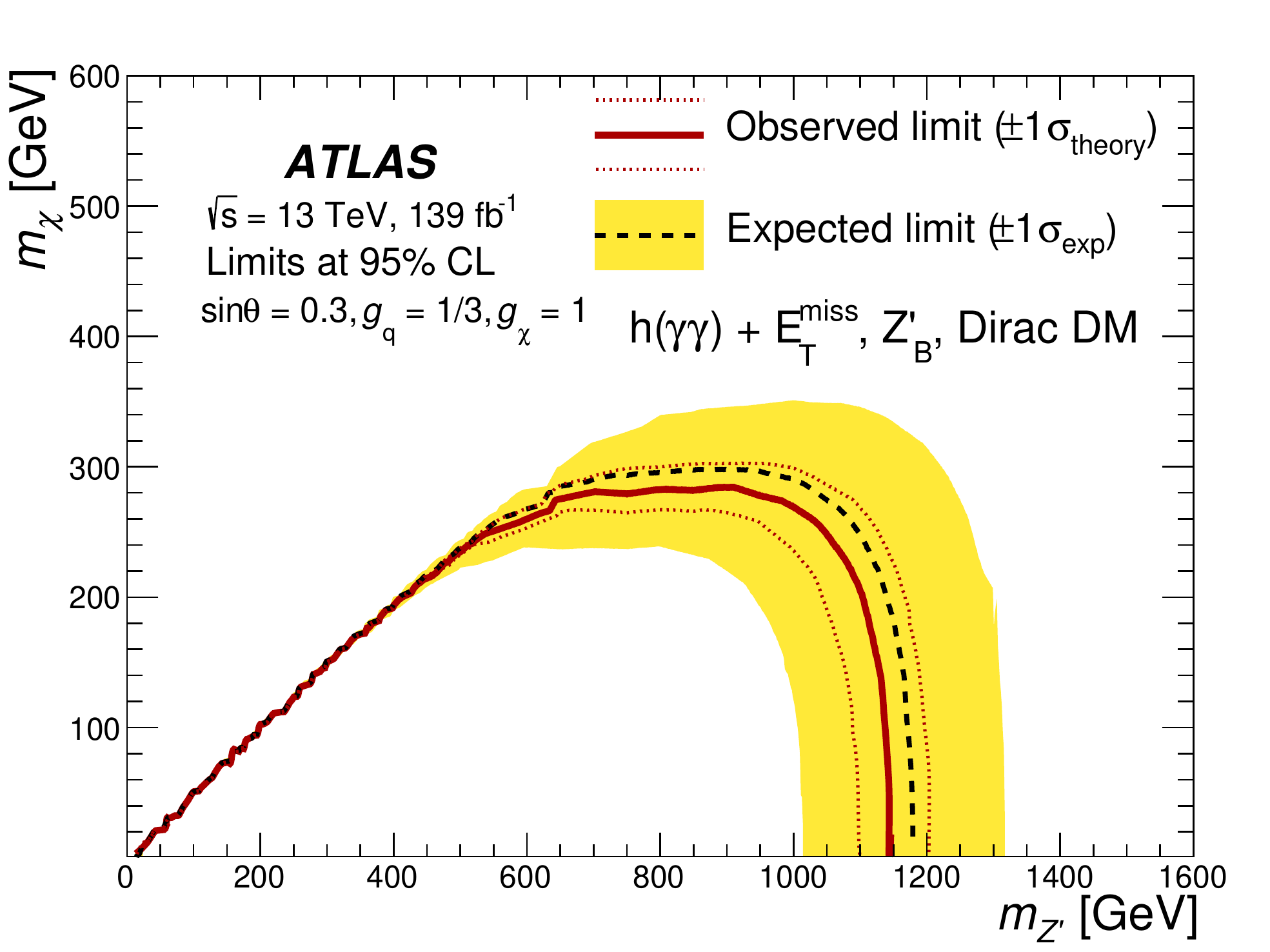}
\caption{
The observed (solid line) and expected (dashed lines) exclusion limit contours at 95\% CL for the $Z'_{B}$ model in the $m_{\chi}$--$m_{Z'}$ plane, for $\sin \theta = 0.3$, $g_q = 1/3$, and $g_{\chi} = 1$.
The dotted lines represent the $\pm1\sigma$ theoretical uncertainty for the observed limit. The $\pm1\sigma$ expected exclusion limit contour is shown as the yellow band.
}
\label {fig:2Dexclusion_ZB}
\end{figure}

The observed and expected exclusion contours at 95\% CL for the $Z'_B$ model in the $m_{\chi}$--$m_{Z'}$ plane are shown in Figure~\ref{fig:2Dexclusion_ZB} for $\sin \theta = 0.3$, $g_q=1/3$, and $g_{\chi} = 1$. Compared to the results of previous $h$+\met searches in the $\gamma\gamma$ decay channel, the limit from the full LHC Run~2 dataset extends up to $1150$~\GeV\ in $m_{Z'}$ while it was lower than $1000$~\GeV\ in the previous ATLAS publication in this search channel~\cite{Aaboud:2017uak} with early Run~2 data. The increase in the limit is only 150~\GeV\ despite the four times larger dataset, because the signal cross section decreases strongly with increasing $Z'$ mass.
In addition, the maximum limit on $m_{\chi}$ increases by more than $100$~\GeV\ to reach $280$~\GeV\ for $m_{Z'}$ in the range $700$--$950$~\GeV.
 
\begin{figure}[!htbp]
\centering
\includegraphics[width=0.68\textwidth]{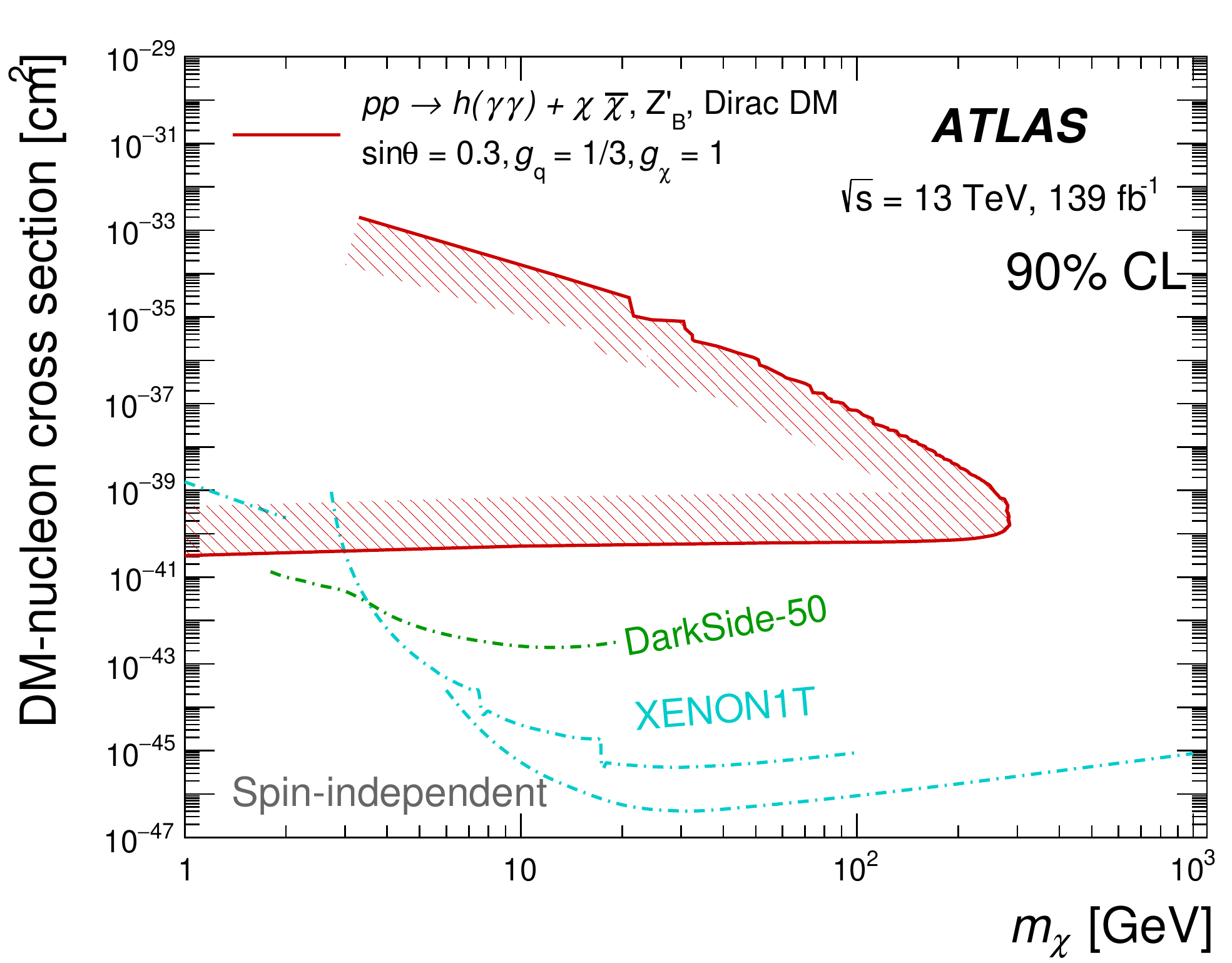}
\caption{
A comparison of the inferred limits with the constraints from direct detection experiments on the spin-independent DM--nucleon cross section in the context of the $Z'_B$ simplified model with vector couplings. Limits are shown at 90\% CL.
The results from this analysis, in which the region on the side of the hatched band inside the contour is excluded, are compared with limits from the XENON~\cite{PhysRevLett.123.251801,PhysRevLett.123.241803,PhysRevLett.121.111302} and
DarkSide-50~\cite{PhysRevLett.121.081307} experiments. The comparison is model-dependent and solely valid in the context of this model, assuming Dirac fermion DM, mixing angle $\sin\theta=0.3$, and the coupling
values $g_q = 1/3$ and $g_{\chi}= 1$. The diagonal upper branch of the limit curve reflects the fact that there are no parameters of the model that predict a cross section larger than this limit.
The impact of renormalization-group evolution effects~\cite{Crivellin:2014qxa,DEramo:2014nmf} when comparing
collider and direct detection limits is not taken into consideration here.
}
\label{fig:monoh_DM_NucleonXS_SI}
\end{figure}

Figure~\ref{fig:monoh_DM_NucleonXS_SI} shows a comparison of the inferred limits at 90\% CL with the constraints from direct detection
experiments on the spin-independent DM--nucleon cross section in the context of the $Z'_B$ simplified model with vector couplings using the relation
 
\begin{equation*}
\sigma^\text{SI}_{N\chi} = \frac{\mu^{2}_{N\chi}}{\pi A^2}\left[Zf_{p}+(A-Z)f_{n}\right]^2 \,,
\end{equation*}
 
in which $\mu_{N\chi}=m_{\chi}m_{N}/(m_{\chi}+m_{N})$ is the reduced mass of the
DM--nucleon system, and ${f_{p}=f_{n}=3g_{q}g_{\chi}/m^{2}_{Z'}}$ are the couplings
between DM particles and protons and neutrons, respectively~\cite{monoh}.
In the above expression, $Z$ and $A$ are the numbers of protons and nucleons in the considered nucleus, set to 1 in this case.
Results from the XENON~\cite{PhysRevLett.123.251801,PhysRevLett.123.241803,PhysRevLett.121.111302} and DarkSide-50~\cite{PhysRevLett.121.081307} direct detection experiments are overlaid for the comparison. The diagonal upper branch of the limit curve reflects the fact that there are no parameters of the model that predict a cross section larger than this limit. Consequently, an observed cross section greater than this limit cannot be interpreted in this model. On the contrary, the horizontal branch delimits the sensitivity of the analysis to this model. This result improves the upper limit on the spin-independent cross section for $m_{\chi} < 2$~\GeV\ by a factor of two relative to the previous publication~\cite{Aaboud:2017uak}. LHC data offers a unique window on low-mass DM candidates that complements direct DM searches in an interesting way but provides results that are more model-dependent than direct search results. The ability of each result to constrain new physics depends crucially on the model parameters. For instance, the results for the $Z'_B$ model, with model parameters $\sin \theta = 0.3$, $g_q=1/3$, and $g_{\chi} = 1$ for this search, are more stringent than direct detection experiments for $m_{\chi}<2$~\GeV\ and extend to DM masses well below 1~\GeV, while in the case of a much lower coupling between DM and SM particles, the direct search limits may provide the more stringent constraints on this possible phenomenology. The impact of renormalization-group evolution effects ~\cite{Crivellin:2014qxa,DEramo:2014nmf} when comparing collider and direct detection limits is not considered here.

\begin{figure}[!htbp]
\centering
\includegraphics[width=0.68\textwidth]{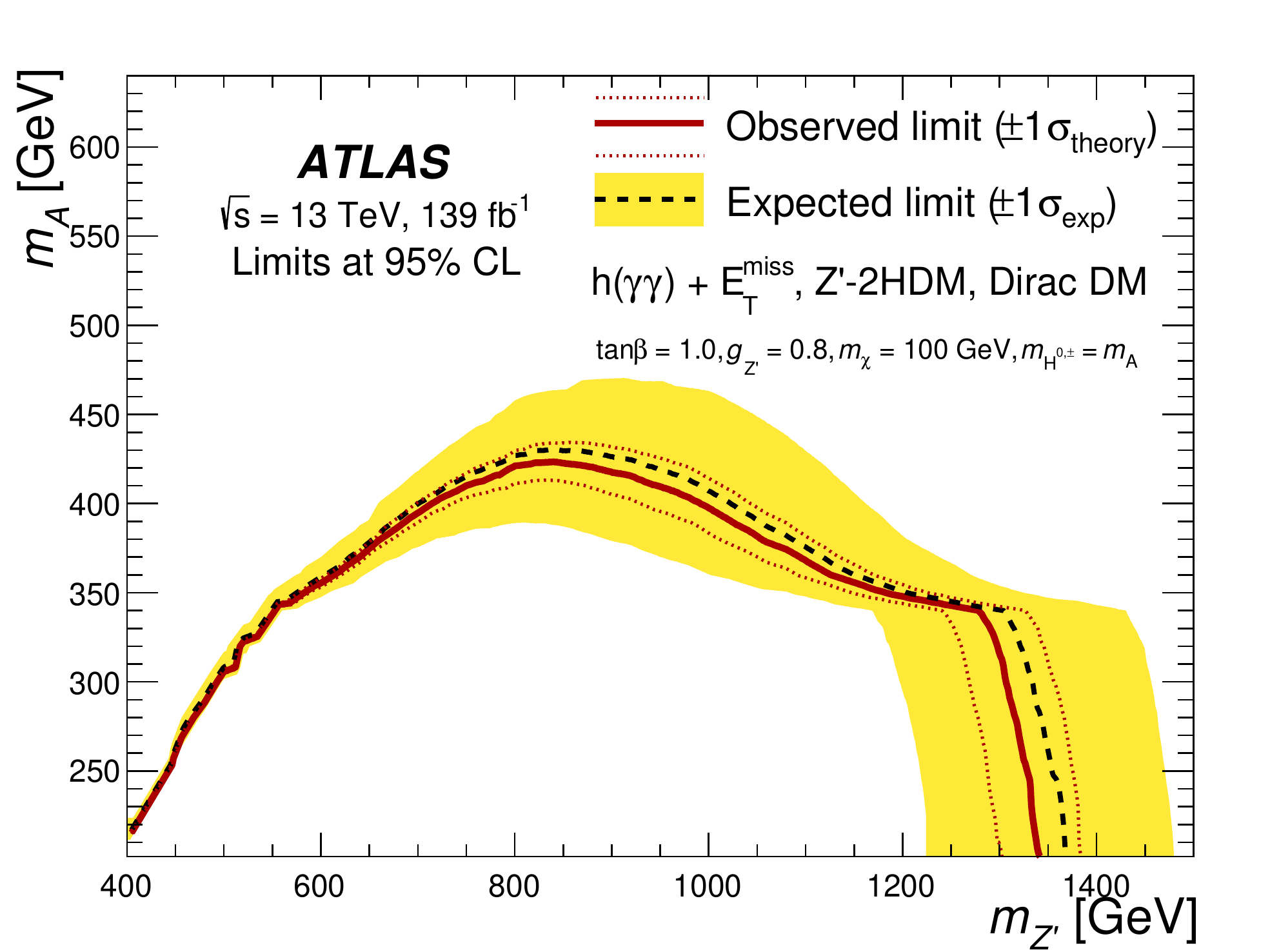}
\caption{
The observed (solid line) and expected (dashed lines) exclusion limit contours at 95\% CL for the $Z'$-2HDM model in the $m_A$--$m_{Z'}$ plane, for $\tan \beta = 1.0$, $g_{Z'} = 0.8$, $m_{\chi} = 100$~\GeV, and $m_{H^{0,\pm}}=m_{A}$.
The dotted lines represent the $\pm1\sigma$ theoretical uncertainty for the observed limit. The $\pm1\sigma$ expected exclusion limit contour is shown as the yellow band. Above $m_A = 350$~\GeV, competing decays from $A \to t\bar{t}$ cause the $A\to\chi\chi$ branching ratio to decrease quickly with increasing $m_A$, resulting in the feature near $m_{Z'} = 1300$~\GeV.
}
\label{fig:2Dexclusion_Zp2HDM}
\end{figure}

The observed and expected exclusion contours at 95\% CL for the $Z'$-2HDM model in the $m_A$--$m_{Z'}$ plane are shown in Figure~\ref{fig:2Dexclusion_Zp2HDM} for $m_{H^{0,\pm}}= m_A$, $m_{\chi} = 100$~\GeV, $\tan\beta = 1.0$, and $g'_Z = 0.8$. The maximum limit on $m_A$ reaches $420$~\GeV\ for a $Z'$ mass of $m_{Z'} = 825$~\GeV. Above $m_A = 350$~\GeV, competing decays from $A \to t\bar{t}$ cause the $A\to\chi\chi$ branching ratio to decrease quickly with increasing $m_A$, resulting in the feature near $m_{Z'} = 1300$~\GeV.
Changes in the DM mass $m_{\chi}$ would not significantly affect these results; such variations would only become important for very high DM masses which are not accessible to the current analysis~\cite{lhcdmf}.
 
\begin{figure}[!htbp]
\centering
\includegraphics[width=0.68\textwidth]{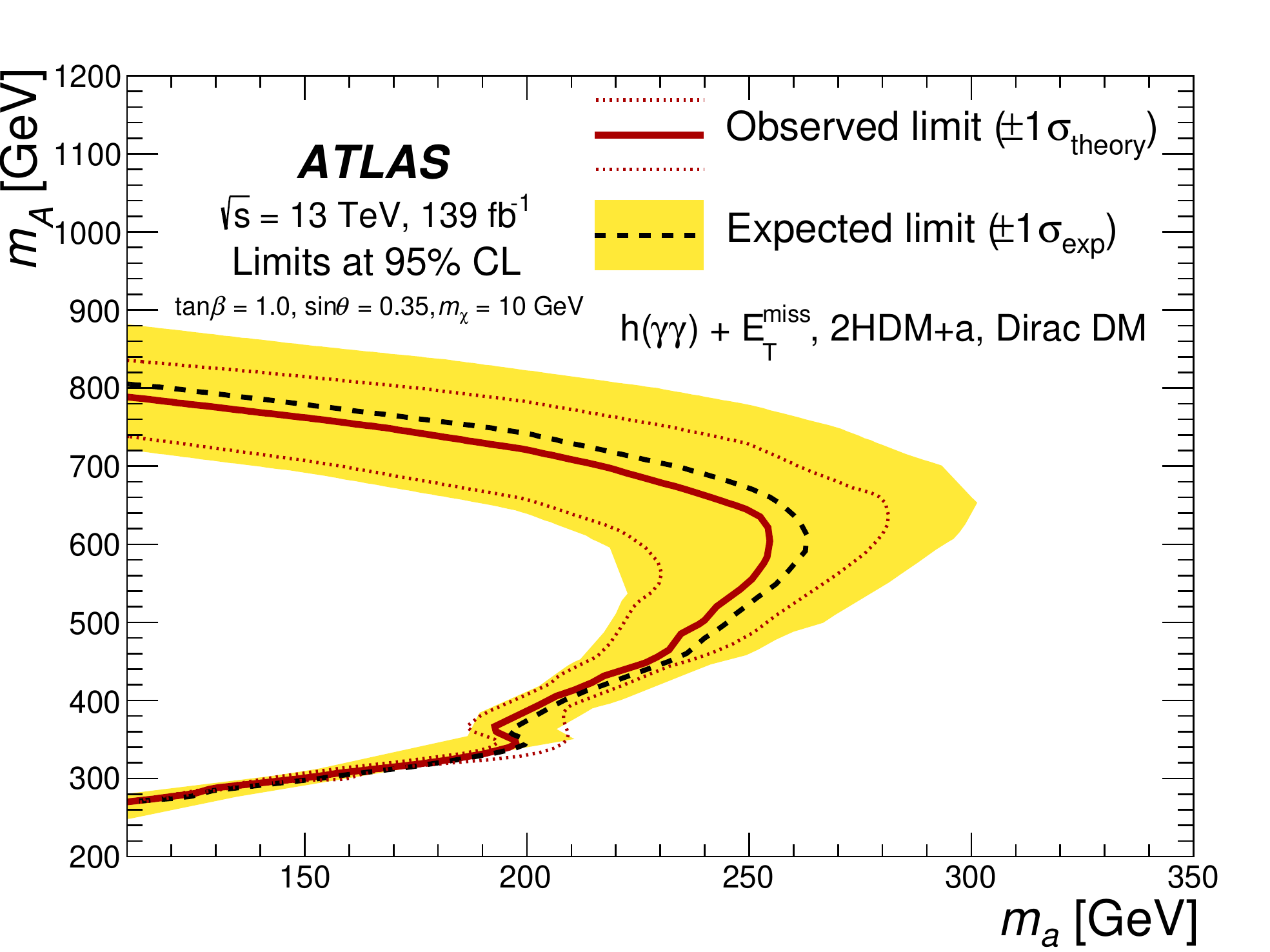}
\caption{
The observed (solid line) and expected (dashed lines) exclusion limit contours at 95\% CL for the 2HDM+a model in the $m_A$--$m_a$ plane, for $\tan \beta = 1.0$, $\sin \theta = 0.35$, and $m_{\chi} = 10$~\GeV.
The dotted lines represent the $\pm1\sigma$ theoretical uncertainty for the observed limit. The $\pm1\sigma$ expected exclusion limit contour is shown as the yellow band. Around the threshold $m_A = 350$~\GeV, the competition between the resonant decay $A \to t\bar{t}$ and $A \to ah$ causes the $A\to a h$ branching ratio to decrease suddenly with a limited increase of $m_A$, resulting in the feature near $m_{a} = 200$~\GeV. Above $m_A = 350$~\GeV, the limit is mainly driven by the increased selection efficiency due to the harder signal \etmiss distribution.
}
\label{fig:2Dexclusion_mAvsma}
\end{figure}

\begin{figure}[!htbp]
\centering
\includegraphics[width=0.68\textwidth]{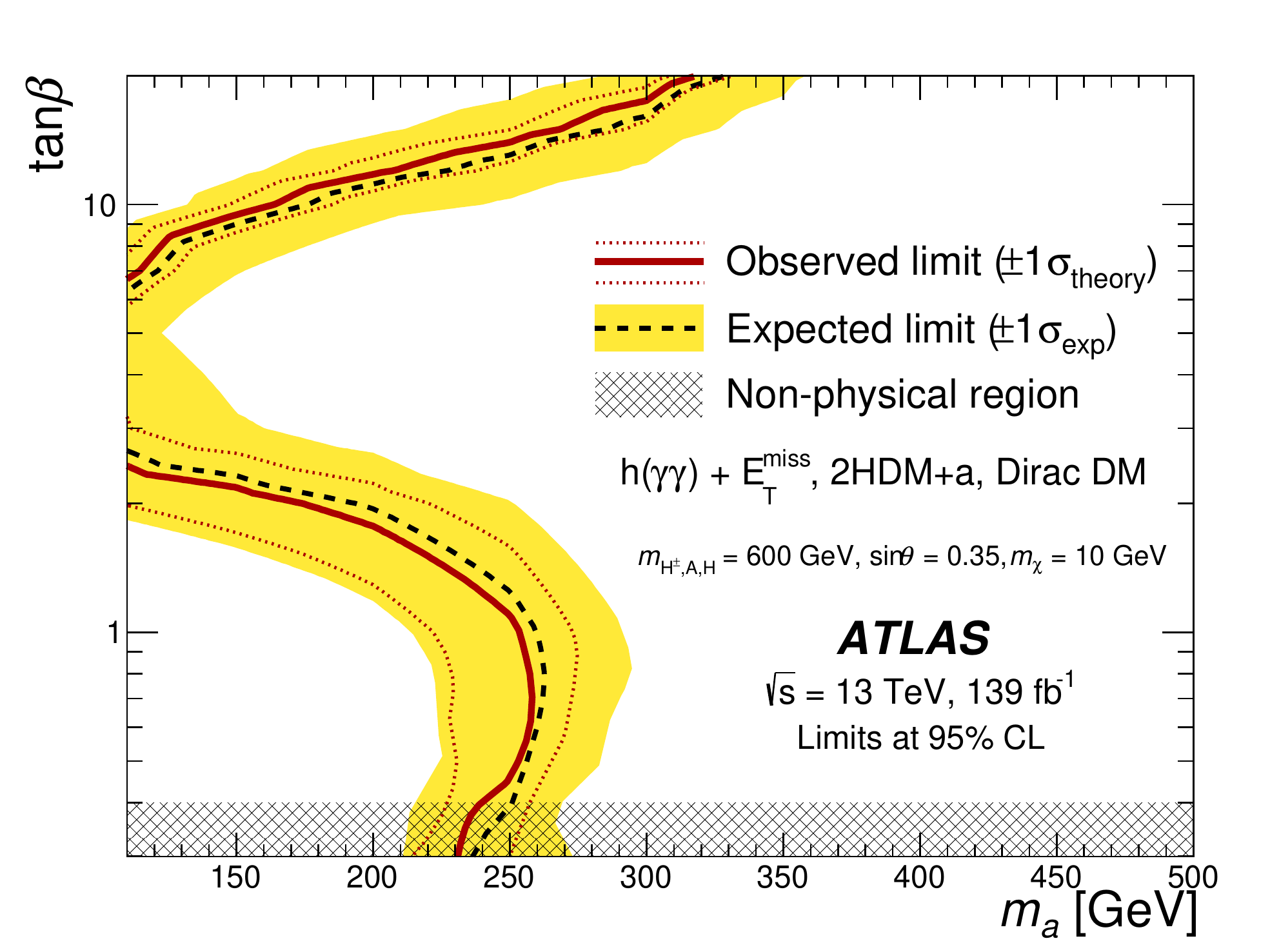}
\caption{
The observed (solid line) and expected (dashed lines) exclusion limit contours at 95\% CL for the 2HDM+a model in the $\tan\beta$--$m_a$ plane, for $m_{A,H^{\pm},H}= 600$~\GeV, $\sin \theta = 0.35$, and $m_{\chi} = 10$~\GeV.
The dotted lines represent the $\pm1\sigma$ theoretical uncertainty for the observed limit. The $\pm1\sigma$ expected exclusion limit contour is shown as the yellow band. The region $\tan\beta<0.4$ is covered with a hatched band because there the decay width of the low-mass Higgs boson is greater than 20\% of its mass, which renders the cross-section calculation unreliable~\cite{Aaboud:2019yqu}. The shape of the limit curve closely follows the signal cross section, which is dominated by ggF for low $\tan\beta$ and $b\bar{b}$-fusion for high $\tan\beta$.
}
\label{fig:2Dexclusion_mavstanBeta}
\end{figure}

\begin{figure}[!htbp]
\centering
\includegraphics[width=0.68\textwidth]{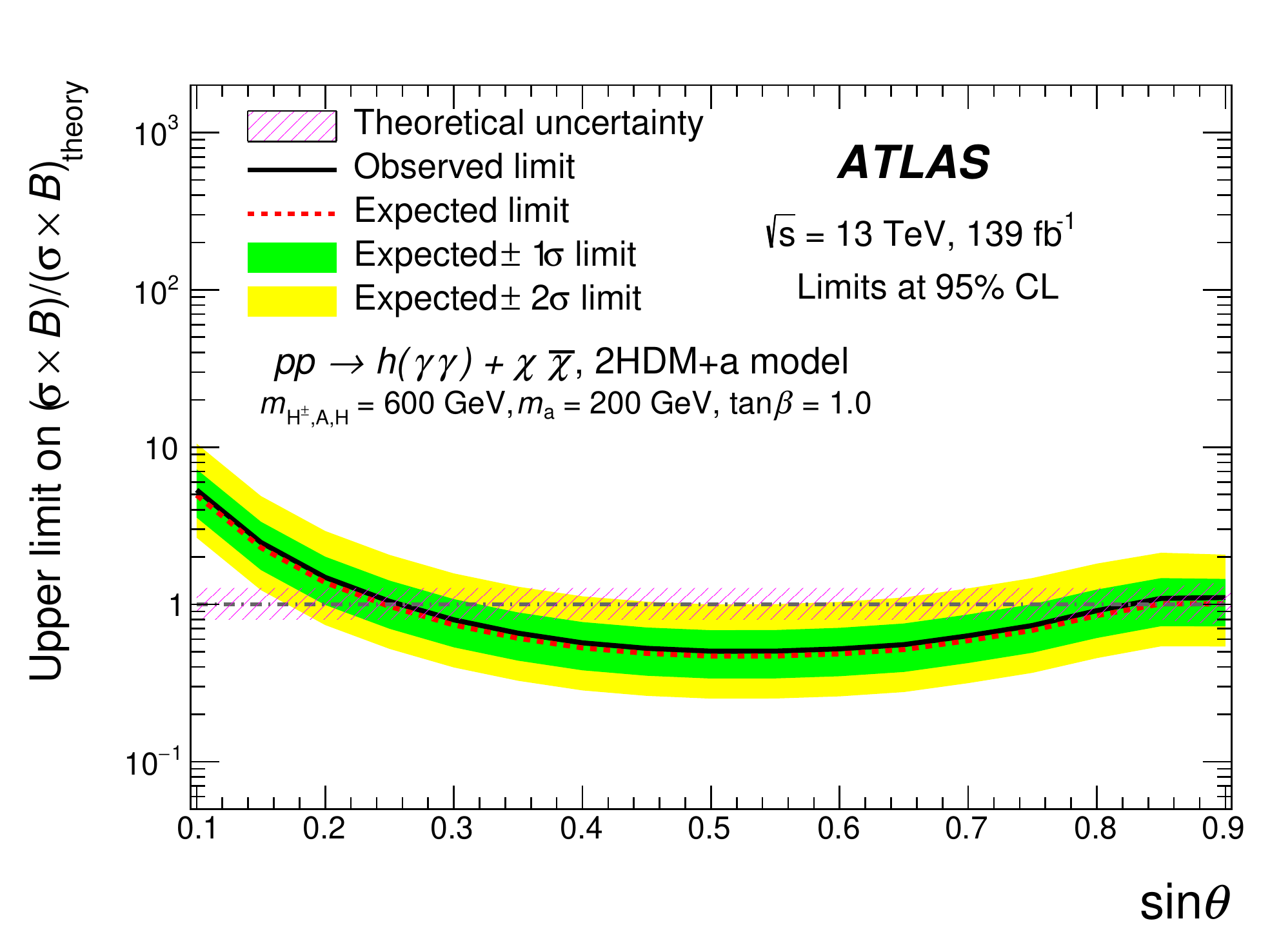}
\caption{ The observed (solid line) and expected (dashed lines) exclusion limits at 95\% CL
for the 2HDM+a model as a function of $\sin \theta$, for $m_{A,H^{\pm},H}= 600$~\GeV,
$m_a = 200$~\GeV, and $\tan \beta = 1.0$. Since the predicted yield for this model vanishes at mixing angles of $\theta = 0$ and $\theta = \pi/2$, $\sin\theta$ is limited to the range $[0.1, 0.9]$.
}
\label{fig:1Dexclusion_2HDMa_sinbeta_mA600ma200}
\end{figure}

Figures~\ref{fig:2Dexclusion_mAvsma} and~\ref{fig:2Dexclusion_mavstanBeta} show the observed and expected exclusion contours at 95\% CL for the 2HDM+a model in the $m_A$--$m_a$ and $\tan\beta$--$m_a$ planes, respectively, for $\sin \theta = 0.35$ and $m_{\chi}=10$~\GeV. In the $m_A$--$m_a$ scan for this benchmark point, the highest excluded $m_A$ is $800$~\GeV\ for $m_{a} = 110$~\GeV\ while the maximum excluded $m_a$ reaches about $260$~\GeV\ for $m_A = 600$~\GeV. The scan of $m_{a}$ starts from 110~\GeV\ to avoid opening the decay of $h\rightarrow a \chi\chi$ for $\chi=10$~\GeV. In the $\tan\beta$--$m_a$ scan, the region $\tan\beta<0.4$ is covered with a hatched band because there the decay width of the low-mass Higgs boson is greater than 20\% of its mass, which renders the cross-section calculation unreliable~\cite{Aaboud:2019yqu}. The shape of the limit curve closely follows the signal cross section, which is dominated by ggF for low $\tan\beta$ and $b\bar{b}$-fusion for high $\tan\beta$. Figure~\ref{fig:1Dexclusion_2HDMa_sinbeta_mA600ma200} shows the observed and expected limits on the 2HDM+a model as a function of $\sin \theta$ for $m_{A,H^{\pm},H}= 600$~\GeV, $m_a = 200$~\GeV, and $\tan \beta = 1.0$, showing that for this benchmark point the ATLAS data excludes a vast domain of possible mixing angle $\theta$~values.
 
\FloatBarrier

\section{Summary}
\label{sec:summary}
A search for dark matter in association with a Higgs boson decaying into two photons is presented. This study is based on data collected with the ATLAS detector during 2015--2018, corresponding to an integrated luminosity of 139~\ifb\ of proton--proton collisions at the LHC at a centre-of-mass
energy of 13~\TeV. No significant excess over the expected background is observed.
Upper limits at 95\% CL are set on the possible contributions to the cross sections times branching fraction of the Higgs boson decaying into
two photons in association with missing transverse momentum for three models: a $Z'_{B}$ model, a $Z'$-2HDM model, and a 2HDM+a model. The $Z'_{B}$ and $Z'$-2HDM models provide interesting information complementary to leptophobic vector or axial-vector $Z'$ models, which are already largely excluded by dijet searches~\cite{Aaboud:2019yqu}.
Limits at 95\% CL are set on the observed signal strength in the $m_{\chi}$--$m_{Z'}$ plane for the $Z'_{B}$ model, the $m_A$--$m_{Z'}$ plane for the $Z'$-2HDM model, and the $m_A$--$m_a$ and $\tan\beta$--$m_a$ planes for the 2HDM+a model. A one-dimensional scan of the mixing parameter $\sin \theta$ for $m_{A,H^{\pm},H}= 600$~\GeV, $m_a = 200$~\GeV\ and $\tan \beta = 1.0$ in the 2HDM+a model is performed as well. Additionally, the results for the $Z'_{B}$ model are interpreted in terms of 90\% CL limits on the DM--nucleon scattering cross section, as a function of the DM particle mass, for a spin-independent scenario. For a DM mass lower than 2~\GeV, the constraint with couplings $\sin \theta = 0.3$, $g_q=1/3$, and $g_{\chi} = 1$ placed on the DM--nucleon cross section is more stringent than limits from direct detection experiments at low DM mass, showing the complementarity between the several approaches trying to unveil the microscopic nature of DM.
 

\section*{Acknowledgements}
 

We thank CERN for the very successful operation of the LHC, as well as the
support staff from our institutions without whom ATLAS could not be
operated efficiently.
 
We acknowledge the support of
ANPCyT, Argentina;
YerPhI, Armenia;
ARC, Australia;
BMWFW and FWF, Austria;
ANAS, Azerbaijan;
SSTC, Belarus;
CNPq and FAPESP, Brazil;
NSERC, NRC and CFI, Canada;
CERN;
ANID, Chile;
CAS, MOST and NSFC, China;
Minciencias, Colombia;
MSMT CR, MPO CR and VSC CR, Czech Republic;
DNRF and DNSRC, Denmark;
IN2P3-CNRS and CEA-DRF/IRFU, France;
SRNSFG, Georgia;
BMBF, HGF and MPG, Germany;
GSRI, Greece;
RGC and Hong Kong SAR, China;
ISF and Benoziyo Center, Israel;
INFN, Italy;
MEXT and JSPS, Japan;
CNRST, Morocco;
NWO, Netherlands;
RCN, Norway;
MNiSW and NCN, Poland;
FCT, Portugal;
MNE/IFA, Romania;
JINR;
MES of Russia and NRC KI, Russian Federation;
MESTD, Serbia;
MSSR, Slovakia;
ARRS and MIZ\v{S}, Slovenia;
DSI/NRF, South Africa;
MICINN, Spain;
SRC and Wallenberg Foundation, Sweden;
SERI, SNSF and Cantons of Bern and Geneva, Switzerland;
MOST, Taiwan;
TAEK, Turkey;
STFC, United Kingdom;
DOE and NSF, United States of America.
In addition, individual groups and members have received support from
BCKDF, CANARIE, Compute Canada and CRC, Canada;
COST, ERC, ERDF, Horizon 2020 and Marie Sk{\l}odowska-Curie Actions, European Union;
Investissements d'Avenir Labex, Investissements d'Avenir Idex and ANR, France;
DFG and AvH Foundation, Germany;
Herakleitos, Thales and Aristeia programmes co-financed by EU-ESF and the Greek NSRF, Greece;
BSF-NSF and GIF, Israel;
Norwegian Financial Mechanism 2014-2021, Norway;
La Caixa Banking Foundation, CERCA Programme Generalitat de Catalunya and PROMETEO and GenT Programmes Generalitat Valenciana, Spain;
G\"{o}ran Gustafssons Stiftelse, Sweden;
The Royal Society and Leverhulme Trust, United Kingdom.
 
The crucial computing support from all WLCG partners is acknowledged gratefully, in particular from CERN, the ATLAS Tier-1 facilities at TRIUMF (Canada), NDGF (Denmark, Norway, Sweden), CC-IN2P3 (France), KIT/GridKA (Germany), INFN-CNAF (Italy), NL-T1 (Netherlands), PIC (Spain), ASGC (Taiwan), RAL (UK) and BNL (USA), the Tier-2 facilities worldwide and large non-WLCG resource providers. Major contributors of computing resources are listed in Ref.~\cite{ATL-SOFT-PUB-2021-003}.
 

\printbibliography

@article{Aaboud:2017phn,
    author = "{ATLAS Collaboration}",
    collaboration = "ATLAS",
    title = "{Search for dark matter and other new phenomena in events with an energetic jet and large missing transverse momentum using the ATLAS detector}",
    eprint = "1711.03301",
    archivePrefix = "arXiv",
    primaryClass = "hep-ex",
    reportNumber = "CERN-EP-2017-230",
    doi = "10.1007/JHEP01(2018)126",
    journal = "JHEP",
    volume = "01",
    pages = "126",
    year = "2018"
}

@article{Aaboud:2018xdl,
    author = "{ATLAS Collaboration}",
    collaboration = "ATLAS",
    title = "{Search for dark matter in events with a hadronically decaying vector boson and missing transverse momentum in $pp$ collisions at $\sqrt{s} = 13$ TeV with the ATLAS detector}",
    eprint = "1807.11471",
    archivePrefix = "arXiv",
    primaryClass = "hep-ex",
    reportNumber = "CERN-EP-2018-083",
    doi = "10.1007/JHEP10(2018)180",
    journal = "JHEP",
    volume = "10",
    pages = "180",
    year = "2018"
}

@article{Aaboud:2017dor,
    author = "{ATLAS Collaboration}",
    collaboration = "ATLAS",
    title = "{Search for dark matter at $\sqrt{s}=13$ TeV in final states containing an energetic photon and large missing transverse momentum with the ATLAS detector}",
    eprint = "1704.03848",
    archivePrefix = "arXiv",
    primaryClass = "hep-ex",
    reportNumber = "CERN-EP-2017-044",
    doi = "10.1140/epjc/s10052-017-4965-8",
    journal = "Eur. Phys. J. C",
    volume = "77",
    number = "6",
    pages = "393",
    year = "2017"
}

@article{Aaboud:2017uak,
    author = "{ATLAS Collaboration}",
    collaboration = "ATLAS",
    title = "{Search for dark matter in association with a Higgs boson decaying to two photons at $\sqrt{s}$ = 13 TeV with the ATLAS detector}",
    eprint = "1706.03948",
    archivePrefix = "arXiv",
    primaryClass = "hep-ex",
    reportNumber = "CERN-EP-2017-079",
    doi = "10.1103/PhysRevD.96.112004",
    journal = "Phys. Rev. D",
    volume = "96",
    number = "11",
    pages = "112004",
    year = "2017"
}

@article{LUCID2,
  author={G. Avoni and others},
  title={The new LUCID-2 detector for luminosity measurement and monitoring in ATLAS},
  journal={JINST},
  volume={13},
  number={07},
  pages={P07017},
  doi="10.1088/1748-0221/13/07/P07017",
  year={2018},
}

@inproceedings{Chen:2016:XST:2939672.2939785,
 author = {Chen, Tianqi and Guestrin, Carlos},
 title = {{XGBoost}: A Scalable Tree Boosting System},
 eprint         = "1603.02754",
 archivePrefix  = "arXiv",
 booktitle = {Proceedings of the 22nd ACM SIGKDD International Conference on Knowledge Discovery and Data Mining},
 series = {KDD '16},
 year = {2016},
 isbn = {978-1-4503-4232-2},
 location = {San Francisco, California, USA},
 pages = {785--794},
 numpages = {10},
 url = {http://doi.acm.org/10.1145/2939672.2939785},
 doi = {10.1145/2939672.2939785},
 acmid = {2939785},
 publisher = {ACM},
 address = {New York, NY, USA},
}

@article{Aad:2019wsl,
      author = "{ATLAS Collaboration}",
      title          = "{Performance of electron and photon triggers in ATLAS during LHC Run 2}",
      collaboration  = "ATLAS",
      journal        = "Eur. Phys. J. C",
      eprint         = "1909.00761",
      archivePrefix  = "arXiv",
      primaryClass   = "hep-ex",
      reportNumber   = "CERN-EP-2019-169",
      doi = "10.1140/epjc/s10052-019-7500-2",
      journal = "Eur. Phys. J. C",
      number = "1",
      pages = "47",
      volume = "80",
      year = "2020",
      SLACcitation   = "%%CITATION = ARXIV:1909.00761;%%"
}

@article{ATLAS:1300517,
        author = "{ATLAS Collaboration}",
        collaboration = "ATLAS",
        title     = "{The simulation principle and performance of the ATLAS
                      fast calorimeter simulation FastCaloSim}",
        note     = "ATL-PHYS-PUB-2010-013",
        year     = "2010",
        url = "http://cdsweb.cern.ch/record/1300517"
}

@Booklet{JVT,
  author = "{ATLAS Collaboration}",
  title         = "{Tagging and suppression of pileup jets with the ATLAS detector}",
  institution   = "CERN",
  address       = "Geneva",
  number        = "ATLAS-CONF-2014-018",
  month         = "05",
  year          = "2014",
  reportNumber  = "ATLAS-CONF-2014-018",
  url           = "https://cds.cern.ch/record/1700870",
  howpublished   = "{ATLAS-CONF-2014-018}",
}

@article{EGAM-2018-01,
    author         = "{ATLAS Collaboration}",
    title          = "{Electron and photon performance measurements with the ATLAS detector using the 2015--2017 LHC proton--proton collision data}",
    journal        = "JINST",
    volume         = "14",
    year           = "2019",
    pages          = "P12006",
    doi            = "10.1088/1748-0221/14/12/P12006",
    reportNumber   = "CERN-EP-2019-145",
    eprint         = "1908.00005",
    archivePrefix  = "arXiv",
    primaryClass   = "hep-ex",
}

@article{Aaboud:2019yqu,
    author = "{ATLAS Collaboration}",
    collaboration = "ATLAS",
    title = "{Constraints on mediator-based dark matter and scalar dark energy models using $\sqrt s = 13$ TeV $pp$ collision data collected by the ATLAS detector}",
    eprint = "1903.01400",
    archivePrefix = "arXiv",
    primaryClass = "hep-ex",
    reportNumber = "CERN-EP-2018-334",
    doi = "10.1007/JHEP05(2019)142",
    journal = "JHEP",
    volume = "05",
    pages = "142",
    year = "2019"
}

@article{Aaboud:2017yyg,
    author = "{ATLAS Collaboration}",
    collaboration = "ATLAS",
    title = "{Search for new phenomena in high-mass diphoton final states using 37 fb$^{-1}$ of proton--proton collisions collected at $\sqrt{s}=13$ TeV with the ATLAS detector}",
    eprint = "1707.04147",
    archivePrefix = "arXiv",
    primaryClass = "hep-ex",
    reportNumber = "CERN-EP-2017-132",
    doi = "10.1016/j.physletb.2017.10.039",
    journal = "Phys. Lett. B",
    volume = "775",
    pages = "105--125",
    year = "2017"
}

@article{Aaboud:2018tkc,
    author = "{ATLAS Collaboration}",
    collaboration = "ATLAS",
    title = "{Performance of missing transverse momentum reconstruction with the ATLAS detector using proton-proton collisions at $\sqrt{s}$ = 13 TeV}",
    eprint = "1802.08168",
    archivePrefix = "arXiv",
    primaryClass = "hep-ex",
    reportNumber = "CERN-EP-2017-274",
    doi = "10.1140/epjc/s10052-018-6288-9",
    journal = "Eur. Phys. J. C",
    volume = "78",
    number = "11",
    pages = "903",
    year = "2018"
}

@article{Bauer2017,
 author="Bauer, Martin and Haisch, Ulrich and Kahlhoefer, Felix",
 title="Simplified dark matter models with two Higgs doublets: I. Pseudoscalar mediators",
 eprint = "1701.07427",
 archivePrefix = "arXiv",
 journal="JHEP",
 year="2017",
 month="05",
 volume="05",
 number="5",
 pages="138",
 issn="1029-8479",
 doi="10.1007/JHEP05(2017)138",
}

@article{2018arXiv181009420A,
       author = {{Abe}, Tomohiro and {Afik}, Yoav and {Albert}, Andreas and
         {Anelli}, Christopher R. and {Barak}, Liron and {Bauer}, Martin and
         {Behr}, J. Katharina and {Bell}, Nicole F. and {Boveia}, Antonio and
         {Brandt}, Oleg and {Busoni}, Giorgio and {Carpenter}, Linda M. and
         {Chen}, Yu-Heng and {Doglioni}, Caterina and {Elliot}, Alison and
         {Fujiwara}, Motoko and {Genest}, Marie-Helene and {Gerosa}, Raffaele and
         {Gori}, Stefania and {Gramling}, Johanna and {Grohsjean}, Alexander and
         {Gustavino}, Giuliano and {Hahn}, Kristian and {Haisch}, Ulrich and
         {Henkelmann}, Lars and {Hisano}, Junji and {Huitfeldt}, Anders and
         {Ippolito}, Valerio and {Kahlhoefer}, Felix and {Landsberg}, Greg and
         {Lowette}, Steven and {Maier}, Benedikt and {Maltoni}, Fabio and
         {Muehlleitner}, Margarete and {No}, Jose M. and {Pani}, Priscilla and
         {Polesello}, Giacomo and {Price}, Darren D. and {Robens}, Tania and
         {Rovelli}, Giulia and {Rozen}, Yoram and {Sanderson}, Isaac W. and
         {Santos}, Rui and {Sevova}, Stanislava and {Sperka}, David and
         {Sung}, Kevin and {Tait}, Tim M.~P. and {Terashi}, Koji and
         {Ungaro}, Francesca C. and {Vryonidou}, Eleni and {Yu}, Shin-Shan and
         {Wu}, Sau Lan and {Zhou}, Chen},
        title = "{LHC Dark Matter Working Group: Next-generation spin-0 dark matter models}",
         year = "2018",
        month = "10",
archivePrefix = {arXiv},
       eprint = {1810.09420},
 primaryClass = {hep-ex},
       adsurl = {https://ui.adsabs.harvard.edu/abs/2018arXiv181009420A}
}

@article{Beenakker:2002nc,
    author = "Beenakker, W. and Dittmaier, S. and Kramer, M. and Plumper, B. and Spira, M. and Zerwas, P.M.",
    title = "{NLO QCD corrections to $t\bar{t}H$ production in hadron collisions}",
    eprint = "hep-ph/0211352",
    archivePrefix = "arXiv",
    reportNumber = "DESY-02-177, EDINBURGH-2002-18, MPI-PHT-2002-70, PSI-PR-02-22",
    doi = "10.1016/S0550-3213(03)00044-0",
    journal = "Nucl. Phys. B",
    volume = "653",
    pages = "151--203",
    year = "2003"
}

@misc{kalogeropoulos2018syscalc,
    title={The SysCalc code: A tool to derive theoretical systematic uncertainties},
    author={Alexis Kalogeropoulos and Johan Alwall},
    year={2018},
    eprint={1801.08401},
    archivePrefix={arXiv},
    primaryClass={hep-ph}
}

@article{Sirunyan_2019,
   title={Search for dark matter produced in association with a Higgs boson decaying to a pair of bottom quarks in proton–proton collisions at $\sqrt{s}=13\,\text {Te}\text {V} $},
   volume={79},
   ISSN={1434-6052},
   DOI={10.1140/epjc/s10052-019-6730-7},
   number={3},
   journal={Eur. Phys. J. C},
   publisher={Springer Science and Business Media LLC},
   author="{CMS Collaboration}",
   collaboration = "CMS",
   year={2019},
   month={Mar},
   pages={280},
   eprint = "1811.06562",
   archivePrefix = "arXiv",
}

@article{Sirunyan:2018fpy,
    author = "{CMS Collaboration}",
    collaboration = "CMS",
    title = "{Search for dark matter produced in association with a Higgs boson decaying to $\gamma\gamma$ or $\tau^+\tau^-$ at $\sqrt{s} =$ 13 TeV}",
    eprint = "1806.04771",
    archivePrefix = "arXiv",
    primaryClass = "hep-ex",
    reportNumber = "CMS-EXO-16-055, CERN-EP-2018-129",
    doi = "10.1007/JHEP09(2018)046",
    journal = "JHEP",
    volume = "09",
    pages = "046",
    year = "2018"
}

@article{PhysRevLett.123.251801,
  title = {Light Dark Matter Search with Ionization Signals in XENON1T},
  author = {Aprile, E. and Aalbers, J. and Agostini, F. and Alfonsi, M. and Althueser, L. and Amaro, F. D. and Antochi, V. C. and Angelino, E. and Arneodo, F. and Barge, D. and Baudis, L. and Bauermeister, B. and Bellagamba, L. and Benabderrahmane, M. L. and Berger, T. and Breur, P. A. and Brown, A. and Brown, E. and Bruenner, S. and Bruno, G. and Budnik, R. and Capelli, C. and Cardoso, J. M. R. and Cichon, D. and Coderre, D. and Colijn, A. P. and Conrad, J. and Cussonneau, J. P. and Decowski, M. P. and de Perio, P. and Depoian, A. and Di Gangi, P. and Di Giovanni, A. and Diglio, S. and Elykov, A. and Eurin, G. and Fei, J. and Ferella, A. D. and Fieguth, A. and Fulgione, W. and Gaemers, P. and Gallo Rosso, A. and Galloway, M. and Gao, F. and Garbini, M. and Grandi, L. and Greene, Z. and Hasterok, C. and Hils, C. and Hogenbirk, E. and Howlett, J. and Iacovacci, M. and Itay, R. and Joerg, F. and Kazama, S. and Kish, A. and Kobayashi, M. and Koltman, G. and Kopec, A. and Landsman, H. and Lang, R. F. and Levinson, L. and Lin, Q. and Lindemann, S. and Lindner, M. and Lombardi, F. and Lopes, J. A. M. and L\'opez Fune, E. and Macolino, C. and Mahlstedt, J. and Manfredini, A. and Marignetti, F. and Marrod\'an Undagoitia, T. and Masbou, J. and Mastroianni, S. and Messina, M. and Micheneau, K. and Miller, K. and Molinario, A. and Mor\aa{}, K. and Mosbacher, Y. and Murra, M. and Naganoma, J. and Ni, K. and Oberlack, U. and Odgers, K. and Palacio, J. and Pelssers, B. and Peres, R. and Pienaar, J. and Pizzella, V. and Plante, G. and Podviianiuk, R. and Qin, J. and Qiu, H. and Ram\'{\i}rez Garc\'{\i}a, D. and Reichard, S. and Riedel, B. and Rocchetti, A. and Rupp, N. and dos Santos, J. M. F. and Sartorelli, G. and \ifmmode \check{S}\else \v{S}\fi{}ar\ifmmode \check{c}\else \v{c}\fi{}evi\ifmmode \acute{c}\else \'{c}\fi{}, N. and Scheibelhut, M. and Schindler, S. and Schreiner, J. and Schulte, D. and Schumann, M. and Scotto Lavina, L. and Selvi, M. and Shagin, P. and Shockley, E. and Silva, M. and Simgen, H. and Therreau, C. and Thers, D. and Toschi, F. and Trinchero, G. and Tunnell, C. and Upole, N. and Vargas, M. and Volta, G. and Wack, O. and Wang, H. and Wei, Y. and Weinheimer, C. and Wenz, D. and Wittweg, C. and Wulf, J. and Ye, J. and Zhang, Y. and Zhu, T. and Zopounidis, J. P.},
  collaboration = {XENON Collaboration},
  journal = {Phys. Rev. Lett.},
  volume = {123},
  pages = {251801},
  numpages = {8},
  year = {2019},
  month = {Dec},
  publisher = {American Physical Society},
  doi = {10.1103/PhysRevLett.123.251801},
  url = {https://link.aps.org/doi/10.1103/PhysRevLett.123.251801},
  eprint = "1907.11485",
  archivePrefix = "arXiv",

}

@article{PhysRevLett.123.241803,
  title = {Search for Light Dark Matter Interactions Enhanced by the Migdal Effect or Bremsstrahlung in XENON1T},
  author = {Aprile, E. and Aalbers, J. and Agostini, F. and Alfonsi, M. and Althueser, L. and Amaro, F. D. and Antochi, V. C. and Angelino, E. and Arneodo, F. and Barge, D. and Baudis, L. and Bauermeister, B. and Bellagamba, L. and Benabderrahmane, M. L. and Berger, T. and Breur, P. A. and Brown, A. and Brown, E. and Bruenner, S. and Bruno, G. and Budnik, R. and Capelli, C. and Cardoso, J. M. R. and Cichon, D. and Coderre, D. and Colijn, A. P. and Conrad, J. and Cussonneau, J. P. and Decowski, M. P. and de Perio, P. and Depoian, A. and Di Gangi, P. and Di Giovanni, A. and Diglio, S. and Elykov, A. and Eurin, G. and Fei, J. and Ferella, A. D. and Fieguth, A. and Fulgione, W. and Gaemers, P. and Gallo Rosso, A. and Galloway, M. and Gao, F. and Garbini, M. and Grandi, L. and Greene, Z. and Hasterok, C. and Hils, C. and Hogenbirk, E. and Howlett, J. and Iacovacci, M. and Itay, R. and Joerg, F. and Kazama, S. and Kish, A. and Kobayashi, M. and Koltman, G. and Kopec, A. and Landsman, H. and Lang, R. F. and Levinson, L. and Lin, Q. and Lindemann, S. and Lindner, M. and Lombardi, F. and Lopes, J. A. M. and L\'opez Fune, E. and Macolino, C. and Mahlstedt, J. and Manenti, M. and Manfredini, A. and Marignetti, F. and Marrod\'an Undagoitia, T. and Masbou, J. and Mastroianni, S. and Messina, M. and Micheneau, K. and Miller, K. and Molinario, A. and Mor\aa{}, K. and Mosbacher, Y. and Murra, M. and Naganoma, J. and Ni, K. and Oberlack, U. and Odgers, K. and Palacio, J. and Pelssers, B. and Peres, R. and Pienaar, J. and Pizzella, V. and Plante, G. and Podviianiuk, R. and Qin, J. and Qiu, H. and Ram\'{\i}rez Garc\'{\i}a, D. and Reichard, S. and Riedel, B. and Rocchetti, A. and Rupp, N. and dos Santos, J. M. F. and Sartorelli, G. and \ifmmode \check{S}\else \v{S}\fi{}ar\ifmmode \check{c}\else \v{c}\fi{}evi\ifmmode \acute{c}\else \'{c}\fi{}, N. and Scheibelhut, M. and Schindler, S. and Schreiner, J. and Schulte, D. and Schumann, M. and Scotto Lavina, L. and Selvi, M. and Shagin, P. and Shockley, E. and Silva, M. and Simgen, H. and Therreau, C. and Thers, D. and Toschi, F. and Trinchero, G. and Tunnell, C. and Upole, N. and Vargas, M. and Volta, G. and Wack, O. and Wang, H. and Wei, Y. and Weinheimer, C. and Wenz, D. and Wittweg, C. and Wulf, J. and Ye, J. and Zhang, Y. and Zhu, T. and Zopounidis, J. P.},
  collaboration = {XENON Collaboration},
  journal = {Phys. Rev. Lett.},
  volume = {123},
  pages = {241803},
  numpages = {7},
  year = {2019},
  month = {Dec},
  publisher = {American Physical Society},
  doi = {10.1103/PhysRevLett.123.241803},
  url = {https://link.aps.org/doi/10.1103/PhysRevLett.123.241803},
  eprint = "1907.12771",
  archivePrefix = "arXiv",
}

@article{PhysRevLett.121.111302,
  title = {Dark Matter Search Results from a One Ton-Year Exposure of XENON1T},
  author = {Aprile, E. and Aalbers, J. and Agostini, F. and Alfonsi, M. and Althueser, L. and Amaro, F. D. and Anthony, M. and Arneodo, F. and Baudis, L. and Bauermeister, B. and Benabderrahmane, M. L. and Berger, T. and Breur, P. A. and Brown, A. and Brown, A. and Brown, E. and Bruenner, S. and Bruno, G. and Budnik, R. and Capelli, C. and Cardoso, J. M. R. and Cichon, D. and Coderre, D. and Colijn, A. P. and Conrad, J. and Cussonneau, J. P. and Decowski, M. P. and de Perio, P. and Di Gangi, P. and Di Giovanni, A. and Diglio, S. and Elykov, A. and Eurin, G. and Fei, J. and Ferella, A. D. and Fieguth, A. and Fulgione, W. and Gallo Rosso, A. and Galloway, M. and Gao, F. and Garbini, M. and Geis, C. and Grandi, L. and Greene, Z. and Qiu, H. and Hasterok, C. and Hogenbirk, E. and Howlett, J. and Itay, R. and Joerg, F. and Kaminsky, B. and Kazama, S. and Kish, A. and Koltman, G. and Landsman, H. and Lang, R. F. and Levinson, L. and Lin, Q. and Lindemann, S. and Lindner, M. and Lombardi, F. and Lopes, J. A. M. and Mahlstedt, J. and Manfredini, A. and Marrod\'an Undagoitia, T. and Masbou, J. and Masson, D. and Messina, M. and Micheneau, K. and Miller, K. and Molinario, A. and Mor\aa{}, K. and Murra, M. and Naganoma, J. and Ni, K. and Oberlack, U. and Pelssers, B. and Piastra, F. and Pienaar, J. and Pizzella, V. and Plante, G. and Podviianiuk, R. and Priel, N. and Ram\'{\i}rez Garc\'{\i}a, D. and Rauch, L. and Reichard, S. and Reuter, C. and Riedel, B. and Rizzo, A. and Rocchetti, A. and Rupp, N. and dos Santos, J. M. F. and Sartorelli, G. and Scheibelhut, M. and Schindler, S. and Schreiner, J. and Schulte, D. and Schumann, M. and Scotto Lavina, L. and Selvi, M. and Shagin, P. and Shockley, E. and Silva, M. and Simgen, H. and Thers, D. and Toschi, F. and Trinchero, G. and Tunnell, C. and Upole, N. and Vargas, M. and Wack, O. and Wang, H. and Wang, Z. and Wei, Y. and Weinheimer, C. and Wittweg, C. and Wulf, J. and Ye, J. and Zhang, Y. and Zhu, T.},
  collaboration = {XENON Collaboration  7},
  journal = {Phys. Rev. Lett.},
  volume = {121},
  pages = {111302},
  numpages = {8},
  year = {2018},
  month = {Sep},
  publisher = {American Physical Society},
  doi = {10.1103/PhysRevLett.121.111302},
  url = {https://link.aps.org/doi/10.1103/PhysRevLett.121.111302},
  eprint = "1805.12562",
  archivePrefix = "arXiv",
}

@article{PhysRevLett.121.081307,
  title = {Low-Mass Dark Matter Search with the DarkSide-50 Experiment},
  author = {Agnes, P. and Albuquerque, I. F. M. and Alexander, T. and Alton, A. K. and Araujo, G. R. and Asner, D. M. and Ave, M. and Back, H. O. and Baldin, B. and Batignani, G. and Biery, K. and Bocci, V. and Bonfini, G. and Bonivento, W. and Bottino, B. and Budano, F. and Bussino, S. and Cadeddu, M. and Cadoni, M. and Calaprice, F. and Caminata, A. and Canci, N. and Candela, A. and Caravati, M. and Cariello, M. and Carlini, M. and Carpinelli, M. and Catalanotti, S. and Cataudella, V. and Cavalcante, P. and Cavuoti, S. and Cereseto, R. and Chepurnov, A. and Cical\`o, C. and Cifarelli, L. and Cocco, A. G. and Covone, G. and D'Angelo, D. and D'Incecco, M. and D'Urso, D. and Davini, S. and De Candia, A. and De Cecco, S. and De Deo, M. and De Filippis, G. and De Rosa, G. and De Vincenzi, M. and Demontis, P. and Derbin, A. V. and Devoto, A. and Di Eusanio, F. and Di Pietro, G. and Dionisi, C. and Downing, M. and Edkins, E. and Empl, A. and Fan, A. and Fiorillo, G. and Fomenko, K. and Franco, D. and Gabriele, F. and Gabrieli, A. and Galbiati, C. and Garcia Abia, P. and Ghiano, Chiara and Giagu, S. and Giganti, C. and Giovanetti, G. K. and Gorchakov, O. and Goretti, A. M. and Granato, F. and Gromov, M. and Guan, M. and Guardincerri, Y. and Gulino, M. and Hackett, B. R. and Hassanshahi, M. H. and Herner, K. and Hosseini, B. and Hughes, D. and Humble, P. and Hungerford, E. V. and Ianni, Al. and Ianni, An. and Ippolito, V. and James, I. and Johnson, T. N. and Kahn, Y. and Keeter, K. and Kendziora, C. L. and Kochanek, I. and Koh, G. and Korablev, D. and Korga, G. and Kubankin, A. and Kuss, M. and La Commara, M. and Lai, M. and Li, X. and Lisanti, M. and Lissia, M. and Loer, B. and Longo, G. and Ma, Y. and Machado, A. A. and Machulin, I. N. and Mandarano, A. and Mapelli, L. and Mari, S. M. and Maricic, J. and Martoff, C. J. and Messina, A. and Meyers, P. D. and Milincic, R. and Mishra-Sharma, S. and Monte, A. and Morrocchi, M. and Mount, B. J. and Muratova, V. N. and Musico, P. and Nania, R. and Navrer Agasson, A. and Nozdrina, A. O. and Oleinik, A. and Orsini, M. and Ortica, F. and Pagani, L. and Pallavicini, M. and Pandola, L. and Pantic, E. and Paoloni, E. and Pazzona, F. and Pelczar, K. and Pelliccia, N. and Pesudo, V. and Pocar, A. and Pordes, S. and Poudel, S. S. and Pugachev, D. A. and Qian, H. and Ragusa, F. and Razeti, M. and Razeto, A. and Reinhold, B. and Renshaw, A. L. and Rescigno, M. and Riffard, Q. and Romani, A. and Rossi, B. and Rossi, N. and Sablone, D. and Samoylov, O. and Sands, W. and Sanfilippo, S. and Sant, M. and Santorelli, R. and Savarese, C. and Scapparone, E. and Schlitzer, B. and Segreto, E. and Semenov, D. A. and Shchagin, A. and Sheshukov, A. and Singh, P. N. and Skorokhvatov, M. D. and Smirnov, O. and Sotnikov, A. and Stanford, C. and Stracka, S. and Suffritti, G. B. and Suvorov, Y. and Tartaglia, R. and Testera, G. and Tonazzo, A. and Trinchese, P. and Unzhakov, E. V. and Verducci, M. and Vishneva, A. and Vogelaar, B. and Wada, M. and Waldrop, T. J. and Wang, H. and Wang, Y. and Watson, A. W. and Westerdale, S. and Wojcik, M. M. and Wojcik, M. and Xiang, X. and Xiao, X. and Yang, C. and Ye, Z. and Zhu, C. and Zichichi, A. and Zuzel, G.},
  collaboration = {DarkSide Collaboration},
  journal = {Phys. Rev. Lett.},
  volume = {121},
  pages = {081307},
  numpages = {8},
  year = {2018},
  month = {Aug},
  publisher = {American Physical Society},
  doi = {10.1103/PhysRevLett.121.081307},
  url = {https://link.aps.org/doi/10.1103/PhysRevLett.121.081307},
  eprint = "1802.06994",
  archivePrefix = "arXiv",
}

@Article{Bernstein,
    author         = "{Bernstein, S.}",
    title          = "{Démonstration du Théorème de Weierstrass fondée sur le calcul des Probabilités}",
    journal        = "Comm. Soc. Math. Kharkov",
    volume         = "13",
    year           = "1912",
    pages          = "1",
}

@article{Aad:2012tfa,
      author         = "{ATLAS Collaboration}",
      title          = "{Observation of a new particle in the search for the
                        Standard Model Higgs boson with the ATLAS detector at the
                        LHC}",
      collaboration  = "ATLAS",
      journal        = "Phys. Lett. B",
      volume         = "716",
      year           = "2012",
      pages          = "1-29",
      doi            = "10.1016/j.physletb.2012.08.020",
      eprint         = "1207.7214",
      archivePrefix  = "arXiv",
      primaryClass   = "hep-ex",
      reportNumber   = "CERN-PH-EP-2012-218",
      SLACcitation   = "%%CITATION = ARXIV:1207.7214;%%"
}

@article{Chatrchyan:2012xdj,
      author         = "{CMS Collaboration}",
      title          = "{Observation of a new boson at a mass of 125 GeV with the
                        CMS experiment at the LHC}",
      collaboration  = "CMS",
      journal        = "Phys. Lett. B",
      volume         = "716",
      year           = "2012",
      pages          = "30-61",
      doi            = "10.1016/j.physletb.2012.08.021",
      eprint         = "1207.7235",
      archivePrefix  = "arXiv",
      primaryClass   = "hep-ex",
      reportNumber   = "CMS-HIG-12-028, CERN-PH-EP-2012-220",
      SLACcitation   = "%%CITATION = ARXIV:1207.7235;%%"
}

@article{Agostinelli:2002hh,
      author         = "Agostinelli, S. and others",
      title          = "{GEANT4 -- A Simulation toolkit}",
      collaboration  = "GEANT4",
      journal        = "Nucl. Instrum. Meth. A",
      volume         = "506",
      year           = "2003",
      pages          = "250-303",
      doi            = "10.1016/S0168-9002(03)01368-8",
      reportNumber   = "SLAC-PUB-9350, FERMILAB-PUB-03-339",
      SLACcitation   = "%%CITATION = NUIMA,A506,250;%%"
}

@article{deFlorian:2016spz,
      author         = "de Florian, D. and others",
      title          = "{Handbook of LHC Higgs Cross Sections: 4. Deciphering the
                        Nature of the Higgs Sector}",
      collaboration  = "LHC Higgs Cross Section Working Group",
      year           = "2016",
      eprint         = "1610.07922",
      archivePrefix  = "arXiv",
      primaryClass   = "hep-ph",
      reportNumber   = "FERMILAB-FN-1025-T",
      SLACcitation   = "%%CITATION = ARXIV:1610.07922;%%"
}

@article{Nason:2004rx,
      author         = "Nason, Paolo",
      title          = "{A new method for combining NLO QCD with shower Monte
                        Carlo algorithms}",
      journal        = "JHEP",
      volume         = "11",
      year           = "2004",
      pages          = "040",
      doi            = "10.1088/1126-6708/2004/11/040",
      eprint         = "hep-ph/0409146",
      archivePrefix  = "arXiv",
      reportNumber   = "BICOCCA-FT-04-11",
      SLACcitation   = "%%CITATION = HEP-PH/0409146;%%"
}

@article{Alwall:2014hca,
      author         = "Alwall, J. and Frederix, R. and Frixione, S. and Hirschi,
                        V. and Maltoni, F. and Mattelaer, O. and Shao, H. -S. and
                        Stelzer, T. and Torrielli, P. and Zaro, M.",
      title          = "{The automated computation of tree-level and
                        next-to-leading order differential cross sections, and
                        their matching to parton shower simulations}",
      journal        = "JHEP",
      volume         = "07",
      year           = "2014",
      pages          = "79",
      doi            = "10.1007/JHEP07(2014)079",
      eprint         = "1405.0301",
      archivePrefix  = "arXiv",
      primaryClass   = "hep-ph",
      reportNumber   = "CERN-PH-TH-2014-064, CP3-14-18, LPN14-066, MCNET-14-09,
                        ZU-TH-14-14",
      SLACcitation   = "%%CITATION = ARXIV:1405.0301;%%"
}

@article{Ball:2012cx,
      author         = "{NNPDF Collaboration, R.~D.~Ball et al.}",
      title          = "{Parton distributions with LHC data}",
      journal        = "Nucl. Phys. B",
      volume         = "867",
      year           = "2013",
      pages          = "244-289",
      doi            = "10.1016/j.nuclphysb.2012.10.003",
      eprint         = "1207.1303",
      archivePrefix  = "arXiv",
      primaryClass   = "hep-ph",
      reportNumber   = "EDINBURGH-2012-08, IFUM-FT-997, FR-PHENO-2012-014,
                        RWTH-TTK-12-25, CERN-PH-TH-2012-037, SFB-CPP-12-47",
      SLACcitation   = "%%CITATION = ARXIV:1207.1303;%%"
}

@article{Aad:2015yga,
      author         = "{ATLAS Collaboration}",
      title          = "{Search for Dark Matter in Events with Missing Transverse
                        Momentum and a Higgs Boson Decaying to Two Photons in $pp$
                        Collisions at $\sqrt{s}=8$ TeV with the ATLAS Detector}",
      collaboration  = "ATLAS",
      journal        = "Phys. Rev. Lett.",
      volume         = "115",
      year           = "2015",
      number         = "13",
      pages          = "131801",
      doi            = "10.1103/PhysRevLett.115.131801",
      eprint         = "1506.01081",
      archivePrefix  = "arXiv",
      primaryClass   = "hep-ex",
      reportNumber   = "CERN-PH-EP-2015-120",
      SLACcitation   = "%%CITATION = ARXIV:1506.01081;%%"
}

@article{Bertone:2004pz,
      author         = "Bertone, Gianfranco and Hooper, Dan and Silk, Joseph",
      title          = "{Particle dark matter: evidence, candidates and
                        constraints}",
      journal        = "Phys. Rept.",
      volume         = "405",
      year           = "2005",
      pages          = "279-390",
      doi            = "10.1016/j.physrep.2004.08.031",
      eprint         = "hep-ph/0404175",
      archivePrefix  = "arXiv",
      reportNumber   = "FERMILAB-PUB-04-047-A",
      SLACcitation   = "%%CITATION = HEP-PH/0404175;%%"
}

@article{Zyla:2020zbs,
    author = "Zyla, P.A. and others",
    collaboration = "Particle Data Group",
    title = "{Review of Particle Physics}",
    doi = "10.1093/ptep/ptaa104",
    journal = "PTEP",
    volume = "2020",
    number = "8",
    pages = "083C01",
    year = "2020"
}

@article{Sirunyan:2017hnk,
      author         = "{CMS Collaboration}",
      title          = "{Search for associated production of dark matter with a
                        Higgs boson decaying to $b\bar{b}$ or $\gamma\gamma$ at $\sqrt{s}$=13 TeV}",
      collaboration  = "CMS",
      year           = "2017",
      eprint         = "1703.05236",
      archivePrefix  = "arXiv",
      primaryClass   = "hep-ex",
      reportNumber   = "CMS-EXO-16-012, CERN-EP-2017-027",
      SLACcitation   = "%%CITATION = ARXIV:1703.05236;%%",
   volume={10},
   ISSN={1029-8479},
   DOI={10.1007/jhep10(2017)180},
   number={10},
   journal={JHEP},
   publisher={Springer Science and Business Media LLC},
      year={2017},
      month={Oct},
      pages={180}}

@article{monoh,
      author         = "Carpenter, Linda and DiFranzo, Anthony and Mulhearn,
                        Michael and Shimmin, Chase and Tulin, Sean and Whiteson,
                        Daniel",
      title          = "{Mono-Higgs-boson: A new collider probe of dark matter}",
      journal        = "Phys. Rev. D",
      volume         = "89",
      year           = "2014",
      number         = "7",
      pages          = "075017",
      doi            = "10.1103/PhysRevD.89.075017",
      eprint         = "1312.2592",
      archivePrefix  = "arXiv",
      primaryClass   = "hep-ph",
      SLACcitation   = "%%CITATION = ARXIV:1312.2592;%%"
}

@article{Aad:2015dva,
      author         = "{ATLAS Collaboration}",
      title          = "{Search for dark matter produced in association with a
                        Higgs boson decaying to two bottom quarks in $pp$
                        collisions at $\sqrt{s} = 8$ TeV with the ATLAS detector}",
      collaboration  = "ATLAS",
      journal        = "Phys. Rev. D",
      volume         = "93",
      year           = "2016",
      number         = "7",
      pages          = "072007",
      doi            = "10.1103/PhysRevD.93.072007",
      eprint         = "1510.06218",
      archivePrefix  = "arXiv",
      primaryClass   = "hep-ex",
      reportNumber   = "CERN-PH-EP-2015-247",
      SLACcitation   = "%%CITATION = ARXIV:1510.06218;%%"
}

@article{Aaboud:2016obm,
      author         = "{ATLAS Collaboration}",
      title          = "{Search for dark matter in association with a Higgs boson
                        decaying to $b$-quarks in $pp$ collisions at $\sqrt
                        s=13$ TeV with the ATLAS detector}",
      collaboration  = "ATLAS",
      journal        = "Phys. Lett. B",
      volume         = "765",
      year           = "2017",
      pages          = "11-31",
      doi            = "10.1016/j.physletb.2016.11.035",
      eprint         = "1609.04572",
      archivePrefix  = "arXiv",
      primaryClass   = "hep-ex",
      reportNumber   = "CERN-EP-2016-181",
      SLACcitation   = "%%CITATION = ARXIV:1609.04572;%%"
}

@article{lhcdmf,
 author         = "Abercrombie, Daniel and others",
 title          = "{Dark Matter Benchmark Models for Early LHC Run-2
                      Searches: Report of the ATLAS/CMS Dark Matter Forum}",
year           = "2015",
eprint         = "1507.00966",
archivePrefix  = "arXiv",
primaryClass   = "hep-ex",
reportNumber   = "FERMILAB-PUB-15-282-CD",
SLACcitation   = "%%CITATION = ARXIV:1507.00966;%%",
journal = "Physics of the Dark Universe",
volume = "27",
pages = "100371",
year = "2020",
issn = "2212-6864",
doi = "https://doi.org/10.1016/j.dark.2019.100371"
}

@article{Aad:2016jkr,
      author         = "{ATLAS Collaboration}",
      title          = "{Muon reconstruction performance of the ATLAS detector in
                        proton-proton collision data at $\sqrt{s}$ =13 TeV}",
      collaboration  = "ATLAS",
      journal        = "Eur. Phys. J. C",
      volume         = "76",
      year           = "2016",
      number         = "5",
      pages          = "292",
      doi            = "10.1140/epjc/s10052-016-4120-y",
      eprint         = "1603.05598",
      archivePrefix  = "arXiv",
      primaryClass   = "hep-ex",
      reportNumber   = "CERN-EP-2016-033",
      SLACcitation   = "%%CITATION = ARXIV:1603.05598;%%"
}

@article{Aaltonen:2008dn,
      author         = "{CDF Collaboration}",
      title          = "{Search for new particles decaying into dijets in
                        proton-antiproton collisions at $\sqrt{s}=1.96$ TeV}",
      journal        = "Phys. Rev. D",
      volume         = "79",
      year           = "2009",
      pages          = "112002",
      doi            = "10.1103/PhysRevD.79.112002",
      eprint         = "0812.4036",
      archivePrefix  = "arXiv",
      primaryClass   = "hep-ex",
      reportNumber   = "FERMILAB-PUB-08-572-E",
      SLACcitation   = "%%CITATION = ARXIV:0812.4036;%%"
}

@article{Aad:2012tba,
      author         = "{ATLAS Collaboration}",
      title          = "{Measurement of isolated-photon pair production in $pp$
                        collisions at $\sqrt{s}=7$ TeV with the ATLAS detector}",
      journal        = "JHEP",
      volume         = "01",
      year           = "2013",
      pages          = "086",
      doi            = "10.1007/JHEP01(2013)086",
      eprint         = "1211.1913",
      archivePrefix  = "arXiv",
      primaryClass   = "hep-ex",
      reportNumber   = "CERN-PH-EP-2012-300",
      SLACcitation   = "%%CITATION = ARXIV:1211.1913;%%"
}

@article{Crivellin:2014qxa,
      author         = "Crivellin, Andreas and D'Eramo, Francesco and Procura,
                        Massimiliano",
      title          = "{New Constraints on Dark Matter Effective Theories from
                        Standard Model Loops}",
      journal        = "Phys. Rev. Lett.",
      volume         = "112",
      year           = "2014",
      pages          = "191304",
      doi            = "10.1103/PhysRevLett.112.191304",
      eprint         = "1402.1173",
      archivePrefix  = "arXiv",
      primaryClass   = "hep-ph",
      reportNumber   = "UCB-PTH-14-01, CERN-PH-TH-2014-016",
      SLACcitation   = "%%CITATION = ARXIV:1402.1173;%%"
}

@article{DEramo:2014nmf,
      author         = "D'Eramo, Francesco and Procura, Massimiliano",
      title          = "{Connecting dark matter UV complete models to direct
                        detection rates via effective field theory}",
      journal        = "JHEP",
      volume         = "04",
      year           = "2015",
      pages          = "054",
      doi            = "10.1007/JHEP04(2015)054",
      eprint         = "1411.3342",
      archivePrefix  = "arXiv",
      primaryClass   = "hep-ph",
      reportNumber   = "UCB-PTH-14-38, UWTHPH-2014-27",
      SLACcitation   = "%%CITATION = ARXIV:1411.3342;%%"
}

@article{z2hdm,
 author         = "Berlin, Asher and Lin, Tongyan and Wang, Lian-Tao",
 title          = "{Mono-Higgs detection of dark matter at the LHC}",
 journal        = "JHEP",
 volume         = "06",
 year           = "2014",
 pages          = "078",
 doi            = "10.1007/JHEP06(2014)078",
 eprint         = "1402.7074",
 archivePrefix  = "arXiv",
 primaryClass   = "hep-ph",
 SLACcitation   = "%%CITATION = ARXIV:1402.7074;%%"
}

@Booklet{ATLAS-TDR-2010-19,
    author         = "{ATLAS Collaboration}",
    title          = "{ATLAS Insertable B-Layer Technical Design Report}",
    howpublished   = "ATLAS-TDR-19",
    url            = "https://cds.cern.ch/record/1291633",
    year           = "2010",
    related        = "ATLAS-TDR-2010-19-add",
    relatedstring  = "Addendum:",
}

@article{Catani:2007vq,
      author         = "Catani, Stefano and Grazzini, Massimiliano",
      title          = "{Next-to-Next-to-Leading-Order Subtraction Formalism in Hadron Collisions and its Application to Higgs-Boson Production at the Large Hadron Collider}",
      journal        = "Phys. Rev. Lett.",
      volume         = "98",
      year           = "2007",
      pages          = "222002",
      doi            = "10.1103/PhysRevLett.98.222002",
      eprint         = "hep-ph/0703012",
      archivePrefix  = "arXiv",
      primaryClass   = "hep-ph",
      SLACcitation   = "%%CITATION = HEP-PH/0703012;%%"
}

@article{Aad:2008zzm,
      author         = "{ATLAS Collaboration}",
      title          = "{The ATLAS Experiment at the CERN Large Hadron Collider}",
      collaboration  = "ATLAS",
      journal        = "JINST",
      volume         = "3",
      year           = "2008",
      pages          = "S08003",
      doi            = "10.1088/1748-0221/3/08/S08003",
      SLACcitation   = "%%CITATION = JINST,3,S08003;%%"                                                                                                                                               
}

@Article{Aaboud:2016leb,
    author         = "{ATLAS Collaboration}",
    title          = "{Performance of the ATLAS Trigger System in 2015}",
    journal        = "Eur. Phys. J. C",
    volume         = "77",
    year           = "2017",
    pages          = "317",
    doi            = "10.1140/epjc/s10052-017-4852-3",
    reportNumber   = "CERN-EP-2016-241",
    eprint         = "1611.09661",
    archivePrefix  = "arXiv",
    primaryClass   = "hep-ex",
}

@article{Mimasu:2015nqa,
      author         = "Mimasu, Ken and Sanz, Veronica and Williams, Ciaran",
      title          = "{Higher order QCD predictions for associated Higgs
                        production with anomalous couplings to gauge bosons}",
      journal        = "JHEP",
      volume         = "08",
      year           = "2016",
      pages          = "039",
      doi            = "10.1007/JHEP08(2016)039",
      eprint         = "1512.02572",
      archivePrefix  = "arXiv",
      primaryClass   = "hep-ph",
      SLACcitation   = "%%CITATION = ARXIV:1512.02572;%%"                                                                                                                                                      
}

@article{Luisoni:2013kna,
      author         = "Luisoni, Gionata and Nason, Paolo and Oleari, Carlo and
                        Tramontano, Francesco",
      title          = "{$HW^{\pm}$/HZ + 0 and 1 jet at NLO with the POWHEG BOX
                        interfaced to GoSam and their merging within MiNLO}",
      journal        = "JHEP",
      volume         = "10",
      year           = "2013",
      pages          = "083",
      doi            = "10.1007/JHEP10(2013)083",
      eprint         = "1306.2542",
      archivePrefix  = "arXiv",
      primaryClass   = "hep-ph",
      SLACcitation   = "%%CITATION = ARXIV:1306.2542;%%"
}

@article{Aad:2014xaa,
      author         = "{ATLAS Collaboration}",
      title          = "{Measurement of the $Z/\gamma^*$ boson transverse
                        momentum distribution in $pp$ collisions at $\sqrt{s}$ = 7 TeV with the ATLAS detector}",
      collaboration  = "ATLAS",
      journal        = "JHEP",
      volume         = "09",
      year           = "2014",
      pages          = "145",
      doi            = "10.1007/JHEP09(2014)145",
      eprint         = "1406.3660",
      archivePrefix  = "arXiv",
      primaryClass   = "hep-ex",
      reportNumber   = "CERN-PH-EP-2014-075",
      SLACcitation   = "%%CITATION = ARXIV:1406.3660;%%"                                                                                                                                                       
}

@article{pythia8,
       author         = "Sj{\"o}strand, Torbjorn and Mrenna, Stephen and Skands, Peter
       Z.",
       title          = "{A brief introduction to PYTHIA 8.1}",
       journal        = "Comput. Phys. Commun.",
       volume         = "178",
       year           = "2008",
       pages          = "852-867",
       doi            = "10.1016/j.cpc.2008.01.036",
       eprint         = "0710.3820",
       archivePrefix  = "arXiv",
       primaryClass   = "hep-ph",
       reportNumber   = "CERN-LCGAPP-2007-04, LU-TP-07-28,
       FERMILAB-PUB-07-512-CD-T",
       SLACcitation   = "%%CITATION = ARXIV:0710.3820;%%"
}

@Article{EvtGen,
    author         = "{D. J. Lange}",
    title          = "{The EvtGen particle decay simulation package}",
    journal        = "Nucl. Instrum. Meth. A",
    volume         = "462",
    year           = "2001",
    pages          = "152",
    doi            = "10.1016/S0168-9002(01)00089-4",
    primaryClass   = "hep-ex",
}

@Article{PhysRevD.101.012002,
  title = {Combined measurements of Higgs boson production and decay using up to 80 $fb^{-1}$ of proton-proton collision data at $\sqrt{s}= 13$ TeV collected with the ATLAS experiment},
  author = "{ATLAS Collaboration}",
  collaboration = {ATLAS Collaboration},
  journal = {Phys. Rev. D},
  volume = {101},
  pages = {012002},
  numpages = {48},
  year = {2020},
  month = {Jan},
  publisher = {American Physical Society},
  doi = {10.1103/PhysRevD.101.012002},
  url = {https://link.aps.org/doi/10.1103/PhysRevD.101.012002},
  eprint={1909.02845},
  archivePrefix={arXiv},
  primaryClass={hep-ex}
}

@Article{cmscollaboration2021measurements,
      title={Measurements of Higgs boson production cross sections and couplings in the diphoton decay channel at $\sqrt{s} = 13$ TeV}, 
      author="{CMS Collaboration}",
      collaboration={CMS Collaboration},
      year={2021},
      eprint={2103.06956},
      archivePrefix={arXiv},
      primaryClass={hep-ex}
}

@article{atlas_tMET,
   title={Search for large missing transverse momentum in association with one top-quark in proton-proton collisions at $\sqrt{s} = 13$ TeV with the ATLAS detector},
   volume={05},
   ISSN={1029-8479},
   url={http://dx.doi.org/10.1007/JHEP05(2019)041},
   DOI={10.1007/jhep05(2019)041},
   number={5},
   journal={JHEP},
   publisher={Springer Science and Business Media LLC},
   author="{ATLAS Collaboration}",
   year={2019},
   month={May},
   eprint={1812.09743},
   archivePrefix={arXiv},
   primaryClass={hep-ex},
   pages={041}

}

@article{atlas_ttbarMET,
      Author = "{ATLAS Collaboration}",
	Journal = {JHEP},
	Number = {4},
	Pages = {165},
	Title = {Search for new phenomena in events with two opposite-charge leptons, jets and missing transverse momentum in pp collisions at $\sqrt{s}= 13$ TeV with the ATLAS detector},
      DOI={10.1007/JHEP04(2021)165},
	Volume = {04},
	Year = {2021},
      eprint={2102.01444},
      archivePrefix={arXiv},
      primaryClass={hep-ex}
}

@article{Sirunyan_2020,
   title={Search for dark matter particles produced in association with a Higgs boson in proton-proton collisions at $\sqrt{s}= 13$ TeV},
   volume={03},
   ISSN={1029-8479},
   url={http://dx.doi.org/10.1007/JHEP03(2020)025},
   DOI={10.1007/jhep03(2020)025},
   number={3},
   journal={JHEP},
   publisher={Springer Science and Business Media LLC},
   author="{CMS Collaboration}",
   year={2020},
   pages={025},
   month={Mar},
   eprint={1908.01713},
   archivePrefix={arXiv},
   primaryClass={hep-ex}}

@Article{Cacciari:2008gp,
     author    = "Cacciari, Matteo and Salam, Gavin P. and Soyez, Gregory",
     title     = "{The anti-\(k_{t}\) jet clustering algorithm}",
     journal   = "JHEP",
     volume    = "04",
     year      = "2008",
     pages     = "063",
     eprint    = "0802.1189",
     archivePrefix = "arXiv",
     primaryClass  =  "hep-ph",
     doi       = "10.1088/1126-6708/2008/04/063",
     SLACcitation  = "%%CITATION = 0802.1189;%%"
}

@Article{Butterworth:2015oua,
      author         = "Butterworth, Jon and others",
      title          = "{PDF4LHC recommendations for LHC Run II}",
      journal        = "J. Phys. G",
      volume         = "43",
      year           = "2016",
      pages          = "023001",
      doi            = "10.1088/0954-3899/43/2/023001",
      eprint         = "1510.03865",
      archivePrefix  = "arXiv",
      primaryClass   = "hep-ph",
      reportNumber   = "OUTP-15-17P, SMU-HEP-15-12, TIF-UNIMI-2015-14,
                        LCTS-2015-27, CERN-PH-TH-2015-249",
      SLACcitation   = "%%CITATION = ARXIV:1510.03865;%%"
}

@article{Ball:2014uwa,
      author         = "Ball, Richard D. and others",
      title          = "{Parton distributions for the LHC run II}",
      collaboration  = "NNPDF",
      journal        = "JHEP",
      volume         = "04",
      year           = "2015",
      pages          = "040",
      doi            = "10.1007/JHEP04(2015)040",
      eprint         = "1410.8849",
      archivePrefix  = "arXiv",
      primaryClass   = "hep-ph",
      reportNumber   = "EDINBURGH-2014-15, IFUM-1034-FT, CERN-PH-TH-2013-253,
                        OUTP-14-11P, CAVENDISH-HEP-14-11",
      SLACcitation   = "%%CITATION = ARXIV:1410.8849;%%"
}

@article{Sjostrand:2014zea,
      author         = "Sj{\"o}strand, Torbj{\"o}rn and Ask, Stefan and Christiansen,
                        Jesper R. and Corke, Richard and Desai, Nishita and Ilten,
                        Philip and Mrenna, Stephen and Prestel, Stefan and
                        Rasmussen, Christine O. and Skands, Peter Z.",
      title          = "{An introduction to PYTHIA 8.2}",
      journal        = "Comput. Phys. Commun.",
      volume         = "191",
      year           = "2015",
      pages          = "159",
      doi            = "10.1016/j.cpc.2015.01.024",
      eprint         = "1410.3012",
      archivePrefix  = "arXiv",
      primaryClass   = "hep-ph",
      reportNumber   = "LU-TP-14-36, MCNET-14-22, CERN-PH-TH-2014-190,
                        FERMILAB-PUB-14-316-CD, DESY-14-178, SLAC-PUB-16122,
                        --FERMILAB-PUB-14-316-CD",
      SLACcitation   = "%%CITATION = ARXIV:1410.3012;%%"
}

@Article{Frixione:2008yi,
     author    = "Frixione, Stefano and Laenen, Eric and Motylinski, Patrick and White, Chris and Webber, Bryan R.",
     title     = "{Single-top hadroproduction in association with a \(W\) boson}",
     journal   = "JHEP",
     volume    = "07",
     year      = "2008",
     pages     = "029",
     eprint    = "0805.3067",
     archivePrefix = "arXiv",
     primaryClass  =  "hep-ph",
     doi       = "10.1088/1126-6708/2008/07/029"
}

@article{Artoisenet:2012st,
      author         = "Artoisenet, Pierre and Frederix, Rikkert and Mattelaer,
                        Olivier and Rietkerk, Robbert",
      title          = "{Automatic spin-entangled decays of heavy resonances in
                        Monte Carlo simulations}",
      journal        = "JHEP",
      volume         = "03",
      pages          = "015",
      doi            = "10.1007/JHEP03(2013)015",
      year           = "2013",
      eprint         = "1212.3460",
      archivePrefix  = "arXiv",
      primaryClass   = "hep-ph",
      reportNumber   = "NIKHEF-2012-021, CERN-PH-TH-2012-329",
      SLACcitation   = "%%CITATION = ARXIV:1212.3460;%%",
}

@Article{Bellm:2015jjp,
      author         = "Bellm, Johannes and others",
      title          = "{Herwig 7.0/Herwig++ 3.0 release note}",
      journal        = "Eur. Phys. J. C",
      volume         = "76",
      year           = "2016",
      number         = "4",
      pages          = "196",
      doi            = "10.1140/epjc/s10052-016-4018-8",
      eprint         = "1512.01178",
      archivePrefix  = "arXiv",
      primaryClass   = "hep-ph",
      reportNumber   = "CERN-PH-TH-2015-289, MAN-HEP-2015-15, IFJPAN-IV-2015-13,
                        HERWIG-2015-01, KA-TP-18-2015, DCPT-15-142, MCNET-15-28,
                        IPPP-15-71, --HERWIG-2015-01",
      SLACcitation   = "%%CITATION = ARXIV:1512.01178;%%"
}

@article{Bellm:2017jjp,
      author         = "Bellm, Johannes and others",
      title          = "{Herwig 7.1 Release Note}",
      eprint         = "1705.06919",
      year           = "2017",
      archivePrefix  = "arXiv",
      primaryClass   = "hep-ph",
      reportNumber   = "CERN-PH-TH-2017-109, MAN/HEP/2017/08, UWTHPH-2017-10, 
                        IFJPAN-IV-2017-7, NIKHEF 2017-026, HERWIG-2017-02, 
                        KA-TP-19-2017, MCnet-17-08, IPPP/17/40",
      SLACcitation   = "%%CITATION = ARXIV:1705.06919;%%"
}

@Article{Frixione:2007vw,
      author         = "Frixione, Stefano and Nason, Paolo and Oleari, Carlo",
      title          = "{Matching NLO QCD computations with parton shower
                        simulations: the POWHEG method}",
      journal        = "JHEP",
      volume         = "11",
      pages          = "070",
      doi            = "10.1088/1126-6708/2007/11/070",
      year           = "2007",
      eprint         = "0709.2092",
      archivePrefix  = "arXiv",
      primaryClass   = "hep-ph",
}

@Article{Alioli:2010xd,
      author         = "Alioli, Simone and Nason, Paolo and Oleari, Carlo and Re,
                        Emanuele",
      title          = "{A general framework for implementing NLO calculations in
                        shower Monte Carlo programs: the POWHEG BOX}",
      journal        = "JHEP",
      volume         = "06",
      pages          = "043",
      doi            = "10.1007/JHEP06(2010)043",
      year           = "2010",
      eprint         = "1002.2581",
      archivePrefix  = "arXiv",
      primaryClass   = "hep-ph",
}

@article{Hamilton:2012rf,
      author         = "Hamilton, Keith and Nason, Paolo and Oleari, Carlo and
                        Zanderighi, Giulia",
      title          = "{Merging H/W/Z + 0 and 1 jet at NLO with no merging
                        scale: a path to parton shower + NNLO matching}",
      journal        = "JHEP",
      volume         = "05",
      year           = "2013",
      pages          = "082",
      doi            = "10.1007/JHEP05(2013)082",
      eprint         = "1212.4504",
      archivePrefix  = "arXiv",
      primaryClass   = "hep-ph",
      reportNumber   = "CERN-PH-TH-2012-356",
      SLACcitation   = "%%CITATION = ARXIV:1212.4504;%%"
}

@article{Hamilton:2012np,
      author         = "Hamilton, Keith and Nason, Paolo and Zanderighi, Giulia",
      title          = "{MINLO: Multi-Scale Improved NLO}",
      journal        = "JHEP",
      volume         = "10",
      year           = "2012",
      pages          = "155",
      doi            = "10.1007/JHEP10(2012)155",
      eprint         = "1206.3572",
      archivePrefix  = "arXiv",
      primaryClass   = "hep-ph",
      reportNumber   = "CERN-PH-TH-2012-166, OUTP-12-11P, MCNET-12-07",
      SLACcitation   = "%%CITATION = ARXIV:1206.3572;%%"
}

@article{Hartanto:2015uka,
      author         = "Hartanto, Heribertus B. and J{\"a}ger, Barbara and Reina,
                        Laura and Wackeroth, Doreen",
      title          = "{Higgs boson production in association with top quarks in
                        the POWHEG BOX}",
      journal        = "Phys. Rev. D",
      volume         = "91",
      year           = "2015",
      number         = "9",
      pages          = "094003",
      doi            = "10.1103/PhysRevD.91.094003",
      eprint         = "1501.04498",
      archivePrefix  = "arXiv",
      primaryClass   = "hep-ph",
      SLACcitation   = "%%CITATION = ARXIV:1501.04498;%%"
}

@article{Frixione:2007nw,
      author         = "Frixione, Stefano and Nason, Paolo and Ridolfi, Giovanni",
      title          = "{A positive-weight next-to-leading-order Monte Carlo for
                        heavy flavour hadroproduction}",
      journal        = "JHEP",
      volume         = "09",
      pages          = "126",
      doi            = "10.1088/1126-6708/2007/09/126",
      year           = "2007",
      eprint         = "0707.3088",
      archivePrefix  = "arXiv",
      primaryClass   = "hep-ph",
      SLACcitation   = "%%CITATION = ARXIV:0707.3088;%%",
}

@Article{Hoeche:2009rj,
      author         = "H{\"o}che, Stefan and Krauss, Frank and Schumann, Steffen and Siegert, Frank",
      title          = "{QCD matrix elements and truncated showers}",
      journal        = "JHEP",
      volume         = "05",
      pages          = "053",
      doi            = "10.1088/1126-6708/2009/05/053",
      year           = "2009",
      eprint         = "0903.1219",
      archivePrefix  = "arXiv",
      primaryClass   = "hep-ph",
}

@Article{Gleisberg:2008fv,
      author         = "Gleisberg, Tanju and H{\"o}che, Stefan",
      title          = "{Comix, a new matrix element generator}",
      journal        = "JHEP",
      volume         = "12",
      pages          = "039",
      doi            = "10.1088/1126-6708/2008/12/039",
      year           = "2008",
      eprint         = "0808.3674",
      archivePrefix  = "arXiv",
      primaryClass   = "hep-ph",
}

@Article{Schumann:2007mg,
      author         = "Schumann, Steffen and Krauss, Frank",
      title          = "{A parton shower algorithm based on Catani--Seymour dipole factorisation}",
      journal        = "JHEP",
      volume         = "03",
      pages          = "038",
      doi            = "10.1088/1126-6708/2008/03/038",
      year           = "2008",
      eprint         = "0709.1027",
      archivePrefix  = "arXiv",
      primaryClass   = "hep-ph",
}

@article{Hoeche:2012yf,
      author         = "H{\"o}che, Stefan and Krauss, Frank and Sch{\"o}nherr, Marek and
                        Siegert, Frank",
      title          = "{QCD matrix elements + parton showers. The NLO case}",
      journal        = "JHEP",
      volume         = "04",
      year           = "2013",
      pages          = "027",
      doi            = "10.1007/JHEP04(2013)027",
      eprint         = "1207.5030",
      archivePrefix  = "arXiv",
      primaryClass   = "hep-ph",
      reportNumber   = "SLAC-PUB-15191, IPPP-12-52, DCPT-12-104, LPN12-081,
                        FR-PHENO-2012-017, MCNET-12-09, --FR-PHENO-2012-017",
      SLACcitation   = "%%CITATION = ARXIV:1207.5030;%%"
}

@article{Bothmann:2019yzt,
      author = "Bothmann, Enrico and others",
      title = "{Event generation with Sherpa 2.2}",
      journal = "SciPost Phys.",
      volume = "7",
      year = "2019",
      number = "3",
      pages = "034",
      doi = "10.21468/SciPostPhys.7.3.034",
      reportNumber = "FERMILAB-PUB-19-218-T, SLAC-PUB-17433, IPPP/19/42, MCNET-19-11",
      eprint = "1905.09127",
      archivePrefix = "arXiv",
      primaryClass = "hep-ph",
}

@article{Hoeche:2011fd,
      author         = "H{\"o}che, Stefan and Krauss, Frank and Sch{\"o}nherr, Marek and
                        Siegert, Frank",
      title          = "{A critical appraisal of NLO+PS matching methods}",
      journal        = "JHEP",
      volume         = "09",
      year           = "2012",
      pages          = "049",
      doi            = "10.1007/JHEP09(2012)049",
      eprint         = "1111.1220",
      archivePrefix  = "arXiv",
      primaryClass   = "hep-ph",
      reportNumber   = "SLAC-PUB-14661, IPPP-11-67, DCPT-11-134, LPN11-58,
                        FR-PHENO-2011-019, MCNET-11-24",
      SLACcitation   = "%%CITATION = ARXIV:1111.1220;%%"
}

@article{Catani:2001cc,
      author         = "Catani, S. and Krauss, F. and Kuhn, R. and Webber, B. R.",
      title          = "{QCD Matrix Elements + Parton Showers}",
      journal        = "JHEP",
      volume         = "11",
      year           = "2001",
      pages          = "063",
      doi            = "10.1088/1126-6708/2001/11/063",
      eprint         = "hep-ph/0109231",
      archivePrefix  = "arXiv",
      reportNumber   = "CERN-TH-2000-367, CAVENDISH-HEP-00-03",
      SLACcitation   = "%%CITATION = HEP-PH/0109231;%%"
}

@article{Cascioli:2011va,
      author         = "Cascioli, Fabio and Maierh{\"o}fer, Philipp and Pozzorini, Stefano",
      title          = "{Scattering Amplitudes with Open Loops}",
      journal        = "Phys. Rev. Lett.",
      volume         = "108",
      year           = "2012",
      pages          = "111601",
      doi            = "10.1103/PhysRevLett.108.111601",
      eprint         = "1111.5206",
      archivePrefix  = "arXiv",
      primaryClass   = "hep-ph",
      reportNumber   = "ZU-TH-23-11, LPN11-66",
      SLACcitation   = "%%CITATION = ARXIV:1111.5206;%%"
}

@article{Buccioni:2019sur,
    author = "Buccioni, Federico and Lang, Jean-Nicolas and Lindert, Jonas M. and Maierh{\"o}fer, Philipp and Pozzorini, Stefano and Zhang, Hantian and Zoller, Max F.",
    title = "{OpenLoops 2}",
    eprint = "1907.13071",
    archivePrefix = "arXiv",
    primaryClass = "hep-ph",
    reportNumber = "IPPP/19/62, FR-PHENO-2019-12, PSI-PR-19-15, ZU-TH 37/19",
    doi = "10.1140/epjc/s10052-019-7306-2",
    journal = "Eur. Phys. J. C",
    volume = "79",
    number = "10",
    pages = "866",
    year = "2019"
}

@article{Denner:2016kdg,
      author         = "Denner, Ansgar and Dittmaier, Stefan and Hofer, Lars",
      title          = "{\textsc{Collier}: A fortran-based complex one-loop library in
                        extended regularizations}",
      journal        = "Comput. Phys. Commun.",
      volume         = "212",
      year           = "2017",
      pages          = "220-238",
      doi            = "10.1016/j.cpc.2016.10.013",
      eprint         = "1604.06792",
      archivePrefix  = "arXiv",
      primaryClass   = "hep-ph",
      reportNumber   = "FR-PHENO-2016-003, ICCUB-16-016",
      SLACcitation   = "%%CITATION = ARXIV:1604.06792;%%"
}

@article{Hamilton:2013fea,
      author         = "Hamilton, Keith and Nason, Paolo and Re, Emanuele and
                        Zanderighi, Giulia",
      title          = "{NNLOPS simulation of Higgs boson production}",
      journal        = "JHEP",
      volume         = "10",
      year           = "2013",
      pages          = "222",
      doi            = "10.1007/JHEP10(2013)222",
      eprint         = "1309.0017",
      archivePrefix  = "arXiv",
      primaryClass   = "hep-ph",
      reportNumber   = "MCNET-13-11, CERN-PH-TH-2013-205, OUTP-13-18P",
      SLACcitation   = "%%CITATION = ARXIV:1309.0017;%%"
}

@article{Campbell:2012am,
      author         = "Campbell, John M. and Ellis, R. Keith and Frederix,
                        Rikkert and Nason, Paolo and Oleari, Carlo and Williams,
                        Ciaran",
      title          = "{NLO Higgs Boson Production Plus One and Two Jets Using
                        the POWHEG BOX, MadGraph4 and MCFM}",
      journal        = "JHEP",
      volume         = "07",
      year           = "2012",
      pages          = "092",
      doi            = "10.1007/JHEP07(2012)092",
      eprint         = "1202.5475",
      archivePrefix  = "arXiv",
      primaryClass   = "hep-ph",
      reportNumber   = "FERMILAB-PUB-12-040-T, CERN-PH-TH-2012-048",
      SLACcitation   = "%%CITATION = ARXIV:1202.5475;%%"
}

@article{Bozzi:2005wk,
      author         = "Bozzi, Giuseppe and Catani, Stefano and de Florian,
                        Daniel and Grazzini, Massimiliano",
      title          = "{Transverse-momentum resummation and the spectrum of the
                        Higgs boson at the LHC}",
      journal        = "Nucl. Phys. B",
      volume         = "737",
      year           = "2006",
      pages          = "73-120",
      doi            = "10.1016/j.nuclphysb.2005.12.022",
      eprint         = "hep-ph/0508068",
      archivePrefix  = "arXiv",
      primaryClass   = "hep-ph",
      SLACcitation   = "%%CITATION = HEP-PH/0508068;%%"
}

@article{deFlorian:2011xf,
      author         = "de Florian, Daniel and Ferrera, Giancarlo and Grazzini,
                        Massimiliano and Tommasini, Damiano",
      title          = "{Transverse-momentum resummation: Higgs boson production
                        at the Tevatron and the LHC}",
      journal        = "JHEP",
      volume         = "11",
      year           = "2011",
      pages          = "064",
      doi            = "10.1007/JHEP11(2011)064",
      eprint         = "1109.2109",
      archivePrefix  = "arXiv",
      primaryClass   = "hep-ph",
      reportNumber   = "ZU-TH-17-11",
      SLACcitation   = "%%CITATION = ARXIV:1109.2109;%%"
}

@article{Nason:2009ai,
      author         = "Nason, Paolo and Oleari, Carlo",
      title          = "{NLO Higgs boson production via vector-boson fusion
                        matched with shower in POWHEG}",
      journal        = "JHEP",
      volume         = "02",
      year           = "2010",
      pages          = "037",
      doi            = "10.1007/JHEP02(2010)037",
      eprint         = "0911.5299",
      archivePrefix  = "arXiv",
      primaryClass   = "hep-ph",
      SLACcitation   = "%%CITATION = ARXIV:0911.5299;%%"
}

@article{Demartin:2016axk,
    author = "Demartin, Federico and Maier, Benedikt and Maltoni, Fabio and Mawatari, Kentarou and Zaro, Marco",
    title = "{tWH associated production at the LHC}",
    eprint = "1607.05862",
    archivePrefix = "arXiv",
    primaryClass = "hep-ph",
    reportNumber = "MCNET-16-30, CP3-16-40, LPSC16158",
    doi = "10.1140/epjc/s10052-017-4601-7",
    journal = "Eur. Phys. J. C",
    volume = "77",
    number = "1",
    pages = "34",
    year = "2017"
}

@Article{Cowan:2010js,
   Author = {Cowan, Glen and Cranmer, Kyle and Gross, Eilam and Vitells, Ofer},
   Title = {{Asymptotic formulae for likelihood-based tests of new physics}},
   Journal = {Eur. Phys. J. C},
   Volume = {71},
   Year = {2011},
   Pages = {1554},
   doi = "10.1140/epjc/s10052-011-1554-0",
   Eprint = "1007.1727",
   Archiveprefix = {arXiv},
   Primaryclass = {physics.data-an},
   related = "Cowan:2010js-err",
   relatedstring  = "Erratum:",
}

@Article{SOFT-2010-01,
    author         = "{ATLAS Collaboration}",
    title          = "{The ATLAS Simulation Infrastructure}",
    journal        = "Eur. Phys. J. C",
    volume         = "70",
    year           = "2010",
    pages          = "823",
    doi            = "10.1140/epjc/s10052-010-1429-9",
    eprint         = "1005.4568",
    archivePrefix  = "arXiv",
    primaryClass   = "physics.ins-det",
}

@Article{HIGG-2013-08,
    author         = "{ATLAS Collaboration}",
    title          = "{Measurement of Higgs boson production in the diphoton decay channel in \(pp\) collisions at center-of-mass energies of \(7\) and \(8\,\text{TeV}\) with the ATLAS detector}",
    journal        = "Phys. Rev. D",
    volume         = "90",
    year           = "2014",
    pages          = "112015",
    doi            = "10.1103/PhysRevD.90.112015",
    reportNumber   = "CERN-PH-EP-2014-198",
    eprint         = "1408.7084",
    archivePrefix  = "arXiv",
    primaryClass   = "hep-ex",
}

@Article{HIGG-2014-14,
    author         = "{ATLAS and CMS Collaborations}",
    title          = "{Combined Measurement of the Higgs Boson Mass in \(pp\) Collisions at \(\sqrt{s} = 7\) and \(8\,\text{TeV}\) with the ATLAS and CMS Experiments}",
    journal        = "Phys. Rev. Lett.",
    volume         = "114",
    year           = "2015",
    pages          = "191803",
    doi            = "10.1103/PhysRevLett.114.191803",
    eprint         = "1503.07589",
    archivePrefix  = "arXiv",
    primaryClass   = "hep-ex",
}

@Article{PERF-2014-07,
    author         = "{ATLAS Collaboration}",
    title          = "{Topological cell clustering in the ATLAS calorimeters and its performance in LHC Run~1}",
    journal        = "Eur. Phys. J. C",
    volume         = "77",
    year           = "2017",
    pages          = "490",
    doi            = "10.1140/epjc/s10052-017-5004-5",
    reportNumber   = "CERN-PH-EP-2015-304",
    eprint         = "1603.02934",
    archivePrefix  = "arXiv",
    primaryClass   = "hep-ex",
}

@Article{PERF-2015-09,
    author         = "{ATLAS Collaboration}",
    title          = "{Jet reconstruction and performance using particle flow with the ATLAS Detector}",
    journal        = "Eur. Phys. J. C",
    volume         = "77",
    year           = "2017",
    pages          = "466",
    doi            = "10.1140/epjc/s10052-017-5031-2",
    reportNumber   = "CERN-EP-2017-024",
    eprint         = "1703.10485",
    archivePrefix  = "arXiv",
    primaryClass   = "hep-ex",
}

@Article{HIGG-2016-08,
    author         = "{ATLAS Collaboration}",
    title          = "{Search for resonances in diphoton events at \(\sqrt{s} = 13\,\text{TeV}\) with the ATLAS detector}",
    journal        = "JHEP",
    volume         = "09",
    year           = "2016",
    pages          = "001",
    doi            = "10.1007/JHEP09(2016)001",
    reportNumber   = "CERN-EP-2016-120",
    eprint         = "1606.03833",
    archivePrefix  = "arXiv",
    primaryClass   = "hep-ex",
}

@Article{PERF-2017-01,
    author         = "{ATLAS Collaboration}",
    title          = "{Electron reconstruction and identification in the ATLAS experiment using the 2015 and 2016 LHC proton--proton collision data at \(\sqrt{s} = 13\,\text{TeV}\)}",
    journal        = "Eur. Phys. J. C",
    volume         = "79",
    year           = "2019",
    pages          = "639",
    doi            = "10.1140/epjc/s10052-019-7140-6",
    reportNumber   = "CERN-EP-2018-273",
    eprint         = "1902.04655",
    archivePrefix  = "arXiv",
    primaryClass   = "hep-ex",
}

@Article{PIX-2018-001,
    author         = "Abbott, B. and others",
    title          = "{Production and integration of the ATLAS Insertable B-Layer}",
    journal        = "JINST",
    volume         = "13",
    year           = "2018",
    pages          = "T05008",
    doi            = "10.1088/1748-0221/13/05/T05008",
    eprint         = "1803.00844",
    archivePrefix  = "arXiv",
    primaryClass   = "physics.ins-det",
}

@Booklet{ATL-SOFT-PUB-2021-003,
    author         = "{ATLAS Collaboration}",
    title          = "{ATLAS Computing Acknowledgements}",
    howpublished   = "{ATL-SOFT-PUB-2021-003}",
    url            = "https://cds.cern.ch/record/2776662",
}

@Booklet{ATLAS-CONF-2019-021,
    author         = "{ATLAS Collaboration}",
    title          = "{Luminosity determination in \(pp\) collisions at \(\sqrt{s} = 13\,\text{TeV}\) using the ATLAS detector at the LHC}",
    howpublished   = "{ATLAS-CONF-2019-021}",
    url            = "https://cds.cern.ch/record/2677054",
    year           = "2019",
}

@Booklet{ATL-PHYS-PUB-2014-021,
    author         = "{ATLAS Collaboration}",
    title          = "{ATLAS Pythia~8 tunes to \(7~\text{TeV}\) data}",
    howpublished   = "{ATL-PHYS-PUB-2014-021}",
    url            = "https://cds.cern.ch/record/1966419",
    year           = "2014",
}

@Booklet{ATL-PHYS-PUB-2016-017,
    author         = "{ATLAS Collaboration}",
    title          = "{The Pythia~8 A3 tune description of ATLAS minimum bias and inelastic measurements incorporating the Donnachie--Landshoff diffractive model}",
    howpublished   = "{ATL-PHYS-PUB-2016-017}",
    url            = "https://cds.cern.ch/record/2206965",
    year           = "2016",
}

@Booklet{ATL-PHYS-PUB-2017-022,
    author         = "{ATLAS Collaboration}",
    title          = "{Electron and photon reconstruction and performance in ATLAS using a dynamical, topological cell clustering-based approach}",
    howpublished   = "{ATL-PHYS-PUB-2017-022}",
    url            = "https://cds.cern.ch/record/2298955",
    year           = "2017",
}

\clearpage 
 
\begin{flushleft}
\hypersetup{urlcolor=black}
{\Large The ATLAS Collaboration}

\bigskip

\AtlasOrcid[0000-0002-6665-4934]{G.~Aad}$^\textrm{\scriptsize 101}$,    
\AtlasOrcid[0000-0002-5888-2734]{B.~Abbott}$^\textrm{\scriptsize 127}$,    
\AtlasOrcid[0000-0002-7248-3203]{D.C.~Abbott}$^\textrm{\scriptsize 102}$,    
\AtlasOrcid[0000-0002-2788-3822]{A.~Abed~Abud}$^\textrm{\scriptsize 36}$,    
\AtlasOrcid[0000-0002-1002-1652]{K.~Abeling}$^\textrm{\scriptsize 53}$,    
\AtlasOrcid[0000-0002-2987-4006]{D.K.~Abhayasinghe}$^\textrm{\scriptsize 93}$,    
\AtlasOrcid[0000-0002-8496-9294]{S.H.~Abidi}$^\textrm{\scriptsize 29}$,    
\AtlasOrcid[0000-0002-8279-9324]{O.S.~AbouZeid}$^\textrm{\scriptsize 40}$,    
\AtlasOrcid{N.L.~Abraham}$^\textrm{\scriptsize 155}$,    
\AtlasOrcid[0000-0001-5329-6640]{H.~Abramowicz}$^\textrm{\scriptsize 160}$
% ...
\bigskip
\\

% ...
$^{29}$Physics Department, Brookhaven National Laboratory, Upton NY; United States of America.\\
% ...
$^{36}$CERN, Geneva; Switzerland.\\
% ...
$^{40}$Niels Bohr Institute, University of Copenhagen, Copenhagen; Denmark.\\
% ...
$^{53}$II. Physikalisches Institut, Georg-August-Universit\"{a}t G\"ottingen, G\"ottingen; Germany.\\
% ...
$^{93}$Department of Physics, Royal Holloway University of London, Egham; United Kingdom.\\
% ...
$^{101}$CPPM, Aix-Marseille Universit\'e, CNRS/IN2P3, Marseille; France.\\
$^{102}$Department of Physics, University of Massachusetts, Amherst MA; United States of America.\\
% ...
$^{127}$Homer L. Dodge Department of Physics and Astronomy, University of Oklahoma, Norman OK; United States of America.\\
% ...
$^{155}$Department of Physics and Astronomy, University of Sussex, Brighton; United Kingdom.\\
% ...
$^{160}$Raymond and Beverly Sackler School of Physics and Astronomy, Tel Aviv University, Tel Aviv; Israel.\\
% ...

\end{flushleft}


\end{document}